\documentclass[letterpaper,11pt]{article}

\usepackage{psfig, amsmath, epsfig, amssymb,array,bbold}
\usepackage[usenames]{color}
\usepackage{sidecap}
\usepackage{graphicx}
\usepackage{bm}
\usepackage{enumerate,textcomp}
\def\ptmiss{p\!\!\!\slash_{T}}

\def\pslash{\not{\hbox{\kern-4pt $p$}}}
\def\qslash{\not{\hbox{\kern-4pt $q$}}}
\def\lv{\not{\hbox{\kern-4pt $L$}}}
\def\lsim{\mathrel{\raise.3ex\hbox{$<$\kern-.75em\lower1ex\hbox{$\sim$}}}}
\def\gsim{\mathrel{\raise.3ex\hbox{$>$\kern-.75em\lower1ex\hbox{$\sim$}}}}
\def\ifmath#1{\relax\ifmmode #1\else $#1$\fi}

\definecolor{DarkRed}{rgb}{0.55,0.00,0.00}

\newcommand{\nc}{\newcommand}
\nc{\postscript}[2]{\setlength{\epsfxsize}{#2\hsize}\centerline{\epsfbox{#1}}}

\nc{\beq}{\begin{equation}}   \nc{\eeq}{\end{equation}}
\nc{\bea}{\begin{eqnarray}}   \nc{\eea}{\end{eqnarray}}
\nc{\baa}{\begin{array}}      \nc{\eaa}{\end{array}}
\nc{\bit}{\begin{itemize}}    \nc{\eit}{\end{itemize}}
\nc{\ben}{\begin{enumerate}}  \nc{\een}{\end{enumerate}}
\nc{\bce}{\begin{center}}     \nc{\ece}{\end{center}}

\nc{\non}{\nonumber}

\begin{document} 

\baselineskip=17pt


\thispagestyle{empty}
\vspace{20pt}
\font\cmss=cmss10 \font\cmsss=cmss10 at 7pt


\hfill

\begin{center}
{\Large \textbf
{Dark Baryogenesis}}
\end{center}

\vspace{15pt}

\begin{center}
{\large Devin G. E. Walker} \\
\vspace{15pt}
\textit{SLAC National Accelerator Laboratory, 2575 Sand Hill Road, Menlo Park, CA 94025, U.S.A. \\ 
and \\ 
Center for the Fundamental Laws of Nature, Jefferson
  Physical Laboratory, Harvard University, Cambridge, MA 02138,
  U.S.A.} 
\end{center}

\vspace{3pt}

\begin{center}
\textbf{Abstract}
\end{center}
\vspace{3pt} {\small \noindent 
We first suggested a scenario in which a generic, dark chiral gauge group undergoes a first order phase transition in order to generate the observed baryon asymmetry in the universe, provide a viable dark matter candidate and explain the observed baryon-to-dark matter ratio of relic abundances~\cite{Agashe:2010gt}.  We now provide a model in which a copy of the electroweak gauge group is added to the Standard Model.  We spontaneously break this new gauge group to the diagonal $Z_2$ center which is used to stabilize a dark matter candidate.  In addition to the dark matter candidate,  anomaly free messenger fermions are included which transform non-trivially under all the gauge groups.  In analogy to electroweak baryogenesis, the model generates an excess of messenger ``baryons."  These ``baryons" subsequently decay to the Standard Model and dark matter to generate an excess of Standard Model baryons.  The baryon-to-dark matter number density ratio is ultimately due to the requirement of gauge anomaly freedom.  Dark sphalerons generate operators which violate baryon minus lepton (B - L) number but preserves baryon plus lepton number (B + L).  This ensures any baryon asymmetry generated by the dark phase transition will not be washed out by the Standard Model.
The model radiatively generates a dark matter mass of order of the electroweak vacuum expectation value suppressed by a loop factor ($\sim v_\mathrm{ew}/16\pi^2$) therefore setting the dark matter-to-baryon relic abundance.  We outline some distinctive experimental signatures~\cite{Agashe:2010gt,Walker:2009en,Walker:2009ei,Agashe:2010tu} and ensure these models are consistent with existing constraints.  Notably, as first discussed in~\cite{Walker:2009en}, these dark matter scenarios feature long-lived particles which can be observed at colliders.  We finally show how approximate global symmetries in the higgs sector stabilize both the dark and electroweak scales thereby mitigating the hierarchy problem.  Thus, the SM and dark higgses can have natural masses that are one- and two-loop suppressed, respectively. 
The latter mass is due to ``overlapping" chiral symmetries which generalizes the Little Higgs mechanism and forces the dark higgs mass to be generated at three loops.  In this model, a light dark higgses are needed to ensure the observed relic abundance. }
\vfill\eject
\noindent


\section{Introduction}
\label{sec:intro}

There is a compelling amount of experimental evidence for the existence of dark matter (DM) in the universe \cite{Bertone:2004pz}.  
A host of astrophysical, cosmological and direct detection experiments provide a basic profile:  A viable DM candidate must be electrically neutral, colorless, non-relativistic at redshifts of $z \sim 3000$, and generate the measured relic abundance of~\cite{Hinshaw:2008kr},
\begin{equation}
h^2\, \Omega_{\mathrm{DM}} = 0.1131\pm0.0034.
\label{eq:DMrelicabundance}
\end{equation}
In addition, a dark matter candidate once in thermal equilibrium with the Standard Model (SM) during the early universe is theoretically well motivated \cite{Bertone:2004pz}.  The abundance for thermally populated dark matter relics is correlated with the annihilation cross section~\cite{Kolb:1990vq} by 
\begin{equation}
h^{2}\,\Omega_\mathrm{DM} \,\simeq \frac{0.1 \,\,\mathrm{pb} \cdot c}{\langle \sigma v \rangle},
\label{eq:annhilcross}
\end{equation}
where $c$ is the speed of light and $\langle \sigma v \rangle$ is the thermally averaged annihilation cross section.  This cross section goes as
\begin{equation}
\langle \sigma v \rangle \simeq \frac{g^4}{ 8\pi } \frac{1}{ M^2},
\end{equation}
for a pair of dark matter particles annihilating into a two-particle massless final state.  So long as $g$ is of order the strength of the weak coupling then $M$ can range from a few GeV to several TeV.  It is thus widely expected that the question of the origin and nature of dark matter is correlated with new physics at the unexplored TeV scale.  Because of this and the general experimental requirements above, many models of new physics at the TeV scale propose dark matter candidates.  Regardless of the model, all of the proposed dark matter candidates share three important features:  Each candidate is stable because it transforms non-trivially under a new symmetry.  In addition, the dark matter candidate is typically the lightest new particle which transforms under the new symmetry.  Finally, the dark matter candidate is neutral or very weakly interacts with SM particles.
\newline
\newline
In many respects, baryons are similar to dark matter candidates.  They are forbidden from decaying into lighter leptons by baryon number conservation.  Moreover, the relic abundance of luminous baryons is  measured as~\cite{Hinshaw:2008kr},
\begin{equation} 
h^2\, \Omega_{\mathrm{b}} = 0.02267^{+0.00058}_{-0.00059},
\end{equation}
which differs from the dark matter relic abundance by almost exactly by a factor of five\footnote{Baryons, as defined by cosmologists, actually comprise all of the nuclei and leptons in the universe.  The leptons, however, are only a small fraction of the overall total mass; and the measured ratio of the baryon density to the critical density $\rho_b/\rho_{cr} \sim \Omega_b$ is dominated by protons and neutrons.}.  Because of the similarity of these numbers, it has been often thought that both the baryon and dark matter relic abundances might have a common origin origin~\cite{Barr:1990ca,Kaplan:1991ah,Dodelson:1991iv}.  
\newline
\newline
Baryons and anti-baryons have equal status under the fundamental equations.  Yet, anti-baryons are not observed in large quantities in the universe~\cite{Kolb:1990vq,Cline:2006ts}.  Given no excess of primordial baryon number, a fundamental process in the early universe that generates a baryon asymmetry is needed.  In the 1960s, Sakharov~\cite{Sakharov:1967dj} outlined the conditions\footnote{For a discussion on reasons behind each of these requirements, please see an especially clear treatment in \cite{Cline:2006ts}. } needed to create such an asymmetry from a baryon-anti-baryon symmetric universe: A violation of baryon number (B), charge conjugation plus parity (CP) and charge conjugation (C) symmetries is required; as well, a departure from thermal equilibrium is needed.
A phase transition which generates the observed baryon asymmetry must satisfy all of the Sakharov conditions. 
\newline
\newline
Naively the weak interactions have all of the elements to satisfy Sakharov's requirements: The electroweak sphalerons, \textit{i.e.}, unstable instanton solutions to the field equations at high temperatures, are unsuppressed and violate baryon number~\cite{Kuzmin:1985mm}.  The weak interactions break C and CP.  To generate the needed departure from thermal equilibrium, a strong first order phase transition is necessary for the electroweak phase transition~\cite{Cohen:1990it,Cohen:1990py}.  Beyond Sakarov's conditions, it is extremely interesting that electroweak baryogenesis can potentially be tested with the CERN Large Hadron Collider (LHC).  There are problems, however, with the electroweak baryogenesis paradigm.  It is well known that additional sources of CP violation beyond the SM are needed~\cite{Cohen:1993nk} to generate the observed baryon asymmetry.  Any additional source of CP violation beyond the SM has been indirectly probed with precision measurements of the electric dipole moments (EDM) of cold neutrons, Thalium and Mercury atoms.  The results have yielded scant evidence for additional CP violating phases~\cite{Pospelov:2005pr}.  Moreover, direct searches of the SM higgs boson at CERN Large Electron Positron Collider (LEP) have yielded a lower bound on its mass of $m > 114.4$ GeV~\cite{Barate:2003sz}.  Indeed, the LHC has recently found hints of the electroweak higgs around 126 GeV~\cite{atlashiggsresults,CMShiggsresults}.  The larger the higgs mass, the less likely it is to obtain a strong first order phase transition~\cite{Cohen:1993nk}.  When considering only the relevant, tree-level terms in the SM higgs potential, a strong first order phase transition can occur only for higgs masses of $m < 32$ GeV~\cite{Cline:2006ts} which is far below the LEP bound.  It has been noted that new physics just beyond the SM can induce an effective higgs potential that can give a first order phase transition~\cite{Grojean:2004xa}.  Such new physics can be (or has already been) probed directly by the LHC and the Fermilab TeVatron Collider (TeVatron).
\newline
\newline
For completeness and to place our current work in historical context, we  briefly note there are other often-discussed scenarios beyond electroweak baryogenesis that propose to satisfy Sakharov's requirements. Grand unified theories feature interactions that explicitly break baryon number.  However, any net baryons produced by these processes at the unification scale, $\Lambda \simeq 10^{16-18}$ GeV, can be erased by the non-perturbative processes generated by the weak interactions down to scales of $\Lambda \simeq 10^2$.   As mentioned before, at high temperatures above the weak scale, these processes are unsuppressed~\cite{Cohen:1993nk}. By far the most popular mechanism for baryogenesis today is leptogenesis~\cite{Fukugita:1986hr}, where an excess of leptons is generated which is then, in turn, converted to baryons by, \textit{e.g.}, electroweak sphalerons.  Often the leptogenesis mechanism occurs at scales far beyond what can be tested by colliders in the foreseeable future.  There have been many attempts to explain weak scale baryogenesis within the context of models which inherently have dark matter candidates.  The minimum supersymmetric standard model (MSSM) has been widely studied for the potential to generate the baryon asymmetry~\cite{Carena:2008vj}.  This scenario requires a light stop particle which has not been observed by the TeVatron or (so far at) the LHC~\cite{CDFstopref,Abazov:2008kz}.  Another supersymmetric scenario is the Affleck-Dine mechanism which generates a baryon number asymmetry though the presence of flat directions in the scalar potential~\cite{Affleck:1984fy}.  This mechanism is thought to be present at scales far above a TeV.  See, \textit{e.g.}, \cite{Cline:2006ts,Dine:2003ax,Buchmuller:2004nz, Enqvist:2003gh,Dolgov:1991fr} for a reviews of all of these topics.  In the past, there have also been attempts to describe the ratio of $\Omega_{DM}$ to $\Omega_b$; these studies add a dark matter candidate plus additional messenger particles so the electroweak phase transition can generate the $\Omega_{DM}/\Omega_b$ ratio~\cite{Kaplan:1991ah,Barr:1990ca,Dodelson:1991iv,Nussinov:1985xr,Chivukula:1989qb,Barr:1991qn}.  The asymmetric dark matter idea of~\cite{Barr:1990ca} has recently been the focus of much recent work.  (See \textit{e.g.}~\cite{Kaplan:2009ag}.)  An earlier phase transition was used to explain the baryon asymmetry in \cite{Shu:2006mm}.  In that scenario, the earlier phase transition violates baryon plus lepton number (B + L) like the electroweak interactions.  Thus, electroweak processes can potentially erase any baryon excess generated by the new phase transition.  Finally, a hidden non-abelian gauge group was used to drive baryogenesis in~\cite{Dutta:2010va}.
\newline
\newline
In a previous paper~\cite{Agashe:2010gt}, we first described a scenario in which a new, generic chiral gauge group, added to the Standard Model, could generate the observed baryon asymmetry as well as produce a viable dark matter candidate.  This new ``dark" gauge group undergoes a phase transition that spontaneously breaks the chiral gauge symmetry to the discrete $Z_N$  center.   The discrete symmetry is used to stabilize a dark matter candidate~\cite{Walker:2009en}.  In comparison to previous attempts to provide an explanation for the $\Omega_{DM}/\Omega_b$ ratio, the addition of this new dark phase transition is novel.  In this work we describe how many of the problems that have constrained (or ruled out)  previous attempts to explain the observed baryon asymmetry, the dark matter relic abundance and the observed ratio $\Omega_{DM}/\Omega_b$ can be simply solved with the addition of the new chiral gauge group.  Moreover, adding this new chiral gauge group, has definitive experimental consequences which can be potentially measured at the LHC~\cite{Walker:2009ei}. 
\newline
\newline
The outline of the paper is as follows:  In the next section we first summarize the basic features of this class of models.  
Next, in Section~\ref{sec:model}, we detail a simple model in which a copy of the electroweak gauge group is spontaneously broken to the $Z_2$ center.  There we also show how chiral symmetries protecting the dark matter mass naturally generate a mass near a GeV.  We also include a discussion on how the model non-perturbatively generates operators which induce a baryon asymmetry; this asymmetry is not erased by background electroweak processes.  The reader who first prefers to focus on model specifics rather than the broad outlines of this class of models may prefer to start with Section~\ref{sec:model} and later read Section~\ref{sec:blue}. In Section~\ref{sec:bDMabn}, we show how the decays of new heavy baryons into the dark matter and the SM can give an explanation for why the ratio of number densities is $n_{DM}/n_b \sim 5$.  As described above, generally models which try to explain the baryon asymmetry near the weak scale invoke large CP violating phases that are testable with EDMs.  We show in Section~\ref{sec:exp} this model is less likely to generate EDMs and is poorly constrained by those experimental constraints.  In this work, we emphasize how having a new chiral gauge group that is spontaneously broken provides new, uniquely testable consequences.   Section~\ref{sec:exp} broadly outlines the distinctive experimental signatures for these models.   Afterwards, we conclude.  Notably, in Appendix B, we demonstrate how approximate symmetries can stabilize the dark and electroweak scales.  This provides a solution to the hierarchy problem.   Appendices A and C support the discussion in Section~\ref{sec:bDMabn}.

\section{The Blueprint}
\label{sec:blue}

In this section we summarize the essential points of the class of models described in this paper.  Later, we review our rationale for constructing models which feature chiral gauge symmetries that are spontaneously broken to discrete symmetries.  All of the points made in this section are discussed in much greater detail in the later sections.
%
\subsection{Model Summary} 
This class of models share the following features.
\begin{enumerate}
\newcounter{enum_saved}
\item A non-abelian chiral gauge group is added to the SM gauge group.  The new dark chiral gauge group is spontaneously broken to the diagonal $Z_N$ center.  The $Z_N$ center is used to stabilize dark matter\footnote{Here $N$ is the dimension of the smallest fundamental representation of all of the simple, non-abelian groups in the chiral gauge group.}.
\setcounter{enum_saved}{\value{enumi}}
\end{enumerate}
In the forthcoming simple model, we will take a copy of the electroweak gauge group to be our ÒdarkÓ gauge group, $SU(2)_D \times U(1)_D$.  We break this chiral symmetry to the diagonal $Z_2$ center.  
\begin{enumerate}
\setcounter{enumi}{\value{enum_saved}}
\item Anomaly free messenger fermions are added which transform non-trivially under the dark and SM gauge groups.  Anomaly free fermions that are charged only under the dark chiral symmetry are also added;  after the dark phase transition, these fermions serve as the dark matter.
\setcounter{enum_saved}{\value{enumi}}
\end{enumerate} 
Additionally, we require:
\begin{enumerate}
\setcounter{enumi}{\value{enum_saved}}
\item The dark gauge group undergoes a strong first order phase transition.
The dark phase transition generates an asymmetry of messenger and dark matter fermions.  
\setcounter{enum_saved}{\value{enumi}}
\end{enumerate}   
The dark phase transition generates expanding bubbles of the asymmetric vacuum within regions of false vacua.  Because of the first order phase transition, a departure from thermal equilibrium is achieved at the bubble interface.  There non-perturbative processes generated by the non-abelian gauge group violate the global symmetries.  In total, this mechanism~\cite{Cohen:1990it,Cohen:1990py,Cohen:1993nk}, generates the dark fermion asymmetry. 
\begin{enumerate}
\setcounter{enumi}{\value{enum_saved}}
\item The dark phase transition violates a linear combination of the global baryon, lepton, and possibly dark matter number so any asymmetry generated is not erased by the weak interactions.
\setcounter{enum_saved}{\value{enumi}}
\end{enumerate}  
We detail a simple model where global dark matter number is explicitly broken to lepton number; the non-perturbative operators generated by the dark gauge group preserve baryon plus lepton number ($B + L$) and violate baryon minus lepton number ($B - L$).  This is the inverse of what happens with the electroweak gauge group; therefore, the generated dark asymmetry is not erased.  Having an asymmetry of stable dark, heavy baryons and leptons does not describe the world we observe today.  We therefore require:
\begin{enumerate} 
\setcounter{enumi}{\value{enum_saved}}
\item The heavy baryons decay to SM baryons and dark matter; similarly, the heavy leptons decay to SM leptons and dark matter.
\item The dark gauge quantum numbers are chosen so that the messenger fermions are charged under the $Z_N$ symmetry after the dark phase transition.
\setcounter{enum_saved}{\value{enumi}}
\end{enumerate} 
Requiring the messenger fermions to be charged under the $Z_N$ ensures (at least) one dark matter particle in the final state of the decay.  This also allows the messenger fermions to be decoupled from the SM (at the one loop level) and thereby minimizes the sensitivity of these fermions to precision electroweak and direct constraints.  Because the heavy messenger particles decay to dark matter and the SM, we finally require:
\begin{enumerate}
\setcounter{enumi}{\value{enum_saved}}
\item The decay of the heavy messenger particles sets the relative number density, $n_\mathrm{DM}/n_\mathrm{baryon}$.  Often this means the messenger particles are relatively long-lived on collider time scales.
\setcounter{enum_saved}{\value{enumi}}
\end{enumerate}
This requirement also means that the contributions to the dark matter number density from sources other than the heavy messenger fermion decay is negligible.  We first described how the observation of long-lived particles plus dark matter at the LHC could provide detailed information on the nature of a dark sector~\cite{Walker:2009en}.  The long-lived particles produced distinct, unique signatures which can be easily seen~\cite{Walker:2009en}.  The coming model is constructed to generate $n_\mathrm{DM}/n_\mathrm{baryon} \sim 5$.  Because,
\begin{equation}
\Omega_\mathrm{DM}/\Omega_\mathrm{baryon} \equiv (n_\mathrm{DM}\,m_\mathrm{DM})/(n_\mathrm{baryon} m_\mathrm{baryon}),
\label{eq:fundamratio}
\end{equation}
the number density ratio implies a GeV dark matter mass.

\subsubsection{Why $n_\mathrm{DM}/n_\mathrm{baryon} = 5$ and $m_\mathrm{DM} \sim 1$ GeV}

The symmetries which govern the decays of the heavy messenger fermions can set $n_\mathrm{DM}/n_\mathrm{baryon} = 5$.  To see this, we first note the SM charges of the messenger fermions must be anomaly free.  It is well known that the SM fits into $SU(5)$ multiplets.  The smallest anomaly-free $SU(5)$ representations are $5\, \oplus\, \overline{5}$ and $5\, \oplus\, \overline{10}$.  Consider the former\footnote{The arguments presented in this section work for any anomaly-free $SU(5)$ multiplet with the fundamental $5$ representation.}.  
%
The messenger fermions must be chiral under the dark gauge group.  We choose the simple non-abelian chiral gauge group, $SU(2)_D \times U(1)_D$, to be the dark gauge group.  The anomaly-free 
messenger fermion content, assuming yukawa couplings to $SU(2)_D$ higgs doublets, 
\textbf{\textit{must}} have the form
\begin{equation}
\psi = (5,2)_0\, \oplus \, (\overline{5},1)_{a} \, \oplus \,(\overline{5},1)_{-a},
\end{equation}
where $a$ are the $U(1)_D$ charges.  The entries in parenthesis are representations of $(SU(5), SU(2)_D)$.  Eliminating the $SU(2)$ Witten anomalies requires minimally adding additional dark matter fermions in the representations
\begin{equation}
\chi = (1,2)_0\, \oplus \, (1,1)_{b} \, \oplus \,(1,1)_{-b}.
\end{equation}
The important point here is that the %
$SU(2)_D$ gauge group generates non-perturbative sphaleron operators involving the $(5,2)_0$ and $(1,2)_0$ fermions.  This brings us to our next point:
%
\begin{enumerate}
\setcounter{enumi}{\value{enum_saved}}
\item Each dark sphaleron generates five messenger fermions (one dark baryon and two dark leptons) which subsequently decays to five dark matter particles and one SM baryon.
\setcounter{enum_saved}{\value{enumi}}
\end{enumerate}
To be complete, two SM leptons (and five anti-leptons) are also generated by the messenger fermions decay.  As mentioned in the previous subsection, we require the decay of the messenger fermions to set the dark matter and baryonic relic abundances.  We show the proper decays of the messenger fermions are essentially determined by assuming electroweak and dark gauge group unification.  We detail the necessary conditions to ensure this mechanism in Section~\ref{sec:bDMabn}.
\newline
\newline
Given the arguments near equation~\ref{eq:fundamratio}, our dark matter mass must have a GeV mass in order for the model to generate the observed ratio of dark matter-to-baryon relic abundance.  
In the following, we detail a simple, dynamical mechanism which generates this mass.  
\begin{enumerate}
\setcounter{enumi}{\value{enum_saved}}
\item The dark matter is protected from getting a mass from the dark phase transition by global chiral symmetries.  The last remnant of this symmetry is broken by the SM higgs vev.
\setcounter{enum_saved}{\value{enumi}}
\end{enumerate}
A consequence of this is the dark matter radiatively generates a loop suppressed mass.  In more detail, the dark matter transforms non-trivially under both the dark and electroweak global chiral symmetries.  The diagram that generates the dark matter mass has messenger leptons as well as an electroweak higgs in the loop; the SM higgs and the dark matter are on the external legs of the diagram.  The messenger leptons get a mass, break the chiral symmetry associated with the dark sector and are integrated out.  The global electroweak chiral symmetry remains and protects the dark matter mass.   Upon electroweak symmetry breaking, this remaining chiral symmetry is broken and the dark matter gets a mass.  In all,
\begin{enumerate}
\setcounter{enumi}{\value{enum_saved}}
\item The dark matter mass is of order $v_\mathrm{ew}/16 \pi^2$.
\setcounter{enum_saved}{\value{enumi}}
\end{enumerate}

\subsubsection{Collider Signatures and Constraints}

Above, we have outlined how a model in which a dark chiral gauge group is spontaneously broken to a discrete symmetry can generate $\Omega_\mathrm{DM}/\Omega_\mathrm{baryon} \sim 5$.  This particular scenario inherently yields distinctive experimental signatures.  Discovery of the new dark gauge bosons would provide incontrovertible evidence.  Because these new gauge bosons are not charged under the SM, their discovery by colliders is dependent on messenger particles charged under both the SM and the dark gauge group.  As we will show in Sections~\ref{bDMabn} and~\ref{sec:exp}, these models require messenger fermions whose lifetime is often long-lived in comparison to the time for a relativistic particle to transverse a detector at the LHC.  In~\cite{Walker:2009en}, we described how long-lived particles could be a bellwether for characterizing the properties of dark matter particles.  We also showed for models of this type:
\begin{enumerate}
\setcounter{enumi}{\value{enum_saved}}
\item A defining signature is the observation of long-lived particles produced along with large amounts of missing momentum at colliders. 
\setcounter{enum_saved}{\value{enumi}}
\end{enumerate}
The missing momentum is generated by dark gauge bosons and dark matter.  The dark gauge bosons are radiated from the long-lived messenger fermions.  It is possible that the dark gauge bosons are heavier than the messenger fermions.  In that case, the colliders would dominantly observe long-lived particles.  This signature is the defining one for this class of models.  We detail the signature in Section~\ref{sec:exp}.
\newline
\newline
If this mechanism is realized in a larger theory such as supersymmetry, then other complementary signatures are possible.  Previously~\cite{Agashe:2010gt,Agashe:2010tu}, we have detailed some prominent collider signatures for SM partners which radiate off massive dark gauge bosons.
\begin{enumerate}
\setcounter{enumi}{\value{enum_saved}}
\item SM partners, which decay promptly into the SM and dark matter, can radiate off dark gauge bosons which generates unique decay topologies with large amounts of missing momentum. 
%
%
\setcounter{enum_saved}{\value{enumi}}
\end{enumerate}
\begin{enumerate}
\setcounter{enumi}{\value{enum_saved}}
\item The decay topologies generate unique signals that can be largely identified.
\setcounter{enum_saved}{\value{enumi}}
\end{enumerate}
For brevity, we do not further consider these signatures since they were discussed in detail in~\cite{Agashe:2010gt,Agashe:2010tu}.
\newline
\newline
\noindent
As described in the previous subsection, we want our dark phase transition to address the observed baryon asymmetry.  In analogy with electroweak baryogenesis, we require new large CP-violating phases.  We include soft CP-violating terms in the dark higgs sector that generate large CP-violating phases for the dark phase transition.  The result is:
\begin{enumerate}
\setcounter{enumi}{\value{enum_saved}}
\item The CP-violating phases in the dark sector are decoupled from the SM.  Neutron, thallium and mercury EDMs experiments therefore have reduced sensitivity to the dark CP violating phases.  
\setcounter{enum_saved}{\value{enumi}}
\end{enumerate}
In Section~\ref{sec:edm}, we provide details that large CP violating phases in the higgs sector are minimally constrained by experiment thereby allowing large phases. 

\subsubsection{Additional Notable Points} 
 
Generically, dark higgses which break the dark chiral symmetry can be arranged to transform as fundamentals.  Thus these models have a  ``complementarity"~\cite{Dimopoulos:1980hn,Fradkin:1978dv} between the higgs and strongly coupled phases.  For the strongly coupled scenario, a dynamical explanation for dark symmetry breaking scale is possible.  As a final note, the model detailed in the the coming sections adds a copy of the electroweak gauge group to the SM.  Additionally new dark higgses complement the SM higgs(-es).  Consider unification of the dark and electroweak gauge groups into a larger gauge group that is spontaneously broken at a high scale.  The framework allows:
\begin{enumerate}
\setcounter{enumi}{\value{enum_saved}}
\item New global and/or discrete symmetries can be added so the one-loop divergences of the SM higgs are cancelled by the dark sector in order to ameliorate the hierarchy problem.
\setcounter{enum_saved}{\value{enumi}}
\end{enumerate}
This can be done in analogy to the little and twin higgs models~\cite{ArkaniHamed:2002qy,Chacko:2005pe}.  Writing a model which does this requires moving out of the primary focus of this paper.  To this end, we outline a scenario to accomplish this in Appendix B.  There we use ``overlapping" chiral divergences to stabilize the dark and electroweak scales.
%
%
%
                 
\subsection{A Motivation for Breaking Chiral Symmetries to the $Z_N$ Center}

It is a general truism in particle physics that global symmetries are only approximately conserved.  
Now, dark matter candidates must have a lifetime at least equal to the age of the universe; and, since global symmetries are almost always explicitly broken, it seems reasonable to focus on gauge symmetries as a first step in considering candidate stabilization symmetries for dark matter.  
\newline
\newline
We should note, however, that \textbf{\textit{all}} known gauge symmetries undergo a phase transition.  Moreover, the conserved symmetries observed at low energies, whether global or gauged, are a consequence of these phase transitions.  
We take seriously this hint and consider the symmetry that stabilizes dark matter to be the result of a phase transition.  In~\cite{Walker:2009en}, we used this hint as motivation to show that a non-abelian gauge theory could be broken to a discrete symmetry\footnote{See \cite{FileviezPerez:2010gw, Batell:2010bp} for recent work in which the dark matter stabilization symmetry was motivated by  gauge theories.} and therefore provide a viable dark matter candidate.  Because of the strong theoretical connection between the weak scale and dark matter, the aim was to eventually use a copy of the electroweak gauge group to generate a dark matter candidate.  In \cite{Agashe:2010gt}, we presented this idea and proposed using the phase transition from dark chiral symmetry breaking to provide an explanation for the existing baryon asymmetry in the universe.  In the current work
, we show this framework provides a simple and efficient mechanism for explaining the ratio of the dark matter to baryon relic abundances as well as the observed dark matter relic abundance.

\section{A Simple Model}
\label{sec:model}

In this section, we describe in detail a particular model in which a new  $SU(2)_D \times U(1)_D$ gauge group is spontaneously broken to the $Z_2$ center.  Although we focus on one model, the results can be extended to other models where a new chiral gauge group with a non-abelian subgroup that is spontaneously broken to a discrete center\footnote{Most of the results can also be applied to models where the center is broken as well.}. 

\subsection{Dark Scalar Sector}
\label{subsec:scalar}

In this subsection we construct the dark scalar sector.  In order to spontaneously break the $SU(2)_D \times U(1)_D$ gauge group to the diagonal $Z_2$ subgroup, we require two new higgs fields
\begin{align}
h_1 = \begin{pmatrix} \eta_2 + i \,\eta_3 \\ \eta_0 + i \,\eta_1 \end{pmatrix} & & h_2 = \begin{pmatrix} \xi_0 + i\, \xi_1 \\ \xi_2 + i \,\xi_3  \end{pmatrix}.
\label{eq:higgscomponents}
\end{align}  
Both higgses are charged under $SU(2)_D \times U(1)_D$ as
\begin{align}
\vec{T}\, h_{1,2} = {\vec{\tau} \over 2} \,h_{1,2} && S\,h_{1,2} = {1 \over 2}\, h_{1,2} 
\label{eq:charges}
\end{align}
where $\vec{\tau}$ are the $SU(2)_D$ Pauli sigma matrices and $S$ the $U(1)_D$ generator\footnote{Because the higgses transform under the fundamental of SU(2), the model has an equivalent ``complementary" picture as a strongly coupled theory~\cite{Dimopoulos:1980hn,Fradkin:1978dv}.  The results in the low energy effective theory are the same whether the UV theory is perturbative or strongly coupled.}.  
To break the $SU(2)_D \times U(1)_D$ gauge group to $Z_2$, we require $h_{1,2}$ to get the following vacuum expectation values (vevs)
\begin{align}
\langle h_1 \rangle = \begin{pmatrix} 0 \\ v_1 \end{pmatrix} & & \langle h_2 \rangle = \begin{pmatrix} v_2 \\ 0  \end{pmatrix}.
\label{eq:higgsvev}
\end{align}
The infinitesimal $SU(2)_D \times U(1)_D$ transformation, 
\begin{eqnarray}
\delta h_{1,2} = \biggl( i \epsilon_a T^a\, h_{1,2} + {i \over 2} \epsilon\,h_{1,2}\biggr) \biggl|_{h_1 = \langle h_1 \rangle, \,\,h_2 = \langle h_2 \rangle} \neq 0 \label{eq:infh}
\end{eqnarray}
makes it clear that the vevs in equation~\ref{eq:higgsvev} completely break $SU(2)_D$ and $U(1)_D$.  The vevs, however, preserve the diagonal $Z_2$ center.  

\subsubsection{Invariance of the Diagonal $Z_2$ Center}
\label{subsec:z2center}

The center of a group is defined as the invariant subgroup that commutes with all of the elements in that particular group.  In analogy with hypercharge and custodial $SU(2)_R$, we presume $U(1)_D$ is a gauged subgroup of a larger SU(2) global symmetry.  Thus, the $Z_2$ center for both $SU(2)_D$ and $U(1)_D$ is 
\begin{equation}
V_k =  \begin{pmatrix} e^{i\pi} & \\  & e^{i\pi} \end{pmatrix}.
\end{equation}
Here the diagonal entries are determined by the requirement that $\det V = 1$.  
Consider a finite $SU(2)_D \times U(1)_D$ transformation 
\begin{equation}
h_{1,2} \to  e^{i S}\, e^{i \epsilon_a T^a}\, h_{1,2} \Rightarrow  V_2\, V_1\, \, h_{1,2} \to e^{i\pi\, \mathbb{1}} e^{i\pi\,\mathbb{1}}\, \,h_{1,2} \to h_{1,2}.
\label{eq:higgscenterinv}
\end{equation}
It is now clear that \textit{individually} both the $SU(2)_D$ and $U(1)_D$ centers are broken by the vevs in equation~\ref{eq:higgsvev}.  Only the diagonal center remains invariant.  Moreover, for the unbroken diagonal center, any state that has an odd number of fundamental $SU(2)_D$ and $U(1)_D$ indices will be odd under this discrete symmetry\footnote{The reader may realize that these same arguments apply to the SM and the electroweak higgs.  The higgs vev breaks the electroweak  gauge group to $SU(2)_L \times U(1)_Y \to U(1)_\mathrm{em} \times Z_2$.  This $Z_2$ makes all of the right-handed fermions odd and is explicitly broken by the top yukawa coupling.  It is often mentioned in the literature that one of the biggest mysteries is why the top quark mass is so much larger than the other quarks.  Using this line of reasoning, the other fermion masses are simply protected by an approximate $Z_2$ discrete symmetry.}.  

%

\subsubsection{Dark Scalar Potential}
\label{subsec:scalarpotential}

The vevs in equation~\ref{eq:higgsvev} require the vacuum to align so as to drive $\langle h_1^\dagger \,h_2 \rangle$ to zero,
\begin{equation}
\langle h_1^\dagger \,h_2 \rangle \to 0.
\label{eq:align2}
\end{equation}
To find the correct vacuum-aligned scalar potential, we first write down all possible relevant terms consistent with the $SU(2)_D \times U(1)_D$ gauge symmetries
\begin{eqnarray}
V &=& {1 \over 2} m_1^2\, h_1^\dagger h_1 + \lambda_1 \,(h_1^\dagger h_1)^2 + {1 \over 2} \,m_2^2\, h_2^\dagger h_2 + \lambda_2 \,(h_2^\dagger h_2)^2 +  \lambda_3\,h_1^\dagger h_1 \,h_2^\dagger h_2 
 \label{eq:origpotential} \\
 &+& {\lambda_4 \over 2}\biggl(h_1^\dagger h_2 \,h_1^\dagger h_2 + h_2^\dagger h_1\,h_2^\dagger h_1 \biggr)+  \lambda_5 \,h_1^\dagger h_2\,h_2^\dagger h_1 \nonumber.
 \end{eqnarray}  
Similar to electroweak two higgs doublet models, we have imposed a parity symmetry
\begin{equation}
h_1 \to - h_1.
\label{eq:h1parity}
\end{equation}
The symmetry simplifies the analysis by forbidding terms 
like $h_1^\dagger\, h_2$,  $h_1^\dagger\, h_2\, h_2^\dagger \,h_2$ and $h_1^\dagger\, h_2\, h_1^\dagger \,h_1$ which generate tadpoles.  
Relaxing the imposition of this parity can give a viable model; however, one needs more effort is needed to ensure the correct vacuum.  If we take $m_i^2$ in equation~\ref{eq:origpotential} to be negative, the general scalar potential has two other possible minima in addition to equation~\ref{eq:higgsvev}
\begin{align}
\langle h_1 \rangle = \begin{pmatrix} 0 \\ v_1 \end{pmatrix} \,\,\,\,\, \langle h_2 \rangle = \begin{pmatrix} 0 \\ i\,v_2 \end{pmatrix}, && \langle h_1\rangle = \begin{pmatrix} 0 \\ v_1 \end{pmatrix} \,\,\,\,\, \langle h_2 \rangle = \begin{pmatrix} 0 \\ v_2 \end{pmatrix}.
\label{eq:othervevs}
\end{align}
We redefine the $\lambda_i$ couplings in equation~\ref{eq:origpotential} to generate the following aligned potential 
\begin{eqnarray}
V &=& {\lambda_1 \over 4} \biggl( h_1^\dagger h_1  - v_1^2 \biggr)^2 + {\lambda_2 \over 4} \biggl( h_2^\dagger h_2  - v_2^2 \biggr)^2 + \lambda_3 \biggl( h_1^\dagger h_1  + h_2^\dagger h_2  - v_1^2 - v_2^2 \biggr)^2 \label{eq:alignedpotent}   \\
 &+&   {\lambda_4 \over 2}\biggl(h_1^\dagger h_2  + h_2^\dagger h_1 \biggr)^2 + \lambda_5 \,h_1^\dagger h_2\,h_2^\dagger h_1 \nonumber
\end{eqnarray}
which makes the minimum in equation~\ref{eq:higgsvev} be the true minimum with the proviso that all $\lambda_i$ are positive.  
\newline
\newline
Because the dark higgs potential breaks $SU(2)_D \times U(1)_D \to Z_2$, there remains four uneaten goldstone bosons which get masses.  We choose to work in a unitary gauge so that the second line in equation~\ref{eq:alignedpotent} disappears.  Explicitly,
\begin{align}
h_1 = \begin{pmatrix} 0 \\ v_1 + \eta \end{pmatrix} & & h_2 = \begin{pmatrix} v_2 + \xi \\ 0  \end{pmatrix}.
\label{eq:gb}
\end{align}  
where $\langle \eta \rangle = \langle \xi \rangle = 0$, $\eta \equiv \eta_0' + i\,\eta_1'$ and $\xi \equiv \xi_0' +i\,\xi_1'$.  The masses of $\eta$ and $\xi$ are described in Section~\ref{subsec:higgsportal}.  We will drop the primes when further discussing these scalars.
\newline
\newline
Ultimately, we want to generate a large baryon asymmetry from a first order dark phase transition.  As described in section~\ref{sec:blue}, we may require large CP violating phases in the dark sector.  To generate the phases, we introduce soft CP violating terms into the dark higgs sector in the next subsection.  
%

\subsubsection{Dark CP Violating Terms}
\label{subsec:darkCPterms}

We now add CP violating terms to the higgs potential.  We define
 \begin{align}
 \tilde{h}_1 = i \sigma_2\, h_1^*  &&  \tilde{h}_2 = -i \sigma_2 \,h_2^*.
 \label{eq:htilde}
 \end{align}
The true minimum with a CP violating phase is  
\begin{align}
\langle h_1 \rangle = \begin{pmatrix} 0 \\ v_1\,e^{i\beta} \end{pmatrix} & & \langle h_2 \rangle = \begin{pmatrix} v_2 \\ 0  \end{pmatrix}.
\label{eq:higgsvevw/CPviol}
\end{align}
 Thus the most general potential with soft CP violation is
\begin{eqnarray}
V_{CP} &=&  \lambda _6 \biggl(\mathrm{Re}[ h^\dagger_1\, \tilde{h}_2 ] - v_1 v_2 \cos{\beta} \biggr)\biggl(\mathrm{Re}[ h^\dagger_2\, \tilde{h}_1 ] - v_1 v_2 \cos{\beta}  \biggr) \label{eq:CPviolation} \\
&+& \lambda _7 \biggl(\mathrm{Im}[ h^\dagger_1\, \tilde{h}_2 ] - v_1 v_2 \sin{\beta} \biggr)\biggl(\mathrm{Im}[ h^\dagger_2\, \tilde{h}_1 ] + v_1 v_2 \sin{\beta}  \biggr) \nonumber
\end{eqnarray}
where $\lambda_{6,7}$ are positive semi-definite.  The $\lambda_6$ and $\lambda_7$ mass terms softly breaks the imposed parity symmetry in equation~\ref{eq:h1parity}.  The $\lambda_7$ breaks CP; all of the rest of the terms in the potential are invariant under CP.  Thus, all radiatively generated operators which violate CP will be proportional to $\lambda_7 \,v_1 v_2$.  
All of the terms in our scalar potential, equations~\ref{eq:alignedpotent} and \ref{eq:CPviolation}, are zero when evaluated at the minimum with the CP violating phase.

\subsubsection{Higgs Portal to the Standard Model}
\label{subsec:higgsportal}

It is straightforward to couple the dark sector with the SM higgs.  The additional contributions to the scalar potential can be written as
\begin{eqnarray}
V' &=&  {\lambda_8 \over 4} \biggl( h_1^\dagger h_1  - v_1^2 \biggr)\biggl(\phi^2 - v^2 \biggr) + {\lambda_9 \over 4} \biggl( h_2^\dagger h_2  - v_2^2 \biggr)\biggl(\phi^2 - v^2 \biggr) \label{eq:drkSMhiggspotential}\\
 &+&  {\lambda_{10} \over 4} \,\biggl(\phi^2 - v^2 \biggr)^2 \nonumber
\end{eqnarray}
where $v$ the electroweak vev and $\phi$ is the SM higgs.  Again, the imposed parity symmetry in equation~\ref{eq:h1parity} simplifies the potential.  It is straightforward to expand the potential to determine the tree-level masses for scalar particles in the higgs multiplet.
However, since scalar particles generate large quadratically divergent corrections to their mass, we postpone discussing the higgs mass spectrum until Appendix B.  There we outline a solution to the hierarchy problem.  Finally, the higgs spectrum will be important as we require a small abundance from the relic dark matter.  As a reminder, we require the dark matter abundance to result from the decays of the messenger fermions.  Thus, the annihilation of the relic dark matter into the SM (before the messenger fermions decay) must be efficient as possible.  In order increase the annihilation efficiency, we require relatively light dark higgses. 
%

\subsection{Fermion Sector}
\label{sec:fermion}

To the SM we add new, messenger chiral fermions which transform as
\begin{eqnarray}
\psi &=&  (5,2)_0 + (\bar{5},1)_{1/2} + (\bar{5},1)_{-1/2}  \label{eq:fermioncontent} \\
        &+& (1,2)_0 + (1,1)_{1/2} + (1,1)_{-1/2}. \nonumber
\end{eqnarray}
Here we just use compact notation where the SM and dark gauge groups are gauged.  In parenthesis, the first entry denotes representations in the $SU(5)$ unification group.\footnote{See for example section 18.6 in \cite{Georgi:1982jb}.}  The second entry and subscript denote representations of $SU(2)_D$ and $U(1)_D$, respectively.  In this notation, it is clear the fermion content is gauge anomaly free.  Going forward, we discuss the fermion content in both the $SU(5)$ representation of equation~\ref{eq:fermioncontent} and also explicitly in the dark and SM gauge representations.  To standardize the notation, the same set of left-handed fermions transform as 
\begin{align}
Q_L &=  (3,1,2)_{(-1/3,0)} & U &= (\bar{3},1,1)_{(1/3,1/2)} & D &= (\bar{3},1,1)_{(1/3,-1/2)} \label{eq:chiralquarks} \\
L_L &=  (1,2,2)_{(1/2,0)} & N &= (1,2,1)_{(-1/2,1/2)} & E &= (1,2,1)_{(-1/2,-1/2)} \label{eq:chirallepton} \\ \nonumber  \\
\chi_L &= (1,1,2)_{(0,0)} & \chi_{R_1} &= (1,1,1)_{(0,1/2)} & \chi_{R_2} &= (1,1,1)_{(0,-1/2)} \label{eq:chiraldark}.
\end{align}
where the entries in parenthesis are representations transforming under $SU(3)_c$, $SU(2)_L$ and $SU(2)_D$, respectively.  The subscripts entries are $U(1)_Y$ and $U(1)_D$, respectively.  All of the new fermions listed above are odd under the $Z_2$ center and left-handed.  As described in Section~\ref{subsec:z2center}, this is because each of the fermions transform non-trivially under either $SU(2)_D$ or $U(1)_D$ but not both.  We assume a mass hierarchy so the $\chi_L$ and $\chi_{R_i}$ fermions will be the lightest of the new states in order to effectively serve as the dark matter.  We will generally assume the new quarks will be heavier than the new leptons.

\subsubsection{Yukawa Couplings}

In this subsection, we consider all possible yukawa couplings and their implications for any preserved global symmetries.  As a reminder, 
 the dark and SM higgses transform as 
\begin{align}
h_{1,2} = (1,1,2)_{(0,1/2)} && \phi = (1,2,1)_{(-1/2,0)} ,
\label{eq:higgscharges}
\end{align}
respectively, in the notation of equations~\ref{eq:chiralquarks}-\ref{eq:chiraldark}.  In section~\ref{subsec:scalarpotential}, we introduced a parity symmetry, $h_1 \to - h_1$, to simply the scalar potential.  It can also be used to simplify the dark yukawa coupling terms.  We therefore also require
\begin{align}
D \to - D && E \to - E && \chi_{R_2} \to - \chi_{R_2}.
\end{align}
The dark yukawa terms are simply
\begin{eqnarray}
\mathcal{L}_1 &=& \lambda_{12}\,Q_L\, h_1 \,D + \lambda_{13}\,Q_L\, h_2 \,U + \lambda_{14}\,L_L\, h_1 \,E + \lambda_{15}\,L_L\, h_2 \,N \label{eq:darkyukawa}  \\
&+& \lambda_{16}\,\chi_L\, h_1 \,\chi_{R_2} + \lambda_{17}\,\chi_L\, h_2 \,\chi_{R_1}  \nonumber 
\end{eqnarray}
which are analogous to the SM yukawa couplings. 
There are additional terms involving the SM higgs that are also possible
\begin{eqnarray}
\mathcal{L}_2 =  \lambda_{18}\,L_L\, \phi \,\chi_L + \lambda_{19}\,E\, \tilde{\phi} \,\chi_{R_1} + \lambda_{20}\,N\, \tilde{\phi} \,\chi_{R_2}. 
\label{eq:yukawahiggs}
\end{eqnarray}
Here we use the standard definition $\tilde{\phi} = i \tau_2\, \phi^*$ where $\tau$ a pauli sigma matrix.  For later simplicity, we will include these terms in our effective field theory.   These terms have an important impact on how the new fermions transform under various approximate global symmetries.
%

\subsubsection{Lepton and Dark Matter Number}
\label{sec:ldmnumber}

The structure of the yukawa couplings is fixed by the gauge symmetries.  This in turn defines the global symmetries of the model.  Naively, from equation~\ref{eq:darkyukawa}, we can assign independent $U(1)$ global symmetries to the dark matter, colored and leptonic messenger fermions.  We naively assign dark lepton and baryon number to the colored and lepton messenger fermions, respectively.  Consider the dimension six operators composed of the SM fermions and the new fermions that are singlets with the SM and dark gauge symmetries.  Two very important operators are
\begin{align}
\mathcal{O}_1 = (q_L^c \, \sigma\, Q_L)\, (l_L^c\,\sigma\,\chi_L)   &&  \mathcal{O}_{2} = (e_R^c\, \sigma\,L_L)\, ( l_L^c\, \sigma\,\chi_L) 
\label{eq:mvoperators}
\end{align}  
which are generated by a gauge boson\footnote{This gauge boson has the quantum numbers needed for dark-electroweak unification.  We discuss this in more detail later.} transforming under $(1,2,2)_{(1/2,0)}$.  A complete list of all the operators consistent with the gauge symmetry are in Appendix A.  It is clear assigning baryon and lepton number (instead of dark baryon and dark lepton number) to the messenger fermions is consistent so long as these operators exist.
Additionally, because we have chosen to keep equation~\ref{eq:yukawahiggs} in our effective lagrangian, the dark matter candidates transform under the same lepton number, global $U(1)_L$, as the messenger leptons.  Interestingly, those couplings force the dark matter and new heavy leptons to have the \textit{opposite} lepton number.  In Table 1, we list the charges for the new fermions under lepton number and baryon number, $U(1)_B$.
\begin{table}[t]  
\begin{center}
  \begin{tabular}{| c | c | c |}
    \hline
    Particle & $U(1)_B$ & $U(1)_L$   \\ \hline \hline
    $Q_L$ & 1/3 & 0  \\ 
    $U$ & -1/3 & 0  \\ 
    $D$ & -1/3 & 0  \\ \hline
    $L_L$ & 0 & 1  \\ 
    $N$ & 0 & -1  \\ 
    $E$ & 0 & -1 \\ \hline
    $\chi_L$ & 0 & -1 \\   
    $\chi_{R_1}$ & 0 & 1  \\    
    $\chi_{R_2}$ & 0 & 1 \\    \hline
  \end{tabular}
  \label{tab:globalcharges}
  \caption{Lepton and baryon number charges for the messenger and dark matter fermions.  If we were to set $\lambda_{14} = 
  \lambda_{15} = \lambda_{16} = 0$ in equation~\ref{eq:yukawahiggs}, then $\chi_L$, $\chi_{R_1}$ and $\chi_{R_2}$ would transform under an independent  global $U(1)_\chi$ as $1$, $-1$ and $-1$, respectively.}
  \end{center}
  \end{table}
  All of the charges of the global symmetries in Table 1 are preserved by the operators\footnote{As we will see the operators in equation~\ref{eq:mvoperators} and Appendix A encode the possible final states for how the messenger fermions can decay to the dark matter and the SM.} in Appendix A.  
 %
%
%
We do remind the reader that the dark matter candidate, $\chi$, is charged under the diagonal $Z_2$ center after spontaneous symmetry breaking.  SM leptons are not.  Because of this, even though they are both charged under lepton number, at low energies the dark matter and the SM leptons will have very different physics.  

\subsection{Gauge Boson Sector}

In this section we detail how the gauge boson transform under the $SU(2)_D \times U(1)_D$ gauge group as well as the  mass spectrum.  To set notation, we denote the charge eigenstate $SU(2)_D$ and $U(1)_D$ gauge bosons, respectively, as 
\begin{align}
V_i &=  (1,1,3)_{(0,0)} & U &= (1,1,1)_{(0,0)} \label{eq:darkgauge}.
\end{align}
Here the entries in parenthesis denote representations transforming under $SU(3)_c$, $SU(2)_L$ and $SU(2)_D$, respectively.  The subscripts entries are $U(1)_Y$ and $U(1)_D$ charges, respectively. 
\newline
\newline
As for the mass spectrum, recall in equation~\ref{eq:higgsvevw/CPviol}, there are two vevs responsible for making the gauge bosons massive.  Together, both vevs completely break $SU(2)_D \times U(1)_D$ gauge group to the $Z_2$ center.  The easiest way to understand the dependence of the masses on $v_1$ and $v_2$ is to recognize that each vev spontaneously breaks the dark gauge group into two different $U(1)$ gauge groups.   Specifically, 
\begin{align}
SU(2)_D \times U(1)_D &\xrightarrow{v_1} U(1)_V, \label{eq:Vbreaking} \\
SU(2)_D \times U(1)_D &\xrightarrow{v_2}  U(1)_A. \label{eq:Abreaking}
\end{align}
Here the generators for $U(1)_V$ and $U(1)_A$ are, using the notation of equation~\ref{eq:charges}, $V = T_3 + S$ and $A = T_3 - S$, respectively.  The $U(1)_A$ and $U(1)_V$ ``photons" get a mass depending only on $v_1$ and $v_2$, respectively.  The other gauge bosons receive equal contributions from both vevs.  Defining the $SU(2)_D$ and $U(1)_D$ gauge couplings as $g$ and $g'$, respectively, the masses are
\begin{align}
&m_1^2 = g^2(v_1^2 + v_2^2)/2  & m_3^2 = (g^2 + g'^2)\,v_2^2/2\\
&m_2^2 = g^2(v_1^2 + v_2^2)/2  & m_4^2 = (g^2 + g'^2)\,v_1^2/2.
\end{align}
Without a loss of generality, we will often consider $v_2 > v_1$.  The gauge bosons in their mass eigenstates are notated as $V'_i$ and $U'_4$ where $i = 1, 3$.  The subscripts correspond to the mass eigenstates.  The mass spectrum is distinctive and is a consequence of these models.  We discuss some experimental signals in Section~\ref{sec:exp}.

\subsubsection{Kinetic Mixing Between the SM and Dark Photons}
\label{sec:photonkm}

Depending on the hierarchy of the dark vevs, the final step in generating the $Z_2$ stabilization symmetry involves spontaneously breaking either the $U(1)_V$ or $U(1)_A$ symmetry.  As described in the previous section, this breaking generates a massive, dark ``photon."  The dark photon can kinetically mix with the SM to generate the effective operator
\begin{eqnarray}
\mathcal{L}_3 &=& {\epsilon \over 2} \,B_{\mu\nu} B^{\mu\nu}_\mathrm{dark} \\
 		       &=& {\epsilon \over 2} \,\biggl(\cos\theta_W F_{\mu\nu} - \sin\theta_W Z_{\mu\nu} \biggr) \,B^{\mu\nu}_\mathrm{dark}.
\end{eqnarray}
Here $F_{\mu\nu}$, $Z_{\mu\nu}$ and $B^{\mu\nu}_\mathrm{dark}$ are the field strengths of the SM photon, $Z$ boson and dark photon, respectively.  $\theta_W$ is the Weinberg angle.  This operator is generated via loop corrections from the messenger fermions listed in equations~\ref{eq:chiralquarks} and \ref{eq:chirallepton}.  It is straightforward to show $\epsilon$ has the value
\begin{equation}
\epsilon = - {g' g'_Y \over 16\,\pi^2}\sum_i \,q_\mathrm{dark} \,q_Y \,\log\biggl({M_i^2 \over \mu^2} \biggr).
\label{eq:epsilon}
\end{equation}
Here $q_i$ are the dark and hypercharge charges.  $M_i$ is the mass of the messenger fermion.  Because of the sum over the charges, this equation is zero when the messenger fermions are in complete anomaly free multiplets.  A way to prevent this is have a regime of parameter space where the messengers have different masses within the multiplet.  Thus, at a certain scale, one could integrate out the heavy messengers so the sum in equation~\ref{eq:epsilon} is not zero.
%
\newline
\newline
Looking forward, we require the relic dark matter abundance at the dark matter decoupling temperature to be very small.  We show in Appendix C the annihilation via the dark photon is subdominant to the annihilation through the higgs portal even when $\epsilon$ is relatively large.  We will, however, use both effects to ensure sufficient dark matter annihilation.  

\subsection{Dynamically Generated Dark Matter Mass}
\label{subsec:DMmass}

\begin{table}[t]  
\begin{center}
  \begin{tabular}{| c | c | c | c | c | c |}
    \hline
    Particle & $SU(2)_L$ & $SU(2)_c$ & $SU(2)_{D}$ & $SU(2)_{c'}$   \\ \hline \hline
                \vspace{-0.1cm}                   & & & & \\
          $h =   \begin{pmatrix}  h_1 & h_2 \end{pmatrix}$ & 1 & 1 & 2 & 2 \\ 
            \vspace{-0.3cm}                   & & & & \\
          $\phi =   \begin{pmatrix}  \phi & \tilde{\phi} \end{pmatrix} $ & 2 & 2 & 1 & 1 \\          
                \vspace{-0.1cm}                   & & & & \\ \hline
                 \vspace{-0.1cm}                   & & & & \\ 
    $Q_L$  &      1              & 1                    &     2                  &  1   \\ 
  \vspace{-0.3cm}                   & & & & \\
    $Q_R = \begin{pmatrix}  U \\ D \end{pmatrix}$ & 1 & 1  & 1 & 2  \\ 
                \vspace{-0.1cm}                   & & & & \\ \hline
                 \vspace{-0.1cm}                   & & & & \\ 
      $L_L$   &      1              & 1                    &     2                  &  1   \\ 
  \vspace{-0.3cm}                   & & & & \\
    $L_R = \begin{pmatrix} N \\ E \end{pmatrix}$ & 1 & 1 & 1 & 2  \\ 
      \vspace{-0.3cm}                   & & & & \\
      $L = \begin{pmatrix} L_L \\ L_R \end{pmatrix}$ & 2 & 1 & 1 & 1  \\ 
                \vspace{-0.1cm}                   & & & & \\ \hline
                 \vspace{-0.1cm}                   & & & & \\ 
    $\chi_L$&     1              & 1                    &     2                  &   1 \\  
          \vspace{-0.3cm}                   & & & & \\
    $\chi_R = \begin{pmatrix} \chi_{R_1}  \\ \chi_{R_2} \end{pmatrix}$ & 1 & 1 & 1 & 2 \\   
              \vspace{-0.3cm}                   & & & & \\ 
     $\chi = \begin{pmatrix} \chi_{L}  \\ \chi_{R} \end{pmatrix}$ & 1 & 2 & 1 & 1 \\    
          \vspace{-0.1cm}                   & & & & \\ \hline 
  \end{tabular}
  \label{tab:globalchiralcharges}
 \caption{The approximate dark and SM custodial symmetry representations for the dark matter and messenger fermions.  Also listed are how these fermions transform under the gauged $SU(2)_L$ and $SU(2)_D$.  Notably, because of equation~\ref{eq:yukawahiggs}, the dark matter and messenger leptons form doublets under the SM chiral symmetry.}  
  \end{center}
  \end{table}
In the effective field theory description of the SM below the weak scale, the light SM fermions have an approximate global chiral symmetry.  Because these fermion masses are so small, it is often said the chiral symmetry ``protects" the masses.  In this section, we construct chiral symmetries to protect the dark matter masses.  These symmetries protect the dark matter mass giving a mass of order the electroweak vev suppressed by a loop factor.
%
%
%

\subsubsection{Dark and Standard Model Chiral Symmetries}

To build up the arguments for constructing a symmetry which protects the dark matter mass, let us again start with the SM.  As described above, below the weak scale the SM has a $SU(2)_L \times SU(2)_R$ global chiral symmetry which protects the SM fermion masses.  The low energy global $SU(2)_L$ symmetry is a remnant of the gauged $SU(2)_L$ symmetry.   At the weak scale, $SU(2)_R$ is known as an approximate ``custodial" $SU(2)$ symmetry\footnote{Despite being an approximate symmetry, custodial $SU(2)$ is essential to the SM as well as in understanding new physics beyond the SM.  Custodial symmetry is explicitly broken by the top quark yukawa coupling (dominantly) and $U(1)$ hypercharge.  Breaking custodial symmetry with new physics is strongly disfavored by precision electroweak measurements.}.  In more detail, the SM fermion sector has the following form at the weak scale
\begin{align}
\mathcal{L}_4 = \lambda_u\, \overline{q}_L \,\phi \,u_R + \lambda_d\, \overline{q}_L \,\tilde{\phi} \,d_R.
\label{eq:SMyukawa}
\end{align}
Here $q_L$ is the left-handed SM quarks; $u_R$ and $d_R$ are the right-handed SM quarks.  In the limit where $\lambda_u = \lambda_d = \lambda_q$, $\mathcal{L}_4$ simplifies to
\begin{equation}
\mathcal{L}'_4 = \lambda_q\,\overline{q}_L \Phi\,q_R
\label{eq:SMyukawasimplied}
\end{equation}
where $q_R$ is a column matrix composed of $u_R$ and $d_R$.  $\Phi$ is 
a $2 \times 2$ matrix composed of the $\phi$ and $\tilde{\phi}$.  In this custodial symmetric limit, the left- and right-handed quarks as 
\begin{align}
q_L \to L\,q_L &&  q_R \to R\,q_R.
\end{align}
It is easy to see $\Phi$ must transform as $\Phi \to L\,\Phi\,R^{\,\dagger}$.  Here $L$ and $R$ are the generators for the familiar $SU(2)_L$ gauge symmetry and $SU(2)_c$ custodial symmetries, respectively.  This gives us two lessons from the SM.  In order to generate a $SU(2)_L \times SU(2)_c$ global chiral symmetry that protects the fermion masses, it is important that
\begin{itemize}
\item the yukawa couplings ($\lambda_u$,  $\lambda_d$ and $\lambda_q$) are sufficiently small so the masses of the fermions will be small compared to the scales of interest.
 
\item violations of custodial $SU(2)$ are not large.  Thus, the resulting mass differences are small compared to the scales of interest.
\end{itemize}
For our purposes, the latter requirement is not as significant as the first.  However, knowing all of the approximate symmetries in the hidden sector is important bookkeeping device.  It will help us to properly estimate the sizes of the various yukawa couplings.
\newline
\newline
Since the dark sector has chiral matter, it naively features an additional independent, approximate global chiral symmetry which acts on the fermions lighter than the dark symmetry breaking scale.  Consider the dark yukawa terms (equation~\ref{eq:darkyukawa}),
\begin{eqnarray}
\mathcal{L}_1 &=& \lambda_{12}\,Q_L\, h_1 \,D + \lambda_{13}\,Q_L\, h_2 \,U + \lambda_{14}\,L_L\, h_1 \,E + \lambda_{15}\,L_L\, h_2 \,N \nonumber  \\
&+& \lambda_{16}\,\chi_L\, h_1 \,\chi_{R_2} + \lambda_{17}\,\chi_L\, h_2 \,\chi_{R_1}.  \nonumber 
\end{eqnarray}
We want to determine all of the approximate symmetries in this sector.  We are particularly interested in how the dark matter transforms.  In this limit of an exact dark ``custodial" symmetry, the above equation becomes
\begin{eqnarray}
\mathcal{L}'_1 &=& \lambda_{12-13}\, Q_L \,H\,Q_R + \lambda_{14-15}\, L_L \,H\,L_R + \lambda_{16-17}\, \chi_L \,H\,\chi_R.
\end{eqnarray}
Here $H$ is a $2 \times 2$ matrix composed of $h_1$ and $h_2$.  Also,
\begin{align}
Q_R = \begin{pmatrix} U \\ D \end{pmatrix} && L_R = \begin{pmatrix} E \\ N \end{pmatrix} && \chi_R = \begin{pmatrix} \chi_{R_2} \\ \chi_{R_1} \end{pmatrix}.
\end{align}
Thus, the dark yukawa sector has an (independent) approximate $SU(2)_{R'}$ global dark symmetries.  It is explicitly broken by $U(1)_D$.  The dark particles transform under $SU(2)_D \times SU(2)_{R'}$  as
\begin{align}
Q_L \to Q_L\,L'^{\,\dagger} && Q_R \to R'^{\,\dagger} \,Q_R, \\
L_L \to L_L\,L'^{'\,\dagger} && L_R \to R'^{\,\dagger} \,L_R, \label{eq:Ltrans} \\
\chi_L \to \chi_L\,L'^{\,\dagger} && \chi_{R} \to R'^{\,\dagger} \,\chi_R. \label{eq:DMtrans}
\end{align}
With, of course, $H$ transforming as $H \to L'\,H\,R'^{\,\dagger}$.  Here  $L'$ and $R'$ are the representations for $SU(2)_D$ and $SU(2)_{R'}$, respectively.
\newline
\newline
There is an \textbf{\textit{important}} additional twist on how the dark matter and messenger fermions transform under the chiral symmetries.  Consider now the other dark yukawa couplings,
\begin{eqnarray}
\mathcal{L}_2 &=&  \lambda_{18}\,L_L\, \phi \,\chi_L + \lambda_{19}\,E\, \tilde{\phi} \,\chi_{R_1} + \lambda_{20}\,N\, \tilde{\phi} \,\chi_{R_2}.
\nonumber
\end{eqnarray}
Because $\Phi \to L\,\Phi\,R^\dagger$ (and $\Phi$ is $2 \times 2$ matrix composed of the $\phi$ and $\tilde{\phi}$),  it is clear both the messenger leptons and the dark matter must transform under both of the dark and electroweak custodial symmetries!  In the limit where \textit{both} the dark and electroweak custodial symmetries are exact, the above equation becomes
\begin{equation}
\mathcal{L}'_2 = \lambda_{L\chi}\,L_{LR} \,\Phi\,\chi
\label{eq:Lchisimple}
\end{equation}
where 
\begin{align}
L_{LR} &= \begin{pmatrix} L_L & L_R \end{pmatrix} &  \chi &= \begin{pmatrix} \chi_{L}  \\ \chi_{R} \end{pmatrix}  
\label{eq:LDMcustsymm}
\end{align}
and again $\Phi$ is defined below equation~\ref{eq:SMyukawasimplied}.  
It is clear $L_{LR}$ and $\chi$ transform as doublets under the $SU(2)_L \times SU(2)_c$ SM symmetries. 
\begin{align}
L_{LR} \to L_{LR} \,L^\dagger && \chi \to R \,\chi.
\label{eq:LDMtrans}
\end{align}
Here $L$ and $R$ are the generators for the gauged $SU(2)_L$ and the custodial $SU(2)$ respectively.  We summarize the how the higgses and messenger fermions transform in Table 2.  Notice for the messenger leptons and dark matter, the approximate dark and SM chiral symmetries are, in a sense,``overlapping."  In Appendix B, we use this concept to outline a mechanism to stabilize both the dark and electroweak scales.

\subsubsection{Loop Suppressed Dark Matter Mass}
\label{seclsdmm}

To explore how the dark matter mass can be dynamically generated, consider the dark yukawa terms in equation~\ref{eq:darkyukawa}
\begin{equation}
\mathcal{L}_1 = \ldots + \lambda_{16}\,\chi_L\, h_1 \,\chi_{R_2} + \lambda_{17}\,\chi_L\, h_2 \,\chi_{R_1}. \nonumber
\end{equation}
As discussed in the previous section, these terms explicitly break both the dark \textit{and SM} custodial symmetries.  The dark matter yukawa sector preserves the approximate chiral symmetries if  $\lambda_{16}$ and  $\lambda_{17}$ are very small.  Having small yukawa couplings is technically natural and occurs in the SM.  In other words, one can say the dark chiral symmetry ``protects" the dark matter from getting a mass at scales below the dark and electroweak symmetry breaking scales.  For the rest of this paper, we assume these yukawa couplings are sufficiently small so that the dominant contribution to the dark matter mass comes from the loop suppressed irrelevant operators below.
\newline
\newline
Generating the dark matter mass now requires the couplings in equation~\ref{eq:yukawahiggs} (equation~\ref{eq:Lchisimple}) above.  
The dark matter can generate a mass at tree-level via the left panel in Figure 1. After the dark phase transition, the operator generated by integrating out the newly massive messenger leptons is 
\begin{equation}
\mathcal{O}_3 = {c' \over  m_{L_L}}\, \bigl(\chi_R\,\tilde{\phi}\bigr)\,\bigl(\phi\,\chi_L\bigr)  
\label{eq:1stmassoperator}
\end{equation}
Here $c' \sim f(\lambda_{18}\,\lambda_{19}\,\lambda_{20})$ and the above equation breaks the SM custodial symmetry.  (The dark custodial symmetry may have been broken by the messenger fermion masses.)  
When both $\phi$ and $\tilde{\phi}$ gets a vev, the dark matter gets a dirac mass.  If $c'$ is $\mathcal{O}(1)$ (or correspondingly the couplings $\lambda_{18}$, $\lambda_{19}$ and $\lambda_{20}$ are appropriately large), then the dark matter gets a mass of order $\sim v_\mathrm{ew}^2/v_{1,2}$. This mass is way too large to explain the ratio of the dark matter to baryonic relic abundance in this model.  Recall,
\begin{equation}
\Omega_\mathrm{DM}/\Omega_\mathrm{baryon} \equiv (n_\mathrm{DM}\,m_\mathrm{DM})/(n_\mathrm{baryon} m_\mathrm{baryon}). \nonumber
\end{equation}
 In Section~\ref{sec:bDMabn}, we show $n_\mathrm{DM}/n_\mathrm{baryon} \sim 5$ for the model under consideration; thus, $m_\mathrm{baryon} \sim 1$ GeV implies $m_\mathrm{DM}$ must also be of order the GeV scale.    We therefore require $c' $ (or the function $f(\lambda_{18}\,\lambda_{19}\,\lambda_{20})$) to be small and therefore protect the dark matter mass  
\newline
\begin{figure}[t]
\begin{center}
\includegraphics[width=8truecm,height=4.3truecm,clip=true]{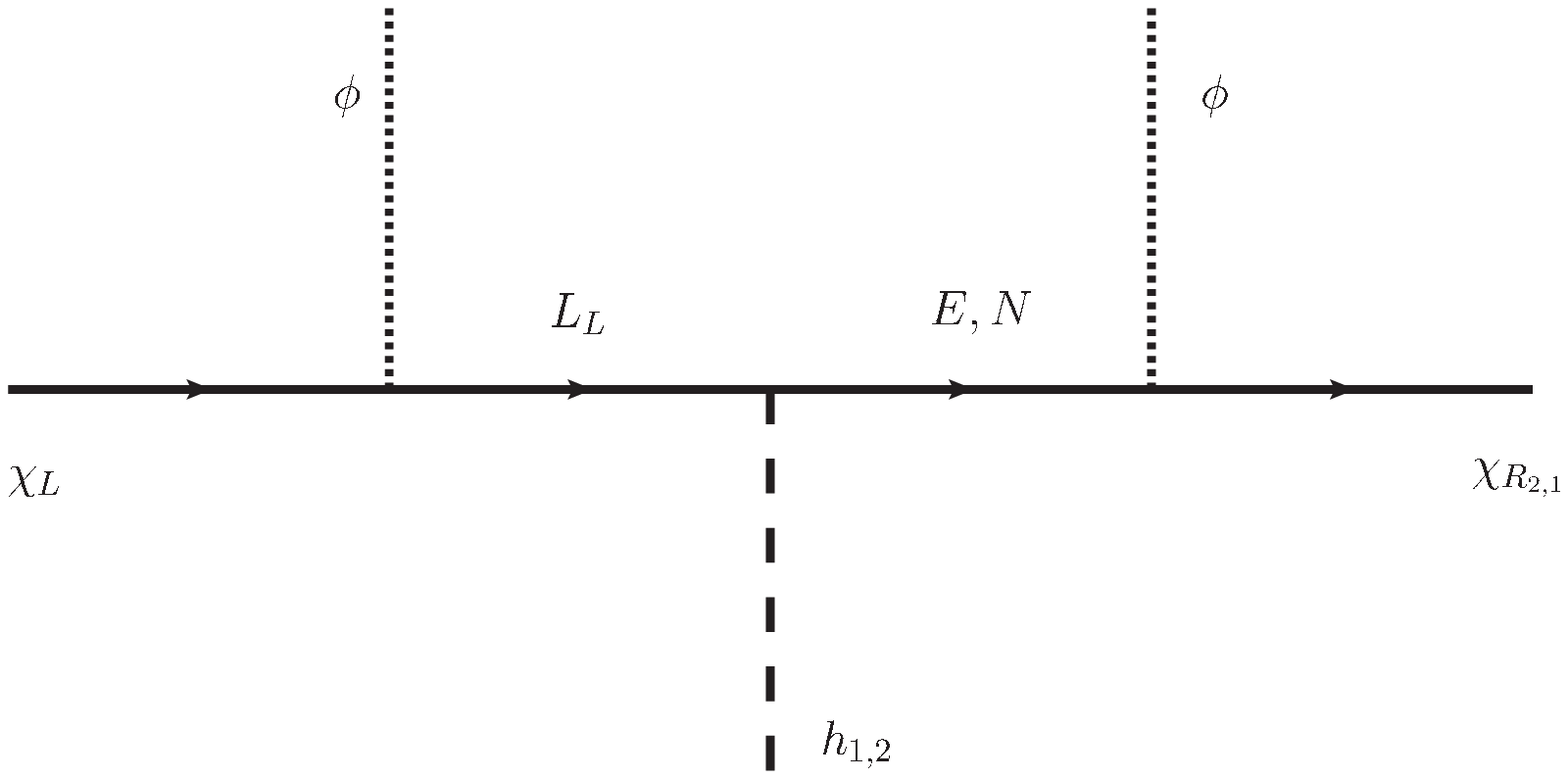}  \hspace{0.2cm}
\includegraphics[width=8truecm,height=6.5truecm,clip=true]{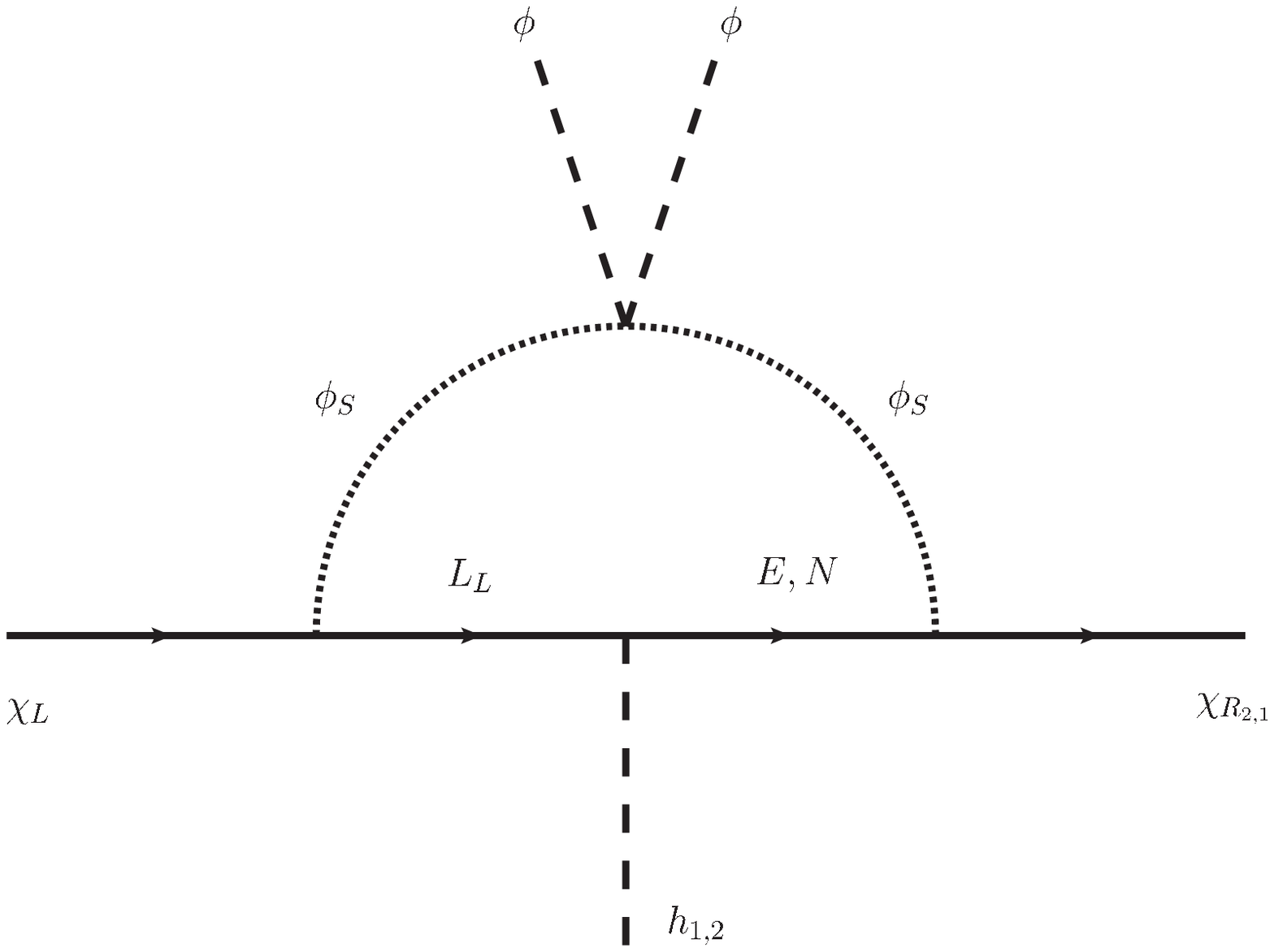}
\caption{Dynamically generated dark matter mass.  The right panel generates the loop suppressed dark matter mass, $v_\mathrm{ew}/16\pi^2$ }
\end{center}
\label{fig:dmmass}
\end{figure}
\newline
In order get the right dark matter mass, we simply add a new electroweak higgs, $\phi_S$, with the same quantum numbers as the SM higgs which does not get an electroweak vev.  Electroweak two-higgs doublet models are very common in the literature; however, it is an uncommon for one of the higgses to be responsible for giving mass to quark, leptons and electroweak gauge bosons and the other to only be a spectator.  It is certainly true, for the generic electroweak two-higgs doublet model, one can always make a transformation so that one higgs gets an electroweak vev and the other higgs has a minimum at zero~\cite{Georgi:1978xz}.  In essence with this model, we are enforcing this choice with the following symmetry
\begin{equation}
\phi_S \to - \phi_S.
\end{equation}
Equation~\ref{eq:yukawahiggs} now has the counterpart
\begin{eqnarray}
\mathcal{L}_2 =  \lambda_{21}\,L_L\, \phi_{S} \,\chi_L + \lambda_{22}\,E\, \tilde{\phi}_{S} \,\chi_{R_2} + \lambda_{23}\,N\, \tilde{\phi}_{S} \,\chi_{R_1}. 
\label{eq:yukawaS}
\end{eqnarray}
where $\tilde{\phi}_S = i \tau_2 \,\phi_S^*$ and $\tau_2$ is a pauli sigma matrix.  We have also required
\begin{align}
 L_L \to - L_L && E \to - E && N \to& - N.
\end{align}
This also ensures equations~\ref{eq:yukawahiggs} (and therefore~\ref{eq:1stmassoperator}) are zero.
Because of the couplings
\begin{align}
\mathcal{O}_4 = \lambda_{S_1}\,\phi^\dagger \phi\,\phi_S^\dagger\, \phi_S && \mathcal{O}_5 = \lambda_{S_2}\,h^\dagger h\,\phi_S^\dagger \,\phi_S, 
\label{eq:Sops}
\end{align}
$\phi_S$ gets a mass.  
Finally, we require the couplings in equation~\ref{eq:yukawaS} are to be $\mathcal{O}(1)$.  The dominant contribution to the dark matter mass is now the diagram in the right panel of Figure 1.  Thus, we have the effective operator
\begin{equation}
\mathcal{O}_6 = {1 \over 16\pi^2}{c \over  m_{L_L}}\,  \bigl(\chi_R\,\tilde{\phi}\bigr)\,\bigl(\phi\,\chi_L\bigr).
\end{equation}
Like before, after electroweak symmetry breaking, this operator explicitly breaks all the chiral symmetry.  The dark matter mass is now 
\begin{align}
m_\chi \sim {v_\mathrm{ew} \over 16\pi^2}\,\biggl({c \,v_\mathrm{ew} \,\over m_{L_L} }\biggr) && c = 2\, \lambda_{\phi^4}^2\,\lambda_{S_1}^2\, \log\biggl[{ v_\mathrm{ew} \over m_{\phi_S}}\biggr]  \label{eq:dmmass} 
\end{align}
We have parametrized the coupling of $\phi^4$ operator to be $\lambda_{\phi^4}$ and taken, for simplicity, $v_1 \sim v_2 \sim v_\mathrm{dark}$.  Choosing $c$ to be significant explicitly breaks the dark and SM custodial symmetries.  For the rest of this paper, we assume the quantity in parenthesis in equation~\ref{eq:dmmass} is $\mathcal{O}(1)$.  
\newline
\newline
Keeping the quantity in parenthesis $\mathcal{O}(1)$, requires a relatively light messenger lepton regardless of the size of dark symmetry breaking vev.  If $m_{\phi_S}$ depends on the dark symmetry breaking vev, 
the messenger lepton's mass can range from $700$ GeV to about $15$ TeV.  The former (latter) value comes from the assumption that the dark symmetry breaking scale is at 1 TeV (the unification scale).  We discuss the experimental consequences of the light messenger leptons in detail in Section~\ref{sec:exp}.
\newline
\newline
In this section, we have shown SM and dark chiral symmetries can protect the dark matter mass.  By construction, this mass is of order the electroweak vev suppressed by a loop factor.  Given a GeV scale dark matter mass, we show in Section~\ref{sec:bDMabn} that this light dark matter annihilate to the SM thereby leaving a very small relic abundance.  Therefore, the decay of the messenger fermions generate dark matter relic abundance observed today.
%

\subsection{Non-Perturbative Operators and Baryon Number Violation}
\label{baryviol}
  
The dark gauge interactions generate global anomalies which violate baryon and lepton number.  Recall the baryon and lepton numbers of the dark matter and messenger fermions were determined in Section~\ref{sec:ldmnumber}.  Interestingly, these anomalies violate $B - L$ number but preserve $B + L$.  This is the opposite of what the weak interactions, and many other models of new physics beyond the SM, preserve.  Given the fermion content in equations~\ref{eq:chiralquarks}-\ref{eq:chiraldark}, we can easily verify these statements.  The anomalies for $B$ and $L$ are
\begin{align}
\mathrm{Tr}[B \,T^a T^a ] = - 1 && \mathrm{Tr}[L \,T^a T^a] = (2 - 1)  = + 1
\label{eq:BLanomaly}
\end{align}
where the $T^a$ are the $SU(2)_D$ generators.  $L_L$ and $\chi_L$ contribute factors of $2$ and $-1$, respectively, to the lepton anomaly.   At zero temperature,  the non-perturbative processes generate an operator which violates $B$, $L$, and $B - L$.  This instanton generated operator is proportional to $\exp(-2\pi/\alpha_{\mathrm{dark}})$.  Here $\alpha_{\mathrm{dark}} = g^2/4\pi$ and $g$ is the $SU(2)_D$ dark coupling.  
Because of the exponential suppression, the size of the instanton operator is effectively zero.  It was realized, however, in \cite{Kuzmin:1985mm} that, at a large finite temperature above the dark phase transition scale, this operator is unsuppressed.  The non-perturbative interactions generate a ratio of heavy quarks to heavy leptons to dark matter candidates as $5:2:1$. We will show in the next section how this ratio is changed to one SM baryon for every five dark matter candidates.  
\newline
\newline
It is worth noting that one generally considers the dark phase transition to happen at an earlier time than the electroweak transition.  Therefore, electroweak sphaleron interactions will always be a background process and potentially can erase any baryon asymmetry generated by the dark interactions.  However, in this model, the dark and electroweak anomalies preserve different global symmetries; and, therefore this is \textbf{not} an issue.  On the whole, models of this type preserve a combination of $B$, $L$ and dark matter number.  In consequence, 
is these models always preserve a different global symmetry than the weak interactions and therefore prevent the electroweak sphalerons from erasing any asymmetry produced by the dark phase transition. 

\section{Baryon and Dark Matter Abundances}
\label{sec:bDMabn}
 
The ratio of dark matter to the baryon abundance is defined as
\begin{equation}
\Omega_\mathrm{DM}/\Omega_\mathrm{baryon} \equiv (n_\mathrm{DM}\,m_\mathrm{DM})/(n_\mathrm{baryon} m_\mathrm{baryon}),
\label{eq:bDMratio}
\end{equation}
where $n_i$ and $m_i$ is the number density and mass for species $i$.  As noted in the introduction, this ratio, $\Omega_\mathrm{DM}/\Omega_\mathrm{baryon}$, has been measured to be roughly five.  In \cite{Barr:1990ca,Kaplan:1991ah}, it was showed that $n_\mathrm{DM}/n_\mathrm{baryon}$ could be explained if there is a common phase transition generating both the dark matter and the baryons.  Importantly, since the baryon mass is about  $1$ GeV, $\Omega_\mathrm{DM}/\Omega_\mathrm{baryon}$ is crucially dependent on the dark matter mass.  Now, unless the new phase transition generates many more baryons than dark matter candidates, the dark matter mass should be of order the baryon mass.  Given a GeV dark matter mass, we use this section to describe a scenario which gives an explanation for the measured dark matter to baryon abundance ratio.

\subsection{Requirements}
\label{bDMabn}

In Section~\ref{baryviol}, we have described a dark phase transition that correlates the production of heavy quarks, heavy leptons and dark matter.  In analogy to weak scale baryogensis, we assume a first order phase transition.  The transition generates three heavy quarks (one heavy baryon) for every two heavy leptons and one dark matter candidate.  (This is instead of producing anti-heavy quarks, anti-heavy lepton and anti-dark matter.)  As a reminder, all of these particles are odd under the $Z_2$ center.  The messenger particles, \textit{i.e.} the heavy quarks and leptons, must decay due to the strict bounds on long-lived particles that are charged under the SM~\cite{Perl:2001xi}.  Thus, upon decay, the heavy quarks and leptons must have final states with at least one dark matter candidate.  In Appendix A, we list all of the distinct, dimension six operators involving the messenger fermions, dark matter and the SM.  These operators encode the simplest possible decays for the heavy messenger fermions as well as the baryon and lepton number assignments.  Of these, we take the simplest and least suppressed to be dominant,  
\begin{align}
Q _L &\to q_L + l_L + \chi_L, \label{eq:hvydecay} \\
L_L  &\to e_R + l_L + \chi_L. \nonumber 
\end{align}
Both of these decays can be mediated by bosons with the quantum numbers of $(1,2,2)_{(1/2,0)}$.  In Appendix B, we show how this mediating particle is naturally a consequence of unifying the dark and electroweak gauge interactions.  If we consider a supersymmetric UV completion, an additional decay channel is opened,  $L_L  \to  d_R + q_L + \chi_L$.   This channel generates a baryon and anti-baryon in the final state which results in net baryon number of zero. Thus, for a generic SUSY completion, the important point is, regardless of the branching fraction, $L_L$ decays to one dark matter candidate.  In order to use these decay channels to explain the $\Omega_{DM}/\Omega_b$ ratio, there are several points which must be treated with care.  We assume the following:
\begin{itemize}
\item We require the heavy fermions to be sufficiently long-lived so the dark matter produced by the dark phase transition is out of thermal equilibrium by the time the messenger particles decay.
\item  Before the messenger fermions decay, we require the remaining dark matter candidates to annihilate efficiently to the SM and not give a significant relic abundance.  Thus, the messenger fermions decay is responsible for the dark matter asymmetry as well as the  relic abundance.
\end{itemize}
In essence these requirements mean all of the observed dark matter in the universe is a result of the \textit{decay} of the heavy quarks and leptons.  Naively the decays in equation~\ref{eq:hvydecay} and the two requirements above alter the ratio of heavy messenger particles to dark matter produced by the dark sphalerons.  The ratio is converted to
\begin{equation}
3:2:1 \to 1:2:5.
\label{eq:ratio}
\end{equation}
Here the order on the left side is heavy quarks to heavy leptons to dark matter.  The order on the right is SM baryons to SM leptons to dark matter.  At this stage, we have a dark matter to SM baryon number density ratio of $5:1$.  Assuming GeV scale dark matter, the resulting factor of  $\Omega_{DM}/\Omega_b \sim 5$ is what we hoped to achieve.  By construction, this factor is ultimately due\footnote{To be more precise, the messenger $Q_L$ and $L_L$ fermions fit into a $5$ of $SU(5)$.  In grand unification, particles that fit into the $5$ of SU(5) are traceless under $U(1)_Y$.  During the dark phase transition, the non-perturbative instanton generated operator must also be a singlet  $U(1)_Y$.  Therefore, this forces three $Q_L$s to be produced with every two $L_L$s.} to the fact that our anomaly free particle content, equation~\ref{eq:fermioncontent}, fits into the fundamental representation of $SU(5)$.
\newline
\newline
To be sure the scenario described in the previous paragraph holds, there are still more items which require care.  We require  
\begin{itemize}
\item  The messenger fermion decays and the dark matter decoupling must not upset nucleosynthesis.
\end{itemize}
Nucleosynthesis provides fairly strong constraints on relativistic new physics~\cite{Nakamura:2010zzi,Sarkar:1995dd,Iocco:2008va} at the MeV scale.  Thus these bounds point to the SM with three neutrino flavors.  Codifying these statements, the requirements generate the following conservative temperature hierarchy
\begin{equation}
1\,\,\mathrm{MeV} \,\,\lesssim \,\,T_\mathrm{decay} \,\,< \,\,T_\mathrm{decoupling}.
\end{equation}
Here $T_\mathrm{decay}$ and $T_\mathrm{decoupling}$  are the characteristic temperatures associated with the messenger fermion decay and the decoupling of the dark matter generated by the dark phase transition, respectively.  Dark matter needs to cold, or non-relativistic, when it decouples from the SM.  Thus, $T_\mathrm{decoupling}$ is typically of order, $M_\chi$, the mass of the dark matter\footnote{We will show $T_\mathrm{decay}$ is basically determined by the dark and electroweak unification scale.}.  
In the next few subsections we use simple estimates to substantiate this hierarchy.  We will see messenger fermion decays at MeV temperatures generates final states that are non-relativistic and therefore do not produce much entropy to upset nucleosynthesis.  Thus the above hierarchy should be regarded as conservative.  Beyond the temperature constraints, we also require
\begin{itemize}
\item The dark phase transition needs to be a highly efficient, strongly first order phase transition.
\end{itemize}
A highly efficient first-order dark phase transition is necessary so that there are almost no anti-heavy baryons around to spoil the ratio in equation~\ref{eq:ratio}.  A strong first order transition implies a large CP violating phase in the dark sector.  In Section~\ref{sec:edm} we show how current EDM constraints are considerably relaxed in the dark sector.  
Finally, it is also important to have
\begin{itemize}
\item The effect of the weak and QCD phase transitions has a negligible effect on the dark matter and baryon asymmetries.
\end{itemize}
This requirement is already built into the model.  The QCD and electroweak phase transitions preserve $B - L$ but violate $B + L$.    As described in Section~\ref{baryviol}, the new dark particle content does the opposite and  preserves $B + L$ but violates $B - L$.  The ratio in equation~\ref{eq:ratio} is preserved in presence of these SM phase transitions.

\subsection{Thermal Decoupling Temperatures and Dark Matter Relic Abundance}

As described Section~\ref{bDMabn}, it is important that the temperature at which the heavy messenger fermions decay is \textit{smaller} than the \textit{relic} dark matter decoupling temperature.  
This satisfies the second requirement in Section~\ref{bDMabn}.  As a reminder, we \textit{want} the dark matter to efficiently annihilate into the SM for as long as possible in order to \textit{minimize} the dark matter relic abundance.  Therefore, the ratio of the dark matter to baryon number densities is set by the decays of the heavy messenger fermions.
\newline
\newline
In Section~\ref{subsec:DMmass}, we showed how the dark matter mass is dynamically generated to be about a GeV.  Thus, it is reasonable to expect that the dark matter annihilations will fall out of thermal equilibrium at temperatures near this mass.  At these low temperatures, the dark matter dominantly annihilates via the higgs portal and, to a lesser extent, via kinetic mixing of the dark and SM photons.  Experiment constrains SM higgs to be heavier than $114.4$ GeV~\cite{Nakamura:2010zzi}.  (Indeed, recent experimental results have given hints of a 126 GeV higgs mass~\cite{atlashiggsresults,CMShiggsresults}.)  Thus, dark matter annihilation through the higgs portal is suppressed by the SM higgs mass which is integrated out.  Because we require the relic dark matter to efficiently annihilate into the SM for as long as possible in order to \textit{minimize} the relic abundance, we therefore require the dark higgses and photons to be light.  
%

\subsubsection{Dark Matter Annihilation through the Higgs/Gauge Boson Portal}

\textbf{Kinetic Mixing Portal:}  As discussed in Section~\ref{sec:photonkm}, these models admit heavy photons which can kinetically mix with the SM photon via equation~\ref{eq:epsilon}:
\begin{align}
\chi_{L}\,\chi_{R} &\to \gamma'/\gamma \to f_i \,f_i \label{eq:dmannhil2}.
\end{align} 
The dark photon sector alone can meet the dark matter annihilation requirements listed in Section~\ref{bDMabn}.  The parameter space, however, can be limited.  In Appendix C, we show the parameter space available to get the right relic abundance.  To open the parameter space up, in the next paragraph, we detail a higgs portal through which the dark matter can also annihilate.  We will include the effects of both portals in the analysis in this section.
\newline
\newline
\textbf{Higgs Portal:}  Ostensively, the dark matter annihilation through the higgs portal would proceed though the following process:
%
\begin{align}
\chi_{L}\,\chi_{R} &\to h_{1,2}/\phi \to f_i \,f_i \label{eq:dmannhil1}
\end{align} 
Here $i$ runs over all of the SM fermion flavors, $u$, $d$, $c$, $s$, $e$ and $\mu$ which are (semi-)relativistic for decoupling temperatures near $1$ GeV.  
However, in Section~\ref{seclsdmm}, we wrote down a mechanism where the dark matter gets a mass of order
\begin{align}
m_\chi \sim {v_\mathrm{ew} \over 16\pi^2}\,\biggl({c \,v_\mathrm{ew} \,\over m_{L_L} }\biggr) && c = 2\, \lambda_{\phi^4}^2\,\lambda_{S_1}^2\, \log\biggl[{ v_\mathrm{ew} \over m_{\phi_S}}\biggr].  \nonumber
\end{align}
where the quantity in parenthesis is taken to be $\mathcal{O}(1)$.  In addition, we required the couplings between the dark matter and $h_{1,2}$ to be technically small so that the above contribution to the dark matter mass is dominant over the tree-level contribution.  This mechanism (and the accompanying set of assumptions) complicates dark matter annihilation story.
\newline
\newline
Below the dark and electroweak symmetry breaking scales, the above mass \textit{implies} that the dark matter has a coupling to the SM higgs that goes as
\begin{equation}
{1 \over 16\pi^2}{c \over  m_{L_L}}\,  \bigl(\chi_R\,\tilde{\phi}\bigr)\,\bigl(\phi\,\chi_L\bigr) \sim {1 \over 16\pi^2} \,\phi_0\,  \chi_L \chi_R.
\label{eq:smallcoupling}
\end{equation}
Here $\phi_0$ is the SM higgs boson leftover after electroweak symmetry breaking.  Of course, on the operator on the RHS of the equation is generated below the dark symmetry breaking scale.  This implies the dark matter annihilation to the SM is suppressed by $(16\,\pi^2)^2$.  Recall, we want the relic dark matter to efficiently annihilate so that the relic abundance is extremely small; thus, the measured dark matter relic abundance today is set by the messenger fermions decay.  Explicit computation of the relic abundance \textit{using this SM higgs coupling} shows the $1/16\,\pi^2$ suppression in equation~\ref{eq:smallcoupling} results in a $\mathcal{O}(10000)$ relic abundance.  This is clearly inconsistent with observation; see equation~\ref{eq:DMrelicabundance}.   
\newline
\newline
It is enough to simply rely on the kinetic mixing sector to meet the dark matter annihilation requirements in Section~\ref{bDMabn}.  However, to open up the available parameter space, we add a new dark higgs, $h_S$, with the same quantum numbers as the dark higgses introduced in equation~\ref{eq:higgscharges}.  The difference is this new dark higgs does not get a vev like $h_1$ and $h_2$.  We require that $h_S$ has $\mathcal{O}(1)$ couplings in
\begin{equation}
\mathcal{L}'_1 = \lambda_{24}\,\chi_L\, h_S \,\chi_{R_2} + \lambda_{25}\,\chi_L\, \tilde{h}_S \,\chi_{R_1}.
\end{equation}
where $\tilde{h}_S = i \tau_2 h_S^*$. The new dark higgs also has the following couplings with the SM and $h_{1,2}$,
\begin{align}
\mathcal{O}_7 &= \lambda_a\,\phi^\dagger\, \phi\,h_S^\dagger\, h_S &  \mathcal{O}_8 &= \lambda_b\,h_S^\dagger\, h_S\,h_{1,2}^\dagger\, h_{1,2} \\
\mathcal{O}_9 &= \lambda_c\,h_S^\dagger\, h_2\,h_2^\dagger\, h_2 &  \mathcal{O}_{10} &= \lambda_d\,h_S^\dagger\, h_2\,\phi^\dagger\, \phi
\end{align}
Recall, to simplify the dark scalar potential, we imposed $h_1 \to - h_1$; therefore there is no corresponding operator similar $\mathcal{O}_{9}$ with $h_1$ substituted for $h_2$.  
After the dark and electroweak symmetry breaking, $\mathcal{O}_{10}$ (along with the hermitian conjugate) gives the term
\begin{equation}
\mathcal{O}_{10} = \lambda_d\,v_\mathrm{ew} v_2 \,\phi_0\,h_{S_1}
\end{equation}
where $h_{S_1}$ is a component in $h_S$.  In the coming work, we assume the simplifying limit, $\lambda_{24} = \lambda_{25} = \lambda_d$, and 
denote it as the ``simplified" higgs coupling.
\newline
\newline
\textbf{Annihilation Cross Section and Relic Abundance:}  Here we provide a basic estimate of the the dark matter annihilation.  We do not account for form factors for dark matter annihilations into SM quarks.  We also do not assume any chemical potentials.  With these simplifications, it is straightforward to compute the low-velocity thermally averaged annihilation cross section~\cite{Kolb:1990vq,Wells:1994qy}.  It is
\begin{eqnarray}
\langle \sigma |v| \rangle &=& a + {6\, T \over m_\chi}\, b + {60 \,T^2 \over m_\chi^2} \,c + \ldots
\end{eqnarray}
where the ``$\ldots$" are additional velocity suppressed terms and $T$ is the dark matter freeze-out temperature.  Definitions for $a$, $b$, and $c$ can be found in~\cite{Kolb:1990vq,Wells:1994qy}.  Explicit computation of $a$ and $b$ for annihilations through the higgs and kinetic mixing portals results in
\begin{eqnarray}
a_\mathrm{higgs} &=& 0, \\
b_\mathrm{higgs} &=& \sum_l\, \frac{\alpha\,\lambda^4\,v_\mathrm{dark}^2 v_\mathrm{ew}^2}{16\,c_W^2\,s_W^2\, m_Z^2\,m_\phi^4 }\,\frac{m_l^2 \left(m_\chi^2-m_l^2\right)^{3/2}}{m_\chi \,\left(4 m_\chi^2 - m_h^2\right)^2 } \\
			     &+& \sum_q \,\frac{3 \alpha\,  \lambda ^4\, v_\mathrm{dark}^2 v_\mathrm{ew}^2}{16 \,c_W^2 s_W^2\, m_Z^2\,m_\phi^4 }\,\frac{m_q^2 \left(m_\chi^2-m_q^2\right)^{3/2}}{m_\chi \left(4 m_\chi^2 - m_h^2 \right)^2},\nonumber \\
a_\mathrm{kinetic} &=& \sum_l\,\frac{8\, \pi\, \epsilon^2 \, \alpha \,\alpha_\mathrm{dark} \,m_\chi \sqrt{m_\chi^2-m_l^2} \left(m_l^2+2 \,m_\chi^2\right) }{\left(m_{\gamma'}^2\, m_\chi - 4\, m_\chi^3\right)^2}  \\
&+& \sum_q\,\frac{32\,\pi\,\epsilon^2 \alpha\,\alpha_\mathrm{dark} \,m_\chi \sqrt{m_\chi^2 -m_q^2} \left(m_q^2+2 m_\chi^2\right) }{3 \left(m_{\gamma'}^2 \,m_\chi - 4 \,m_\chi^3\right)^2},\nonumber \\
b_\mathrm{kinetic} &=& \sum_l \frac{\pi \epsilon^2 \,\alpha\, \alpha_\mathrm{alpha}\left(64\,m_\chi^6 +8 m_\chi^4(m_{\gamma'}^2-4 m_l^2) -4 m_l^2 m_\chi^2(m_{\gamma'}^2+17 m_l^2) + 5 m_{\gamma'}^2 m_l^4\right)}{3 m_\chi \sqrt{m_\chi^2-m_l^2} \left(m_{\gamma'}^2-4 m_\chi^2\right)^3}\\
&+& \sum_q \frac{4\pi\epsilon^2\, \alpha\,\alpha_\mathrm{dark} \left(64 \,m_\chi^6+8 m_\chi^4(m_{\gamma'}^2-4 m_q^2)-4 m_q^2 m_\chi^2\,(m_{\gamma'}^2+17 m_q^2 ) + 5 m_{\gamma'}^2 m_q^4\right) }{9 m_\chi \sqrt{m_\chi^2-m_q^2} \left(m_{\gamma'}^2-4 m_\chi^2\right)^3}.  \nonumber
\end{eqnarray}
where again $\epsilon$ is the kinetic mixing parameter, $s_W$ and $c_W$ are $\sin \theta_W$ and $\cos \theta_W$ with SM Weinberg angle.  $m_{\gamma'}$ and $m_h$ is the mass of the heavy photon and new dark higgs, $h_S$.  $v_1 = v_2 = v_\mathrm{dark}$ for simplicity.  Of course, the summation is over the final state quarks and leptons.  $\lambda$ is the simplified higgs coupling.  Now with the thermally averaged cross section in hand, we can estimate the freeze-out temperature and the dark matter relic abundance.
\newline
\label{sec:higgsann}
\begin{figure}[t]
\centering
\includegraphics[width=7.8truecm,height=5.0truecm,clip=true]{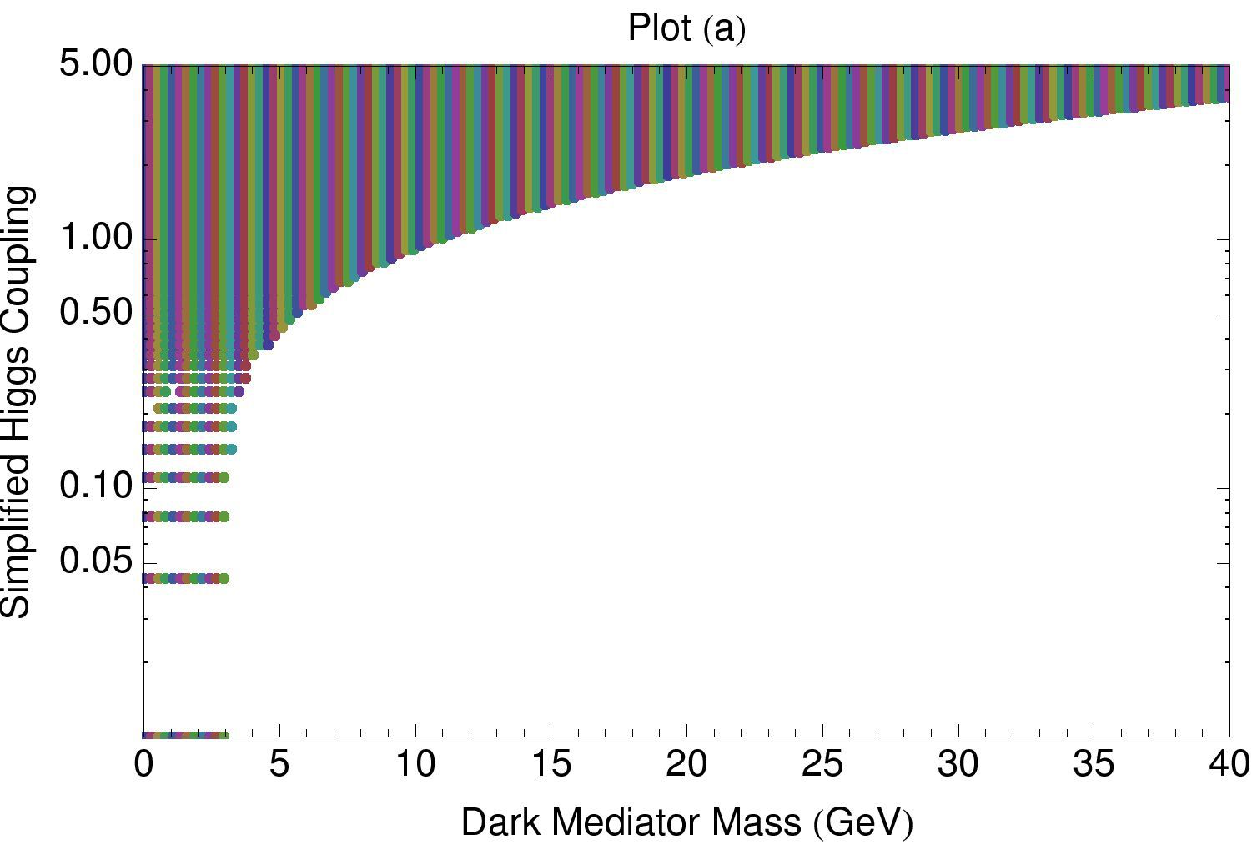}  \hspace{0.2cm}
\includegraphics[width=8.2truecm,height=5.1truecm,clip=true]{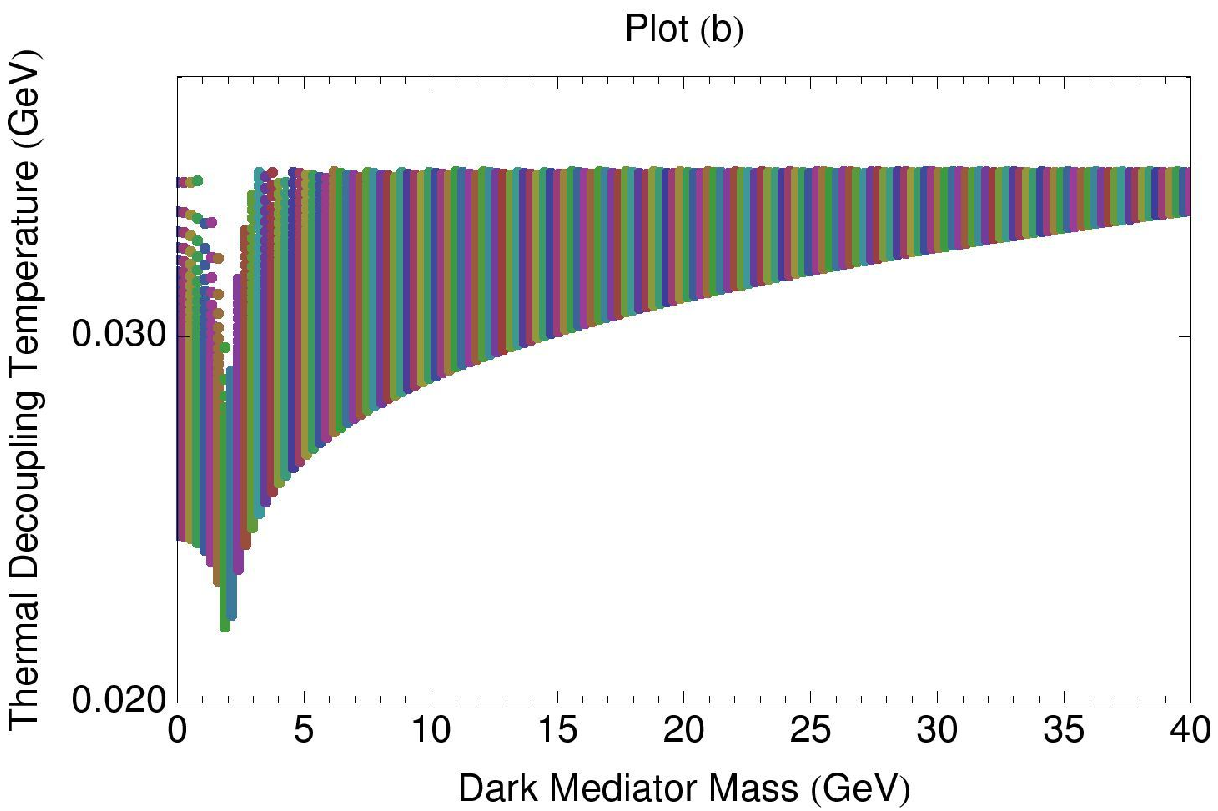}  \vspace{0.3cm}  \\
\includegraphics[width=7.9truecm,height=5.0truecm,clip=true]{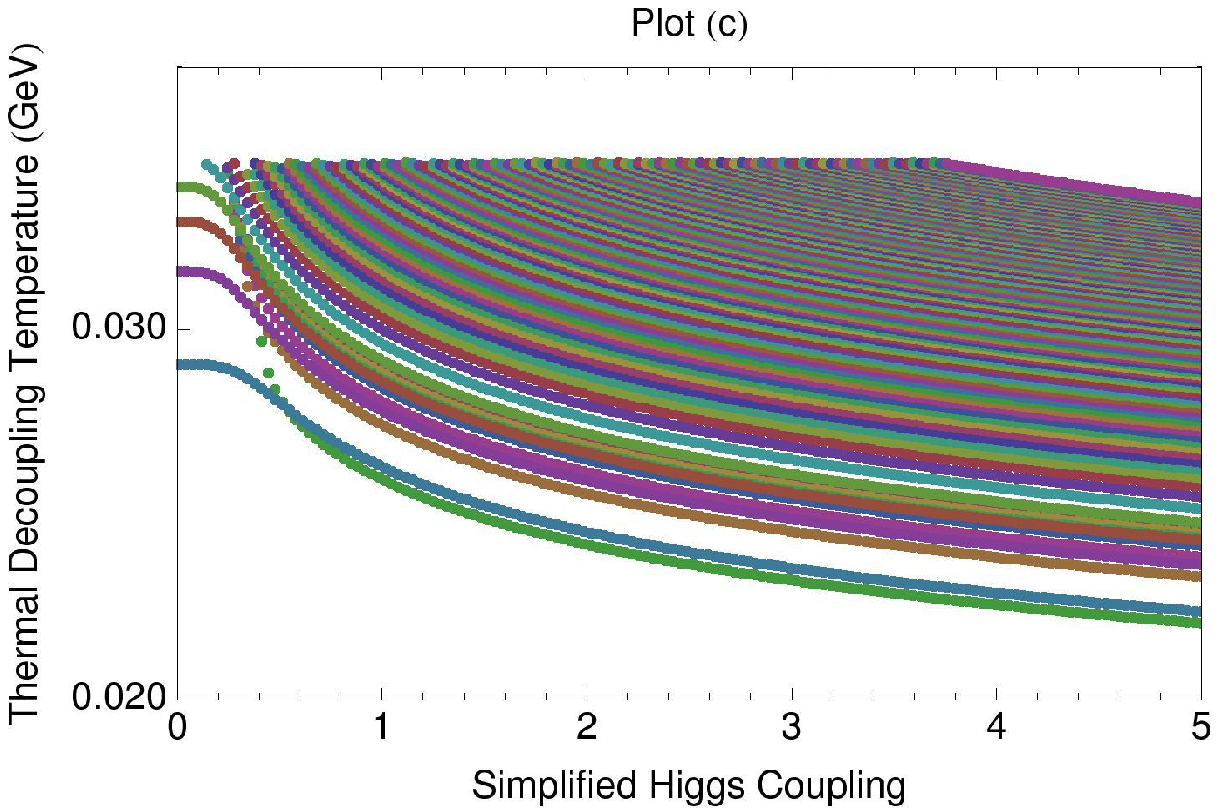}  \hspace{0.2cm}
\includegraphics[width=8.0truecm,height=5.1truecm,clip=true]{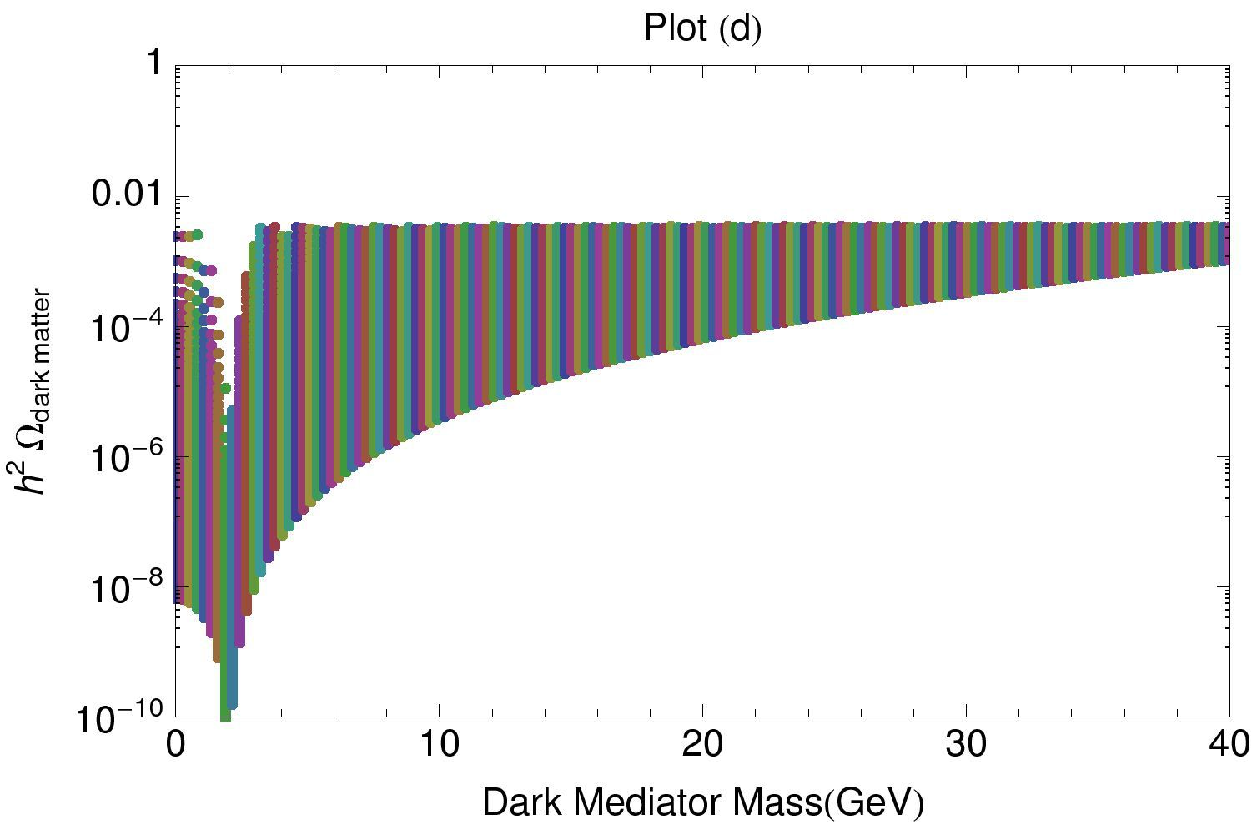}  
\caption{Allowed parameter points for the relic dark matter abundance that does \textit{\textbf{not}} originate from the messenger fermion decays.  Each point represents a point in parameter space that yields an abundance of $h^2 \Omega < 0.0034$.  The dark photon and dark higgs masses are set to be equal; and, to be conservative, we take the dark vevs $v_1 = v_2 = 1$ TeV.  
The kinetic mixing parameter is taken to be $\epsilon = 10^{-3}$.}
\label{fig:annhilationplots}
\end{figure}
\newline
An estimate of the freeze out temperature can be determined when the dark matter annihilation rate drops below the expansion of the universe:  $\Gamma < H$ where $H \sim T^2/M_\mathrm{pl}$ during the radiation dominated universe.  Here
\begin{align}
\Gamma = n\,\langle \sigma |v| \rangle && n = g\, \biggl({m_\chi T \over 2\pi} \biggr)^{3/2} e^{-m_\chi/T}
\end{align}
where $g$ is the degrees of freedom of the dark matter.  $n$ is the dark matter equilibrium density in the non-relativistic limit.
%
The freeze out temperature can be determined by finding the roots of~\cite{Kolb:1990vq}
\begin{equation}
x_F = \ln\biggl(c(c + 2)\sqrt{{45 \over 8}} {g \over 2 \pi^3}{m_\chi M_\mathrm{pl}(a + 6 \,b /x_F) \over g_*^{1/2} x_F^{1/2}  }  \biggr)
\end{equation}
where $x_F \equiv m_\chi/T$.  $T$ is the freeze-out temperature; and, $c$ is usually taken to be a half.  $g_*$ is all of the relativistic  degrees of freedom in the universe.
\newline
\newline
With all of this, the dark matter relic abundance is simple to determine.  It is
 \begin{eqnarray}
\Omega_\chi\,h^2 \sim {1.04 \times 10^9 \over M_\mathrm{pl}}\,{x_F \over \sqrt{g_*}}\,{1 \over a + 3b/x_F}
\end{eqnarray}
We want $\Omega_\chi\,h^2$ to be as small as possible; thus, the observed abundance today would be a result of the decay of the messenger fermions.  The measured dark matter relic abundance is
\begin{equation}
h^2\, \Omega_{\mathrm{DM}} = 0.1131\pm0.0034. \nonumber
\end{equation}
We therefore make the conservative requirement that the relic abundance, that does \textbf{\textit{not}} come from the messenger fermion decays, is
\begin{equation}
h^2\, \Omega_{\mathrm{relic\,\,DM}}  < 0.0034.
\end{equation}
In Figure 2, we show some points in parameter space that are consistent with the above dark matter abundance requirement.  As mentioned before, in these plots, the dark matter annihilations are mediated by the dark photon and higgses.  For simplicity, we fixed both dark photon and dark higgs masses to be the same.  The annihilations via the dark higgs dominate.  In Appendix C, we show the same plots as Figure 2 but for the dark photon only.  This shows the need for a light dark higgs.  
A final point:  The allowed parameter space in Figure 2 can expand slightly if we allowed the dark photon and higgs masses were allowed to vary independently.

\subsection{Messenger Fermion Lifetime and Decay Temperatures}
\label{sec:mfldt}

Implementing the requirements in Section~\ref{bDMabn} is straightforward.  We simply need to ensure the messenger quarks and leptons are sufficiently long-lived so that the relic dark matter in the universe annihilates to produce a minuscule relic abundance.  This ensures the dark matter relic abundance measured today dominantly comes from the messenger fermion decays.  As discussed in Section~\ref{bDMabn}, the messenger fermions decay via new, massive gauge bosons needed for dark-electroweak unification.  Therefore, the dark-electroweak unification scale sets the mass of these gauge bosons and introduces a free parameter into the (lifetime and) characteristic temperature in which the messenger fermion decay.   In order to generate a small relic abundance, above in Figure 2 we showed the dark matter decoupling temperature is around $0.035$ GeV.  In this section, we estimate the parameter space available for the messenger fermion decays.
\newline
\label{fig:lifetimeplots}
\begin{figure}[t]
\centering
\includegraphics[width=7.8truecm,height=5.0truecm,clip=true]{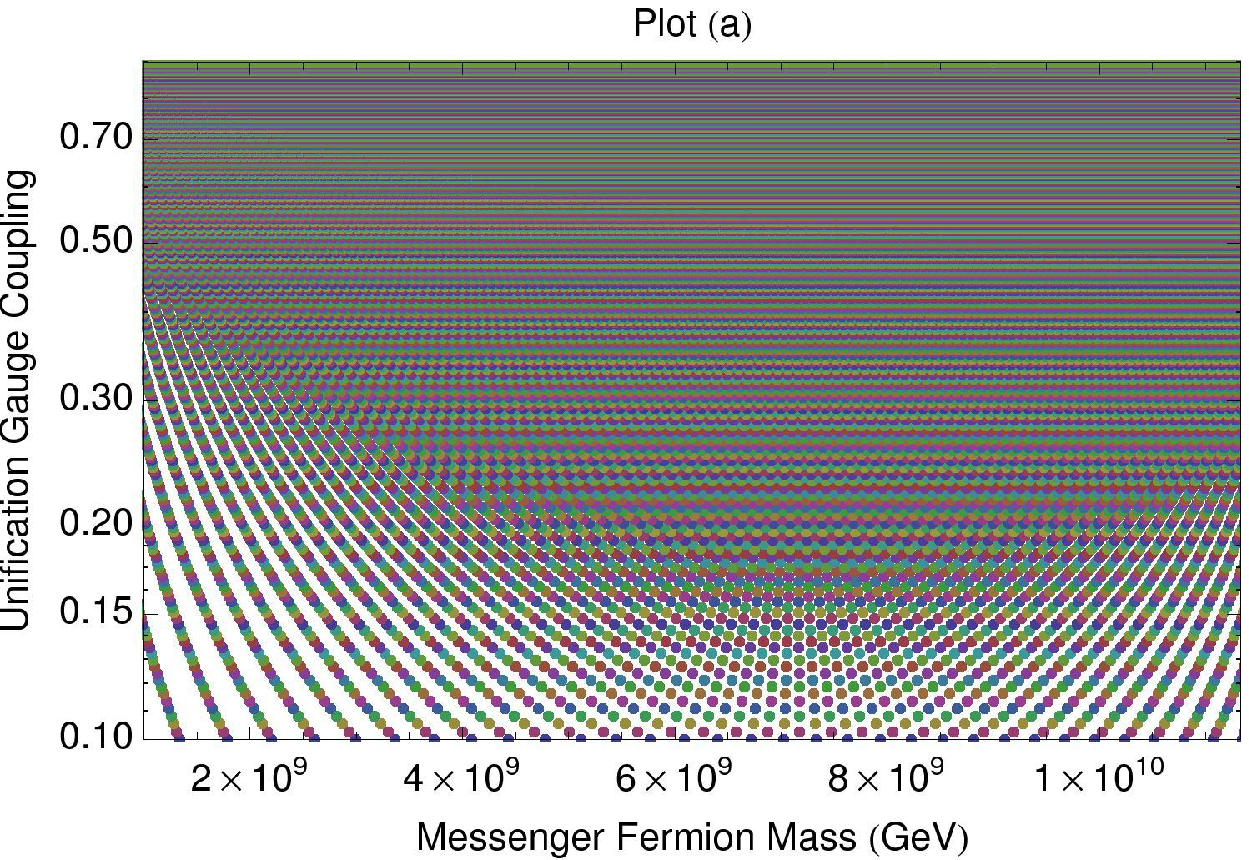}  \hspace{0.2cm}
\includegraphics[width=8.2truecm,height=5.1truecm,clip=true]{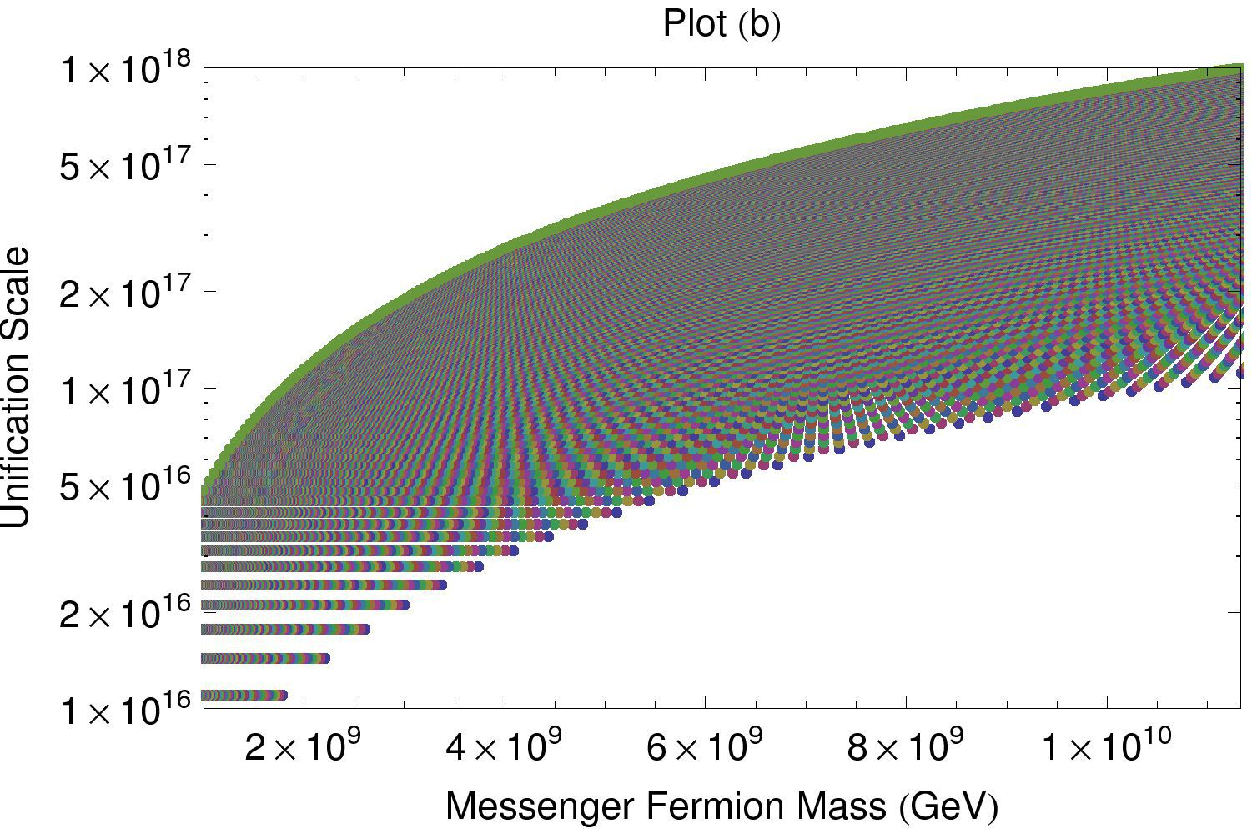}  \vspace{0.3cm}  \\
\includegraphics[width=7.9truecm,height=5.0truecm,clip=true]{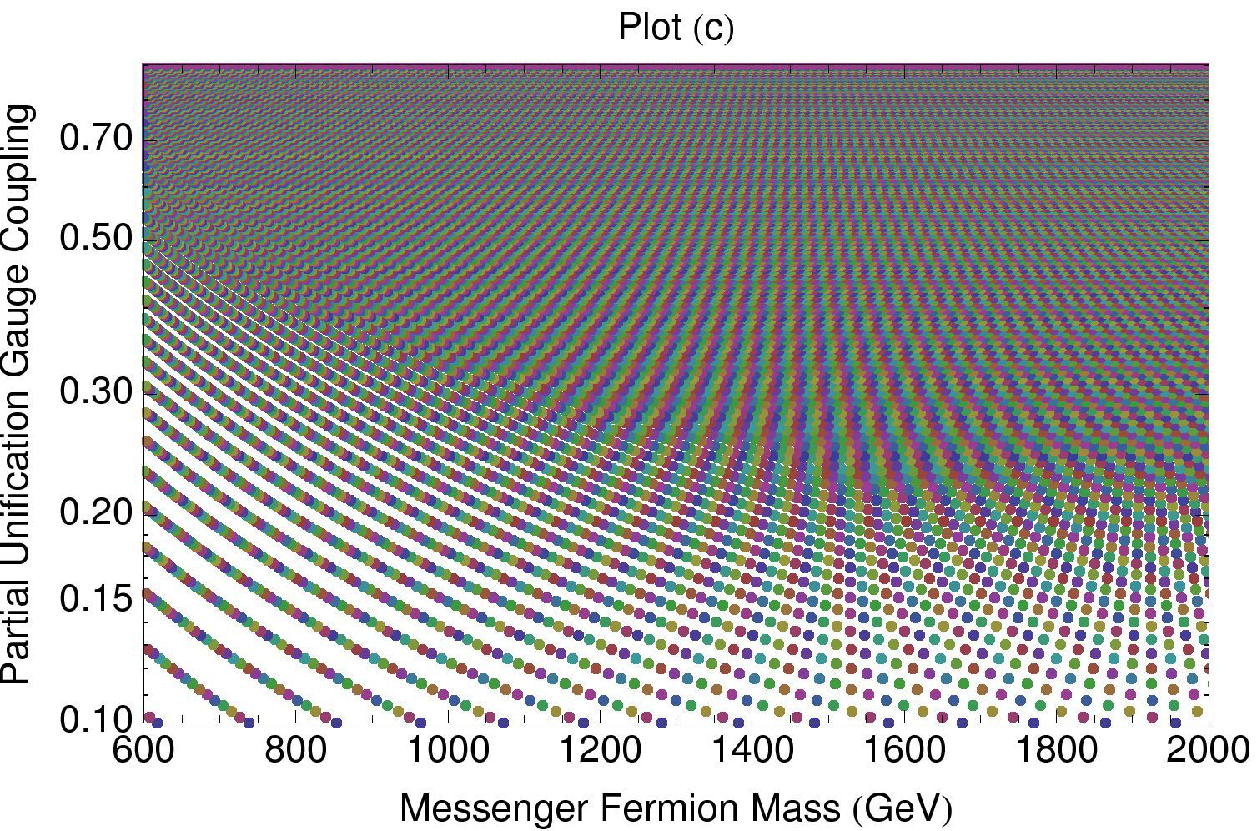}  \hspace{0.2cm}
\includegraphics[width=8.0truecm,height=5.1truecm,clip=true]{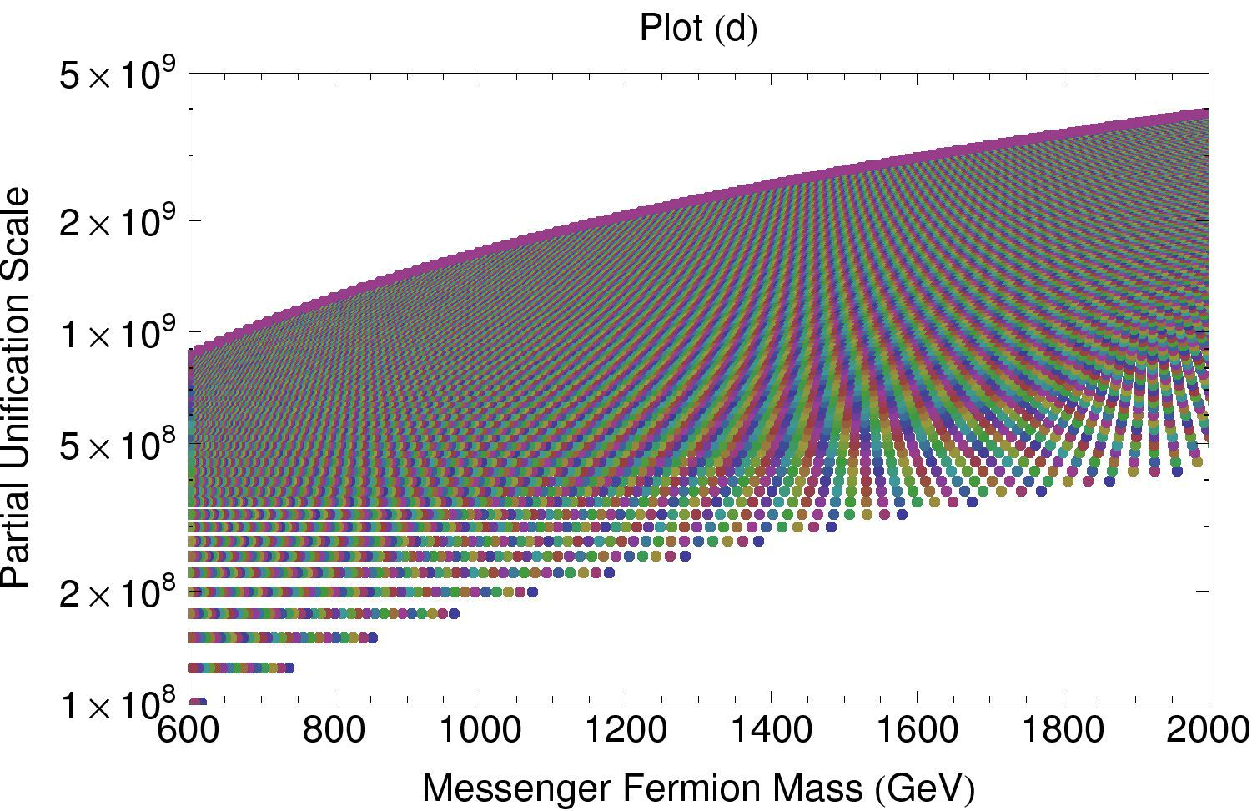}  
\caption{A sample of the parameter space available for the messenger fermion decay into the SM and dark matter.  
We require the decay temperature in equation~\ref{eq:decaytemp} to be $T \sim 0.03$ GeV with the simplification, $M_Q \sim M_L$.  The top two plots, Plots (a) and (b), are the available parameter space points for dark-electroweak unification scale around the grand unification scale.  Allowing a partial dark-electroweak unification scale, Plots (c) and (d), potentially allows the messenger fermions to be observed at the LHC. }
\end{figure}
\newline
Nucleosynthesis occurs at temperatures near 1 MeV and therefore and can potentially constrain the parameter space for the messenger fermion decays.  We remind the reader that particles which decay far out-of-equilibrium into the SM can 
 introduce a significant amount of entropy into the baryon-photon plasma.  This  in turn can cause the universe to cool less slowly and significantly delay the onset of nucleosynthesis.  In Section~\ref{bDMabn}, we showed heavy messenger quarks formed a heavy ``baryon" which decays to a SM baryon, three dark matter candidates and three anti-leptons.  Also, the heavy messenger lepton decays into two SM leptons and one dark matter candidate.  All of final state particles resulting for the messenger fermion decay, except with the possible exception of final state neutrinos from tau decays, are (or rapidly becoming) non-relativistic around the nucleosynthesis scale at $1$ MeV.  Besides relativistic final state neutrinos would dump entropy in the neutrino sector which is decoupled at those temperatures from the SM.  Assuming zero chemical potential, all non-relativistic (and therefore pressure-less) final state particles have an exponentially suppressed energy density~\cite{Kolb:1990vq}.  Thus, the increased entropy from the non-relativistic final states is also exponentially suppressed.
\newline
\newline
With these facts in hand, estimating the temperature of the universe when the messenger fermions decay is now straightforward.  The messengers decay at a time $t_{L,Q} \sim H^{-1} \sim \tau_{L,Q}$.  $\tau_{L,Q}$ is the easily computable lifetime of the messenger leptons and quarks.
\begin{align}
\tau_L \sim {1 \over  g^4}\,{M^4 \over M^5_L} && \tau_Q \sim {1 \over  g^4}\,{M^4 \over M^5_Q}
\end{align}
where $g$ is the unification gauge coupling at dark-electroweak unification.  The temperature is
\begin{align}
T_L \sim \biggl({g \over M} \biggr)^2\,\biggl(M_L^5\,M_\mathrm{pl}\biggr)^{1/2}  &&  T_Q \sim \biggl({g \over M} \biggr)^2\,\biggl(M_Q^5\,M_\mathrm{pl}\biggr)^{1/2} 
\label{eq:decaytemp}
\end{align}
Here $M$ is the mass of the $(1,2,2)_{(1/2,0)}$ unification boson introduced below equation~\ref{eq:hvydecay}.  As emphasized in the first paragraph, we require this temperature to be less than around 0.035 GeV (from Figure 2).  Solving both sets of equations, the lifetime\footnote{This lifetime is extremely long compared to the time (in the lab frame) for particles at relativistic speeds to transverse the ATLAS and CMS detectors at the LHC.  This lifetime is key to identify this physics cleanly.  We discuss this in detail in Section~\ref{sec:exp}.}  is
\begin{align}
\tau \gtrsim 5 \times 10^{-3}\,\,\mathrm{seconds}
\label{eq:mlifetime}
\end{align}
with the requirement $T \sim 0.03$ GeV.  Again, we have taken $M_L \sim M_Q$.  In Figure 3, we sample some of the available parameter space for this decay.  In this plot, we focused on the parts of parameter space that can generate messenger fermion signatures at the LHC.  More on potential signals with long-lived particles at the LHC in Section~\ref{sec:exp} below.

\section{Experimental Considerations}
\label{sec:exp}
 
In this section we discuss experimental constraints and signatures for this class of models.  We focus on constraints from (chromo-)electric dipole moments (EDMs) and signatures of these models at the LHC.  We choose to estimate the contributions to the (chromo-)EDMs because the assumption that they are small is unwarranted in dark baryogenesis models.  At the LHC, One ``smoking gun" signature is the observation of  long-lived particles (with large missing momentum).  This signature was first considered by us in preparation for the this paper~\cite{Walker:2009ei}.  In the next paragraph, we give only a very brief overview on potential constraints from direct and indirect measurements.
\newline
\newline
First consider direct detection experiments:  Exempting the recent debates over dark matter discovery claims, direct detection experiments (such as Xenon100~\cite{Aprile:2010um}) roughly have ruled out dark matter-SM scattering cross sections of order $\sigma \sim 10^{-44}$ cm$^2$ for a range of dark matter masses.  Unfortunately, this cross section is largely insensitive to the GeV dark matter which is featured in this model.  (See, \textit{e.g.}, the scattering cross section vs.~dark matter plots in~\cite{Aprile:2010um}.) Further, dark matter scattering with the SM via the higgs portal
 generically has a much smaller cross section.  Direct detection experiments such as CoGeNT~\cite{Aalseth:2011wp} are searching for dark matter in this mass range.  As for indirect detection experiments, many experimental collaborations (such as the Fermi-LAT telescope~\cite{Ackermann:2011wa}) are looking, \textit{e.g.}, for cosmic gamma rays which could originate from dark matter annihilations.  A particular focus is gamma ray fluxes from dwarf spheroidal galaxies; such galaxies are dominantly composed of dark matter and therefore deviations from expected fluxes from known cosmological sources provide a litmus test.  In this model, annihilation cross section is parametrically dependent on $\epsilon^4$ where $\epsilon$ is the kinetic mixing parameter.  However $\epsilon$ can be very small in this model.  Now the dark matter in this theory does have ``self-interactions."  Here the self-interactions  is defined by the exchange of a light dark higgs or a dark photon.  Constraints on dark matter self-interactions come from gravitational lensing of galaxies such as the bullet cluster; this, in turn, places a constraint on the dark matter interaction cross section today.  The generic bounds~\cite{MiraldaEscude:2000qt,Markevitch:2003at,Randall:2007ph} are for a self-interaction cross section of order $\sigma/m_\chi \lesssim 10^{-25}$ cm$^2$ GeV$^{-1}$.  For almost all of the parameter points in Figure 2, the interaction cross section is of order $\sigma/m_\chi \lesssim 10^{-30}$ cm$^2$ GeV$^{-1}$.  Finally, as the LHC continues to explore new physics at the TeV scale, precision electroweak measurements have placed  constraints up to 5 TeV on new physics beyond the SM.   Because the new physics is charged under the dark gauge group, the precision electroweak observables tend to be suppressed by multiple loops.  Finally, the LHC has not seen large missing momentum signatures which could potentially constrain these models~\cite{Rajaraman:2011wf}.  Like the direct detection experiments, these signals have a reduced sensitivity for GeV scale dark matter.
  
\subsection{Constraints from Electric Dipole Moments}
\label{sec:edm} 
 
In section~\ref{subsec:darkCPterms}, we added CP violating terms to the dark higgs potential.  For a recent review of the effect of CP violating phases on electric dipole moments see, \textit{e.g.},~\cite{Pospelov:2005pr}.   As a reminder, those terms are (from equation~\ref{eq:CPviolation})
\begin{eqnarray}
V_{CP} &=&  \lambda _6 \biggl(\mathrm{Re}[ h^\dagger_1\, \tilde{h}_2 ] - v_1 v_2 \cos{\beta} \biggr)\biggl(\mathrm{Re}[ h^\dagger_2\, \tilde{h}_1 ] - v_1 v_2 \cos{\beta}  \biggr) \nonumber \\
&+& \lambda _7 \biggl(\mathrm{Im}[ h^\dagger_1\, \tilde{h}_2 ] - v_1 v_2 \sin{\beta} \biggr)\biggl(\mathrm{Im}[ h^\dagger_2\, \tilde{h}_1 ] + v_1 v_2 \sin{\beta}  \biggr). \nonumber
\end{eqnarray}
In this section, we estimate whether a large CP violating phase from these terms can generate observably large electric dipole moments (EDMs).  
Communicating CP violation from the dark sector to the SM higgs can occur via the messenger fermions (equations~\ref{eq:chiralquarks} and~\ref{eq:chirallepton}) or through the SM higgs.  Since we are ultimately interested in SM quark (chromo-)EDMs the former generically requires extra loops.  We restrict our attention to CP violation through the higgs portal.
%
\newline
\newline
The lowest dimensional operator connecting the SM higgs and dark CP violating terms is 
\begin{eqnarray}
V_\mathrm{CP}' &=&  {\lambda _{17} \over 4\,\Lambda^2} \,\phi^\dagger \phi\,\biggl(\mathrm{Re}[ h^T_1\, \tilde{h}_2] - v_1 v_2 \cos{\beta} \biggr)\,\biggl(\mathrm{Re}[ h^T_2\, \tilde{h}_1] - \,v_1 v_2 \cos{\beta} \biggr) \label{eq:SMdarkCPviol} \\
&+& {\lambda _{18} \over 4\,\Lambda^2}\,\phi^\dagger \phi\, \biggl(\mathrm{Im}[ h^T_1\, \tilde{h}_2] - v_1 v_2 \sin{\beta} \biggr)\,\biggl(\mathrm{Im}[ h^T_2\, \tilde{h}_1] + v_1 v_2 \sin{\beta} \biggr). \nonumber
\end{eqnarray}
which is dimension-six and suppressed by the scale $\Lambda$.  Here $\Lambda$ is a new physics scale above the dark symmetry breaking scale.  Throughout  this text, we have alluded to dark-electroweak unification as the source of the structure in the model.  $\Lambda$ can fit nicely as the unification scale within this framework.  As a reminder, $\phi$ is the SM higgs(es) and the $\lambda_{18}$ term generates the CP violation.  We also remind the reader of the dark higgs component notation  
\begin{align}
h_1 = \begin{pmatrix} \eta_2 + i \,\eta_3 \\ \eta_0 + i \,\eta_1 \end{pmatrix} & & h_2 = \begin{pmatrix} \xi_0 + i\, \xi_1 \\ \xi_2 + i \,\xi_3  \end{pmatrix}. \nonumber
\end{align}
Going to unitary gauge requires $\eta_2$, $\eta_3$, $\xi_2$ and $\xi_3$ to be eaten.  The remaining $\eta_1$ and $\xi_1$ terms are CP odd while $\eta_0$ and $\xi_0$ are CP even.  
After symmetry breaking, the most important term is
\begin{equation}
V'_\mathrm{CP} \sim {\lambda_{18} \over \Lambda^2}\,v_{ew}\,\phi_0\,\biggl(\cos\beta\,(v_1\,\xi_1 - v_2\,\eta_1)  - \sin\beta\,(v_2\,\eta_0 + v_1\,\xi_0)  \biggr),
\end{equation}
which is relevant.  Here $\beta$ is the CP violating phase.  
%

\subsubsection{Comparison with Data}

The most stringent constraints on models with additional sources of CP violation beyond the SM comes from precision measurements of the electric dipole moments (EDM) of the neutron, thalium and mercury atoms~\cite{Baker:2006ts,Regan:2002ta,Griffith:2009zz}:
\begin{eqnarray}
|d_{Hg}| &<& 3.1 \times 10^{-29}\,\,e\,cm\hspace{0.1cm}(95\%\,\,\mathrm{C.L.}),\\
|d_n| &<& 2.9 \times 10^{-26}\,\,e\,cm\hspace{0.1cm}(90\%\,\,\mathrm{C.L.}), \label{eq:neutrondata}\\
|d_{Tl}| &<& 9.0 \times 10^{-25}\,\,e\,cm\hspace{0.1cm}(90\%\,\,\mathrm{C.L.}). 
\end{eqnarray}
The connection of these measured dipole moments to more fundamental particles which can potentially interact with the new sector is more readily apparent when knowing the following relations~\cite{Pospelov:2005pr}
\begin{eqnarray}
d_{Tl} &=& -585\,d_e - e\,43\,\mathrm{GeV} \times ( C^{(0)}_S - 0.2\,C_S^{(1)}) \sim -585\,d_e, \label{eq:thalliumredm} \\
d_{Hg} &=& -1.8 \times 10^{-4}\,\mathrm{GeV}^{-1}\,\,e\,\bar{g}^{(1)}_{\pi N N} + 10^{-2}\,\,d_e + 3.5 \times 10^{-3}\,\mathrm{GeV}\,e\,C^{(0)}_S,  \label{eq:mercuryedm}  \\
d_n &\sim& 3 \times 10^{-16}\,\theta_\mathrm{QCD} + 0.7\,(d_d - {1 \over 4} \,d_u) + 0.6\,e\,(\tilde{d}_d +  {1 \over 2}\,\tilde{d}_u).  \label{eq:neutronedm}
\end{eqnarray}
Here $C_S$ and $\bar{g}^{(1)}_{\pi N N}$ are the CP-odd electron-nucleon and pion-nucleon couplings, respectively.  Determining both requires complex many-body computations.  Also, $d_i$ and $\tilde{d}_i$ are the EDM and chromo-EDM for particle $i$.  Finally, $\theta_\mathrm{QCD}$ is the coefficient of the well-known operator in the QCD lagrangian
\begin{equation}
{g_s \over 32\pi^2}\,\,\theta_\mathrm{QCD}\, G_{\mu\nu}\tilde{G}^{\mu\nu},
\end{equation}
where $G_{\mu\nu}$ is the gluon field strength.  When doing our estimates, we assume a Peccei-Quinn~\cite{Peccei:1977hh} mechanism so $\theta_\mathrm{QCD}$ has a natural explanation for why it is vanishingly small.  Below, we focus on the $d_q$ and $\tilde{d}_q$ (chromo-)EDMs.  Our estimates are sufficiently general so as to apply to the electron ($d_e$) EDMs.  To make an example, below we make a direct comparison to the neutron EDM.
\newline
\begin{figure}[t]
\centering
\includegraphics[width=9.4truecm,height=12.83truecm,clip=true]{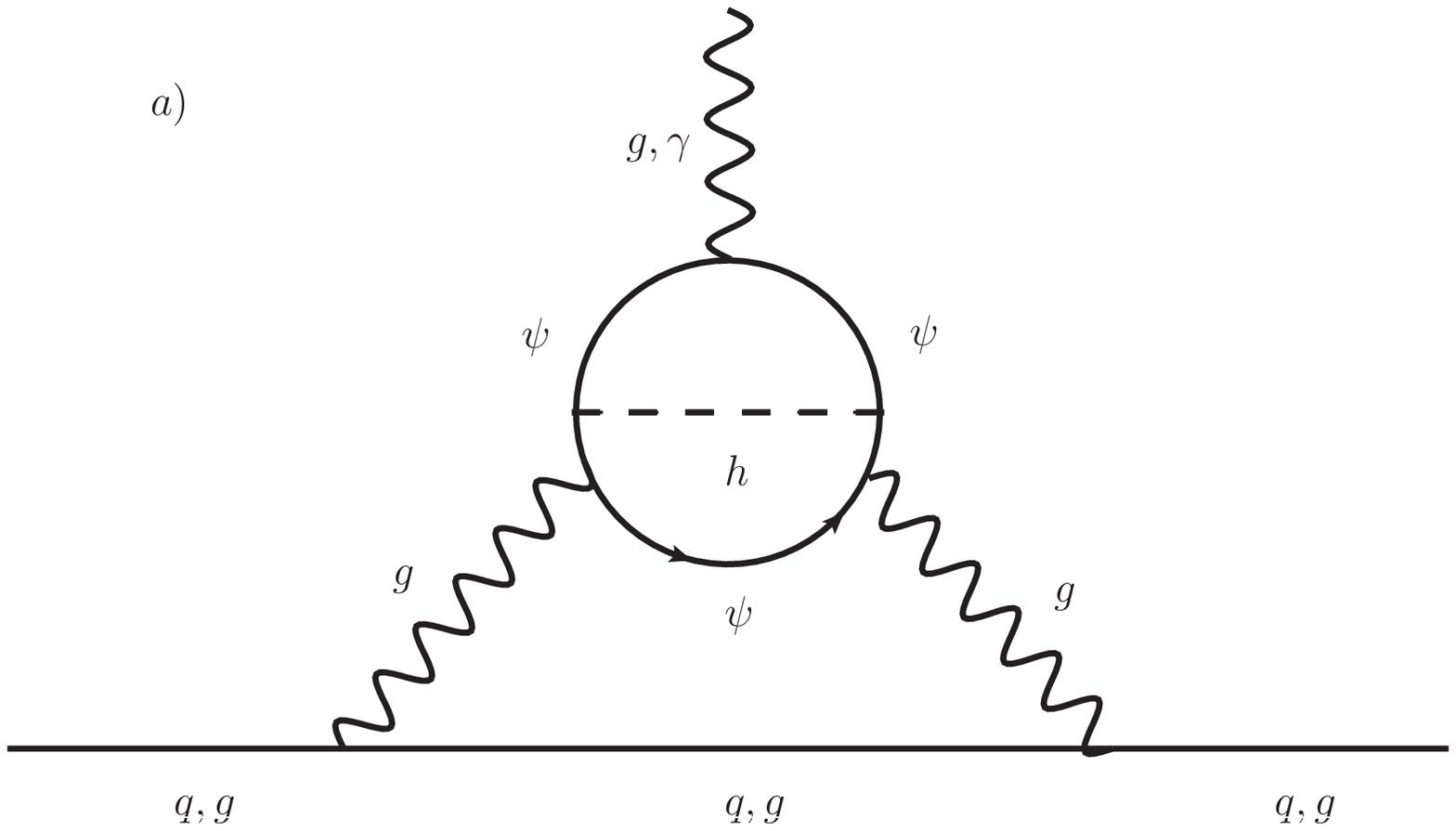} \hspace{-2.6cm}
\includegraphics[width=9.4truecm,height=12.9truecm,clip=true]{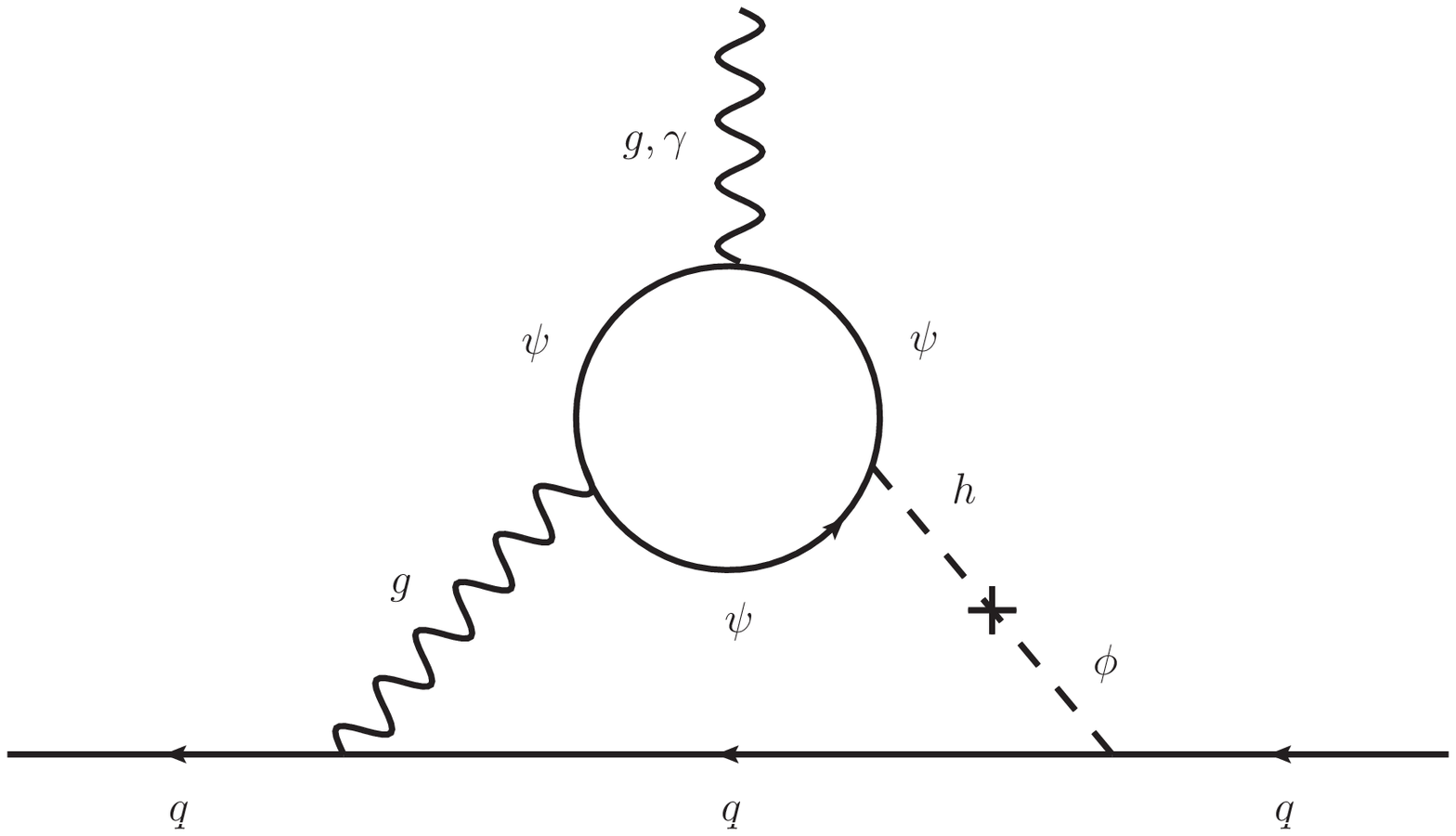} \\ \vspace{-7cm} \centering
\includegraphics[width=9.4truecm,height=12.8truecm,clip=true]{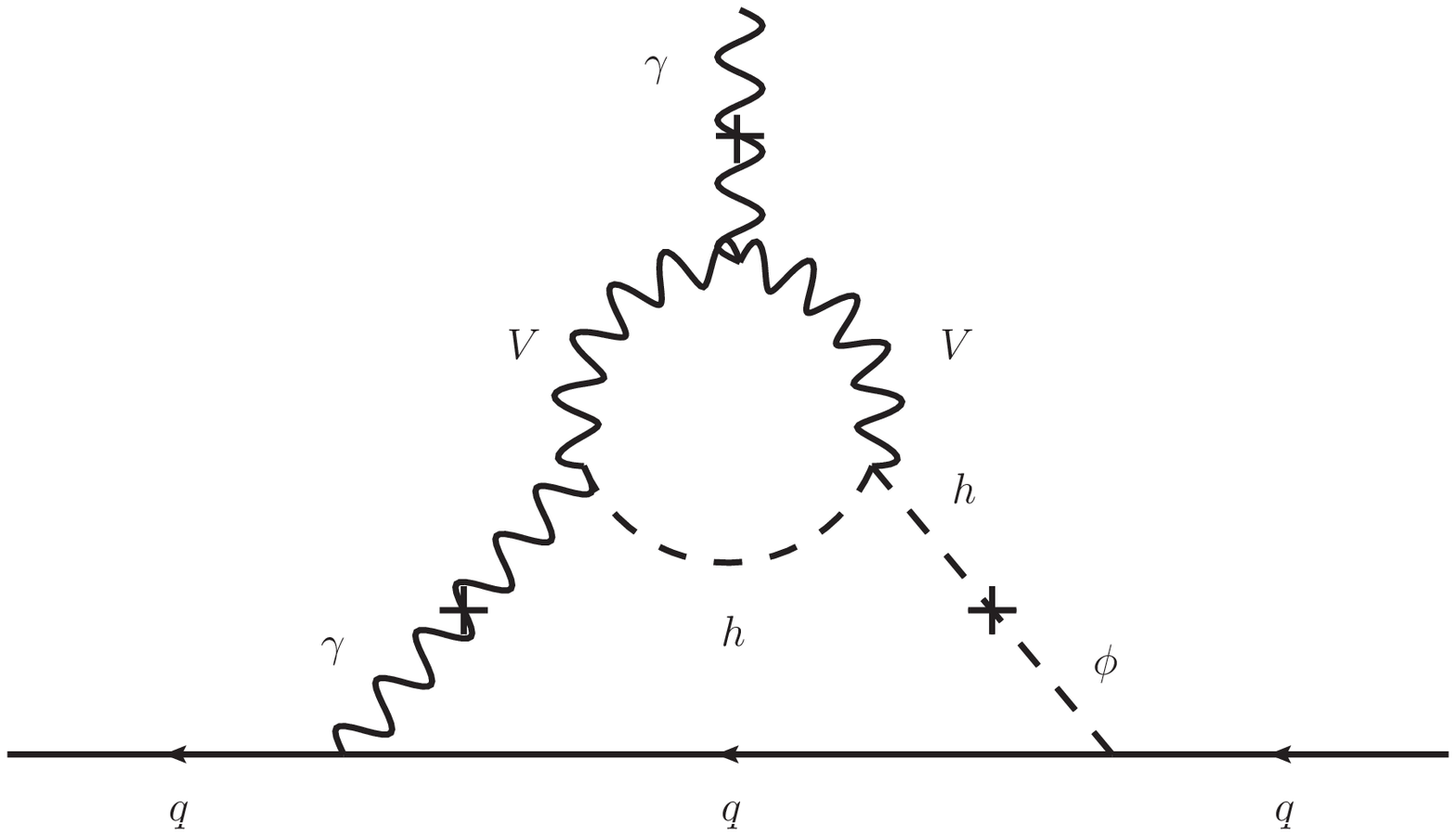} \hspace{-2.6cm}
\includegraphics[width=9.4truecm,height=12.83truecm,clip=true]{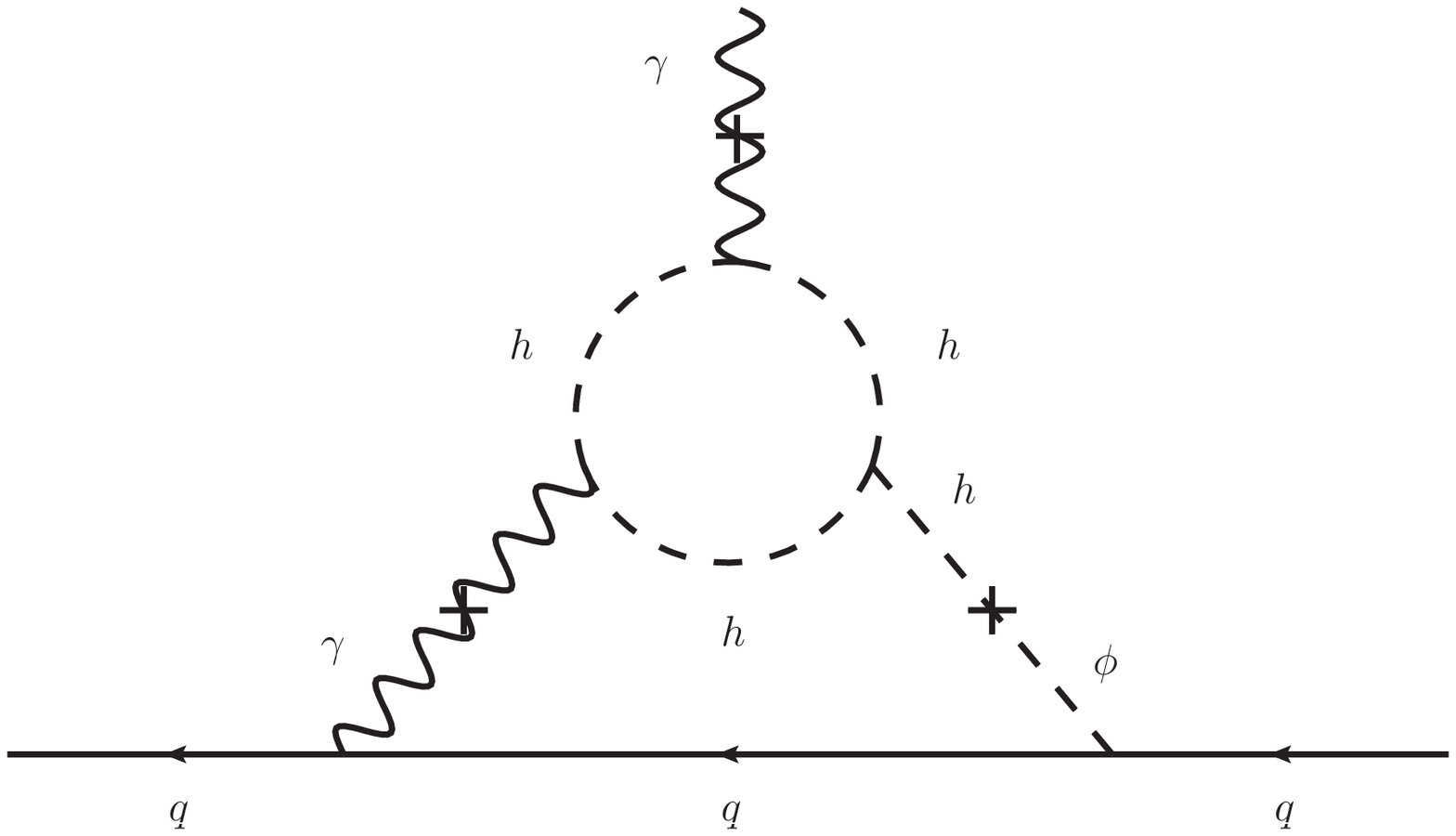} 
\vspace{-7.2cm}
\label{fig:CPvioldiag}
\caption{The dominant Weinberg-type (top left) and Barr-Zee-type (others) contributions to various EDMs and chromo-EDMs.  The Weinberg-type diagrams~\cite{Weinberg:1989dx} are loop suppressed in comparison to the SM diagrams.  The Barr-Zee diagrams~\cite{Barr:1990vd} feature dark fermions (top diagrams), dark higgses and gauge bosons (bottom diagrams) running in the loop.  $\phi$ is the SM higgs and $h$ is short dark higgses.}
\end{figure}
\newline
As discussed above, the CP violation in the dark higgs sector can be transmitted to the SM through the higgs portal.   The dominant EDMs and chromo-EDMs are shown in Figure 4.  The upper-left diagram in  Figure 4 was first identified as a EDM source by Weinberg~\cite{Weinberg:1989dx}.  It is loop suppressed in comparison to the Barr-Zee diagrams~\cite{Barr:1990vd} in the figure.  We no longer consider diagrams of the Weinberg type.  It is straightforward to estimate the contributions to the quark  Barr-Zee (chromo-)EDM from the diagrams shown in Figure 4.
\begin{eqnarray}
\biggl({d_q \over e} \biggr)_{\psi\,\,\mathrm{loop}} &\sim& {4\pi \alpha \over (16\,\pi^2)^2}\,\biggl({m_q \over v_\mathrm{dark}^2} \biggr)\,\biggl({v_\mathrm{dark} \over \Lambda} \biggr)^2\,f\biggl({m_\psi^2 \over m_h^2},{m_\psi^2 \over m_\phi^2}\biggr)\,\mathrm{Im}Z\\
\biggl({\tilde{d}_q \over g_s} \biggr)_{\psi\,\,\mathrm{loop}} &\sim& {4\pi \alpha_s \over (16\,\pi^2)^2}\,\biggl({m_q \over v_\mathrm{dark}^2} \biggr)\,\biggl({v_\mathrm{dark} \over \Lambda} \biggr)^2\,f\biggl({m_\psi^2 \over m_h^2},{m_\psi^2 \over m_\phi^2}\biggr)\,\mathrm{Im}Z \\
\biggl({d_q \over e} \biggr)_{V\,\,\mathrm{loop}} &\sim&  {4\pi \alpha \over (16\,\pi^2)^2}\,\biggl({m_q \over v_\mathrm{dark}^2} \biggr)\,\biggl({v_\mathrm{dark} \over \Lambda} \biggr)^2\,f\biggl({m_V^2 \over m_h^2},{m_V^2 \over m_\phi^2}\biggr)\,\mathrm{Im}Z  \\
\biggl({d_q \over e} \biggr)_{h\,\,\mathrm{loop}} &\sim& {4\pi \alpha \over (16\,\pi^2)^2}\,\biggl({m_q \over v_\mathrm{dark}^2} \biggr)\,\biggl({v_\mathrm{dark} \over \Lambda} \biggr)^2\,f\biggl({m_\phi^2 \over m_h^2}\biggr)\,\mathrm{Im}Z.
\end{eqnarray}
Here $g_s$ is the strong coupling constant.  We have assumed the lightest dark higgs is lighter than the SM higgs.  Here $\mathrm{Im}Z$ contains all of the CP violating phases.  It is clear the EDMs are parametrically suppressed by $\Lambda$.  Following~\cite{Barr:1990vd} we conservatively take the phases to be maximal and therefore $\mathrm{Im}Z \sim \mathcal{O}(1)$.  $f$ calculated can explicitly; but, we do not do so here.  Interestingly, $v_\mathrm{dark}$ does not have an impact on our estimates except through the masses of the dark fermions and higgses.  It is now clear why it is important to do these estimates.  $\Lambda$ can be small depending on the model.  If we conservatively assume, however, that $f \sim  \mathcal{O}(1)$ and $\Lambda \sim 10$ TeV, then the quark chromo-EDM and EDM are of order 
\begin{align}
\biggl({d_q \over g_s} \biggr) \sim 1 \times 10^{-29} && \biggl({d_q \over e} \biggr) \sim 7 \times 10^{-31}.
\end{align}
In comparison to equations~\ref{eq:neutrondata} and~\ref{eq:neutronedm}, CP violation in the dark sector can be consistent with the data due to the parametric suppression of $\Lambda$.  
 
 \subsection{Long-Lived Particles at the LHC}
 
The mechanism discussed in this paper relies on long-lived messenger particles that remain after the dark phase transition.  Our messenger fermions have a lifetime in excess of $10^{-3}$ seconds.  (See Section~\ref{sec:mfldt} for further discussion.)  In general, experimental searches for visible long-lived particles can be divided into two categories:  (1) Searches for visible particles whose lifetime is long enough such that the particles are around today and (2) searches with particle accelerators~\cite{Perl:2001xi}.  Since we require our messenger particles decay, the former does not apply.  Accelerator searches are limited by the particle mass range and coupling to the SM.  Present bounds from LHC generally roughly long-lived colored particles like the gluino with masses less than 600 GeV~\cite{Aad:2011hz,Khachatryan:2011ts,Khachatryan:2010uf}.  Although as we discuss below, these bounds can be a bit porous (or stronger) depending on the particle lifetime and electric charge~\cite{Aad:2011mb}.  
%
%
%
%
%
%
In this section we overview some distinctive features of these models which can appear at the LHC.  We particularly focus on signatures of long-lived particles with and without large amounts of missing transverse momentum.  This is a defining signature for this class of models and was the subject of a previous paper~\cite{Walker:2009ei}.  The results of that paper are summarized in this section.
\newline
\newline
After production at the LHC, the messenger quarks can form color singlets with valence SM quarks.  
In detail, the production process at the LHC is 
\begin{align}
p + p \to hQ + \overline{hQ} && p + p \to mQ + \overline{mQ} 
\end{align}
where $hQ$ and $mQ$ are the colored singlets
\begin{align}
hQ =  Q_L\,q\,q && mQ =  Q_L\,q.
\label{eq:csinglets}
\end{align}
Here $q$ are SM quarks.  Of course, the heavy messenger quark, $Q_L$ can form additional color singlets like $Q_L \,Q_L$, $Q_L \,Q_L \,q$ and $Q_L \,Q_L \,Q_L$.  However, these states require heavier quarks to be created out of the vacuum to form the desired color singlet and are disfavored.  We focus only on the singlets in equation~\ref{eq:csinglets}.  $hQ$ and $mQ$ have the electric charges $Q = 0, \pm 1$.  As for the messenger leptons, the production process at the LHC is
\begin{align}
p + p \to L_L + \overline{L}_L 
\end{align}
which is produced by an off-shell photon or $Z$ boson.  The messenger leptons have the same electric charge as $hQ$ and $mQ$.  Given messenger quarks and leptons of the same mass, the messenger quarks have a larger cross section because of the strongly coupled QCD production.  
\newline
\newline
The ATLAS detector is about twice the size of the CMS detector.  Since the time it takes a relativistic particle to transverse the central part of the larger ATLAS detector is on the order of 10-100 nanoseconds, it is clear our messenger particles with a lifetime of $10^{-3}$ seconds are long-lived on detector time scales.  It is well-known long-lived particles can generate dramatic signatures at the LHC.  For a review of the signatures of long-lived particles, please see~\cite{Fairbairn:2006gg}.  
In summary, the experimental reasons are:
\begin{enumerate}
\item Observation of particles that are ``out-of-time" with the rest of the event.  This includes particles that are stopped in the detector and decay far after the initial event.
\item Observation of ``heavy muons" in the muon detector. 
\item Observation of events with heavily ionized tracks in the electronic calorimeter.
\item Observation of events with sizable displaced vertices.
\item Observation of particles which oscillate between positively and negatively charged in the detector.
\item Observation of particles which oscillate between charged and neutral in the detector.  This allows the distinct signature possible where a fraction of the events have two back-to-back tracks and others with only one one track.  
\end{enumerate}
%
%
All of these signatures are possible for the messenger quarks.  The long-lived leptons can produce only the first three items.  Long-lived particles are out-of-time with the rest of the event because the travel at velocities, $\beta$, that are subluminal.  The ATLAS and CMS collaborations are sensitive to velocities in the range of  $0.7 < \beta < 0.9$ and $0.5 < \beta < 0.8$, respectively\footnote{Here the speed of light is $\beta = 1$.}.  The range in velocities is mostly due to the fact that the ATLAS detector is so much large than the CMS detector.  See, \textit{e.g.}~\cite{Walker:2009ei}, for the velocities of different long-lived particles with different masses.  These slow moving particles punch through the calorimeters like ``cannonballs" and, if charged, arc in the muon detector like a slow, heavy muon.  When transversing the electronic calorimeter, slow moving charged long lived particles generate the heavily ionized tracks.  The most stringent bounds on the mass of long-lived particles are due to particles with large electric charges (that subsequently generate prominent tracks)~\cite{Aad:2011mb}.  To date, our messenger fermions have a small enough charge to evade this most stringent bound.  As mentioned above, colored long-lived particles pick up valence quarks.  The exchange of valence SM quarks with the detector material in the calorimeters generates the distinctive last three signatures in the above list.
\newline
\newline
Many models can produce long-lived particles at the LHC.  A defining signature, however, for this class models is the production of long-lived particles plus large amounts of missing transverse momentum ($\ptmiss$).  This occurs via the following processes
\begin{align}
p + p \to hQ + \overline{hQ}  + V && p + p \to mQ + \overline{mQ}  + V && p + p \to L_L + \overline{L}_L  + V
\label{eq:signalproc}
\end{align}
where $V$ is a dark gauge boson.  The dark gauge boson decays, with a large branching fraction, into the GeV dark matter.  This generates the $\ptmiss$.  The branching fraction into the SM goes as $\sim \epsilon^2 e^2/g_\mathrm{dark}^2$ which  is suppressed and can be miniscule.

\subsubsection{Results}

In this section,  we simulate the production of a 1 TeV heavy messenger quark at the LHC for a 14 TeV center-of-mass energy.  Included are events with and without emission of a 50 GeV dark gauge boson.  Importantly, a detector simulator was created and implemented which accounts for ATLAS and CMS's detector geometry as well as nuclear interactions of the long-lived colored particles with the detector material.  In addition to the velocity cuts described in the previous paragraph, the following were taken to ensure the analysis would be consistent with both ATLAS and CMS collaborations~\cite{Aad:2009wy,CMScuts,privateZ}. See \cite{Walker:2009ei} for additional details.  The $p_T$ of the long-lived particle is required to have
\begin{align}
p_{T\,\,\mathrm{ATLAS}} > 250\,\,\mathrm{GeV}  && p_{T\,\,\mathrm{CMS}} > 30\,\,\mathrm{GeV}.
\end{align}
We also require events to be in the center of the detector.
\begin{equation}
|\eta| \leq 2.4.
\end{equation}
For the CMS collaboration, we also required the averaged reconstructed particle mass to be
\begin{equation}
m_\mathrm{CMS} > 100\,\,\mathrm{GeV}.
\end{equation}
ATLAS also requires no hard jets in the hadronic calorimeter to come within
\begin{equation}
\Delta R_\mathrm{ATLAS} \leq 0.4
\end{equation}
of the muon tracks.  Here $\Delta R \equiv \sqrt{\Delta \eta^2 + \Delta \phi^2}$ where $\eta$ ($\phi$) is the psuedorapidity  (transverse angle) of the two tracks in question.  We finally veto all of the events with only one track in the detector.  This allows computation of the $\ptmiss$.  Formally, it is defined as 
\begin{equation}
\ptmiss = -\sum_i p_{T_i}
\label{eq:ptmiss}
\end{equation}
where $i$ runs over all of the visible particles.  For the ATLAS cuts, the collaboration finds a signal to background ratio of $S/B = 2.6 \times 10^3$ for an integrated luminosity of 1 fb$^{-1}$.  Similarly, for the CMS cuts that collaboration find a ``background free region" for 100 pb$^{-1}$.  Thus, for this analysis we do not consider on any SM backgrounds.  There is however an irreducible background with which we must be careful.  The $Z$ boson can be produced along with the long-lived messenger quarks and decay to neutrinos (equation~\ref{eq:signalproc} where the $V$ boson is replaced by the $Z$ boson).  This generates our signal.  The $Z$ boson also decays to leptons and jets.  Thus, the invisible decay width of the Z can be determined and the number of ``signal" events generated by the irreducible background ascertained.  To be safe, we require the signal cross section to be no less than 10\% for this irreducible background.
\newline
\begin{figure}[t]
\centering
\includegraphics[width=7truecm,height=6truecm,clip=true]{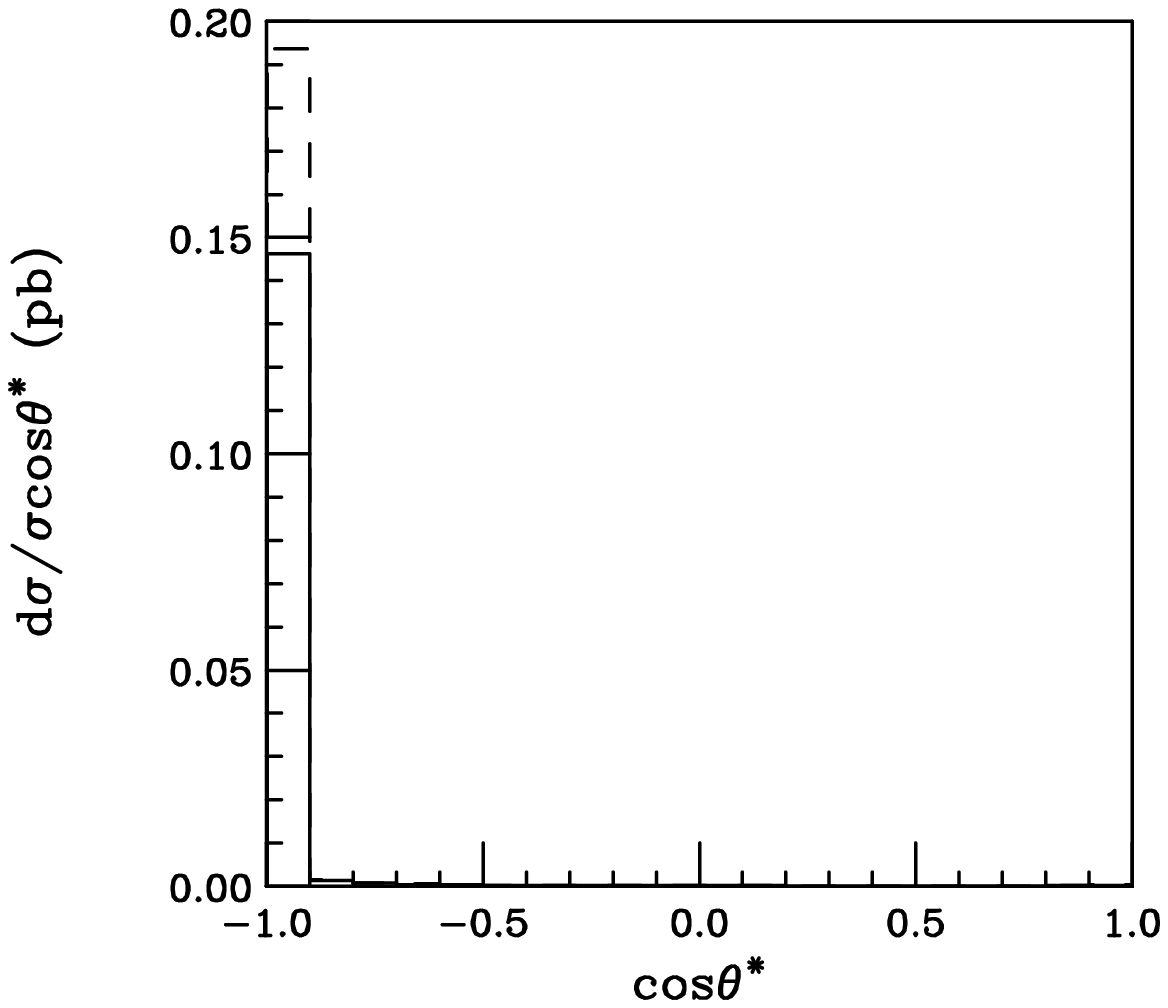}    \hspace{1.0cm}
\includegraphics[width=7truecm,height=6truecm,clip=true]{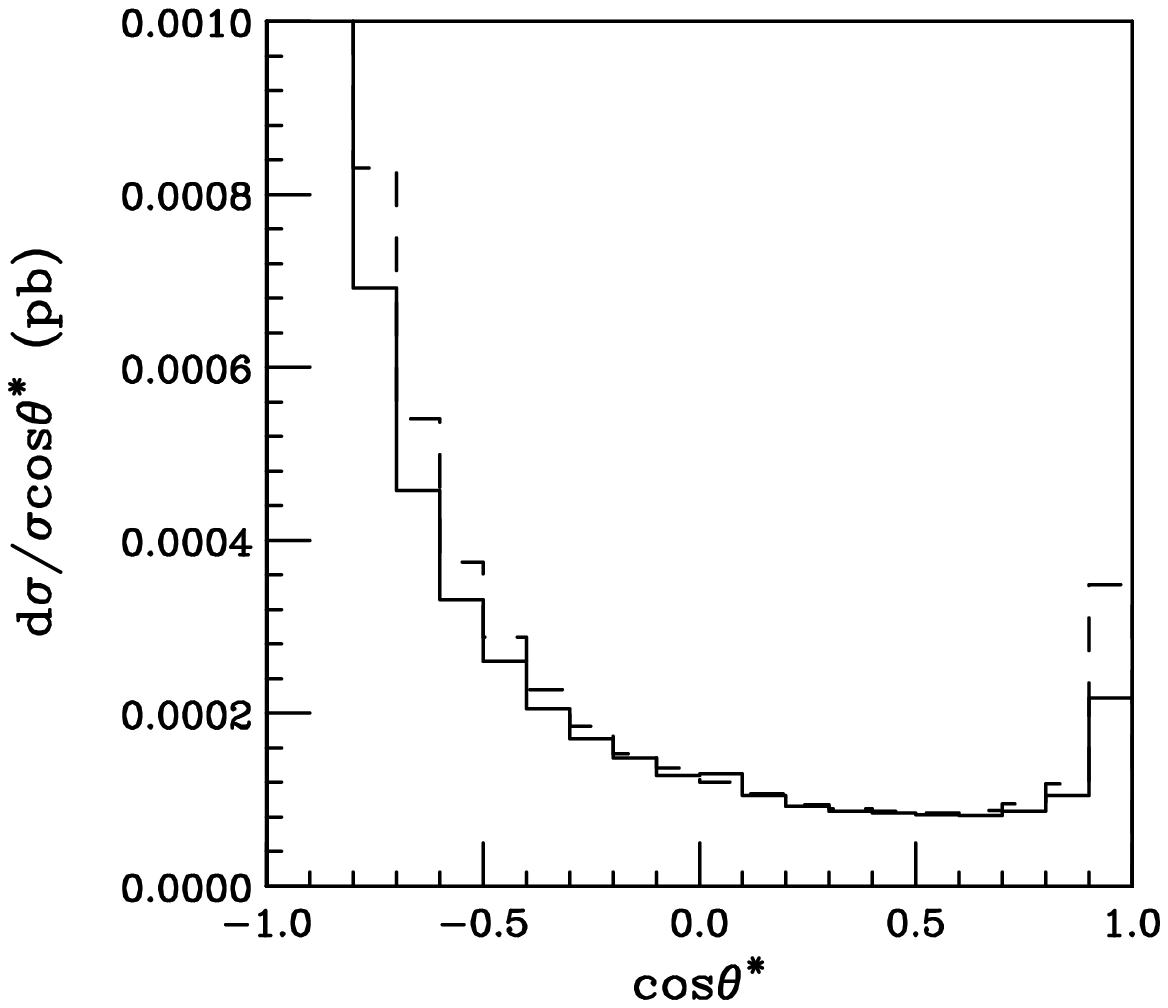} \\ 
	 	 \label{fig:longlived}
\caption{Reconstructed transverse angle between the two tagged 1 TeV long-lived color particles.  Included are events with and without the emission of a 50 GeV dark gauge boson.  A detector simulation which reproduces the GEANT4 response to colored meta-stable particles is implemented.  (This simulation takes into account the ATLAS and CMS detector topology and composition.)  In the left panel, it is clear most of the events are back-to-back.  The right panel zooms in on the non back-to-back events that are generated by the emission of the dark gauge boson.  The dashed (solid) lines are ATLAS (CMS).} 
\end{figure}
\newline
In Figures 4 and 5, we place the results of our simulation.  In the left panel of Figure 4, it is clear the processes that pair produce only the colored long-lived particles dominate over the processes that produce both the long-lived particles and the $\ptmiss$.  These events are dominantly back-to-back.  The right panel in Figure 4 is the same as the left panel but emphasizes the events that are not back-to-back.  These events are created by the emission of a dark boson in addition to the colored long-lived particles.  Thus, to search for the signal events with large $\ptmiss$, we additionally require
\begin{equation}
\cos \phi > -0.9
\label{eq:cutforptmiss}
\end{equation}
where $\phi$ is the transverse angle between the two colored long-lived particles.  This effectively eliminates the events with only pair production of the colored long-lived particles.  In Figure 6, we plot the resulting missing energy from the emission of the dark gauge bosons.
\newline
\begin{figure}[t]
\centering
\includegraphics[width=8truecm,height=7truecm,clip=true]{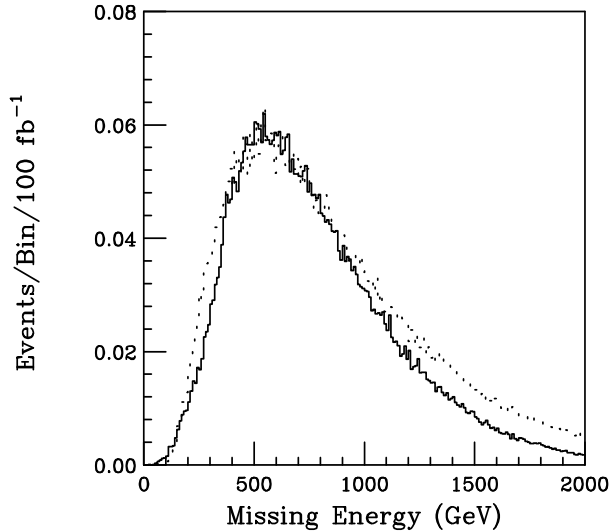}
\label{fig:longlived2}
\caption{The resulting missing momentum after the cut in equation~\ref{eq:cutforptmiss} is applied.  Like Figure 5, two tagged 1 TeV long-lived color particles are simulated with and without the emission of a 50 GeV dark gauge boson.  The latter event remain after equation~\ref{eq:cutforptmiss} is applied.  The kinematic cuts in the text are applied along with a simulation of the GEANT4 detector response to colored meta-stable particles.  This simulation takes into account the ATLAS and CMS detector topology and composition.  The dotted (solid) lines are ATLAS (CMS).}
\end{figure}
\newline
\textbf{Other Potential Signatures of Dark Gauge Bosons:}  It is possible that the long-lived particles are too heavy to observe at the LHC.  If the mechanism presented in this paper is realized in a larger theory (such as supersymmetric UV completion), other complementary signatures are possible.  Typically dark matter searches at the LHC feature searching for distributions with kinematic edges (in addition to those signatures have large amounts of $\ptmiss$).  Radiating of dark gauge bosons from the SUSY partners (or the equivalent) can add additional sources of $\ptmiss$.  Some distinctive signatures are detailed in~\cite{Agashe:2010gt,Agashe:2010tu}.
\newline
\newline
Like the electroweak interactions, discovering the $W$ and $Z$ bosons confirmed the $SU(2)_L \otimes U(1)_Y \to U(1)_{em}$ symmetry breaking pattern.  This is also true for the dark gauge group symmetry breaking.  The signature discussed in the previous paragraph is defining.  More definitive evidence can be generated if the dark gauge bosons produced by the process in equation~\ref{eq:signalproc} have a sizable branching fraction into the SM.  Assuming enough events, the invariant mass of the resulting SM products could then be reconstructed.  This is possible because tagging the two long-lived particles effectively eliminates all backgrounds.  A mass peak way from the Z-boson mass (see related discussion below equation~\ref{eq:ptmiss}) would give a positive identification of the dark gauge boson mass.  Since the system is fully reconstructable, the spin can be determined thereby giving positive identification of the dark gauge boson's identity.  

%
%

\section{Conclusions and Outlook}

In this paper, we presented a model in which a generic, dark chiral gauge group undergoes a first order phase transition in order to generate the observed baryon asymmetry in the universe, provide a viable dark matter candidate and explain the observed baryon-to-dark matter ratio of relic abundances~\cite{Agashe:2010gt}.  We focused a model in which a dark $SU(2)_D \times U(1)_D$ gauge group is spontaneously broken to the diagonal $Z_2$ center.  This symmetry is used to stabilize a dark matter candidate.  Beyond the dark matter candidate, messenger leptons and quarks (charged under both the dark and SM gauge groups) were included.  The details of the particle content of this model is featured in Section~\ref{sec:model}.  In analogy to electroweak baryogenesis, we showed the model generates an excess of heavy messenger quarks and leptons.  These fermions are long-lived and eventually decay to the SM and dark matter to generate a ratio of five dark matter candidates for every one SM baryon.  Ultimately, this ratio is due simply to the requirement of gauge anomaly freedom.  The dark gauge group symmetry breaking scale can be much larger than the weak scale.  This is because dark phase transition violates $B - L$ but preserves $B + L$.  Therefore, the excess of messenger quarks (heavy baryons) and leptons is not erased by the background SM processes.  Eventually, the excess of heavy baryons decay to the Standard Model and dark matter to generate an excess of Standard Model baryons.  
\newline
\newline
We showed new, large CP violating phases could be permitted in the dark sector sector.  The contributions to the various electric dipole moment (EDM) effective operators
are at most
three-loop and parametrically suppressed.  Additionally briefly outlined some distinctive experimental signatures previously described in~\cite{Agashe:2010gt,Walker:2009en,Walker:2009ei,Agashe:2010tu}.  Notably, as first discussed in~\cite{Walker:2009en}, these dark matter scenarios feature long-lived particles which can be observed at colliders.
\newline
\newline
Getting the right baryon-to-dark matter ratio requires a relatively light dark higgs boson.  In order to avoid fine-tuning of these higgses, we outlined how approximate symmetries in the higgs sector stabilize the dark and electroweak scales thereby mitigating the hierarchy problem.  Thus, given new physics at the scale $f$
, the SM and dark higgses can have natural masses of order $\lambda^4 f/(4\pi)$ and  $\lambda^8 f/(4\pi)^2$, respectively.  The latter mass is due to ``overlapping" chiral symmetries which force the dark higgs mass to be generated at three loops.  In addition, given a chiral symmetry which protects the dark matter mass, this model radiatively generates a dark matter mass of order of the electroweak vacuum expectation value (vev) suppressed by a loop factor ($\sim v_\mathrm{ew}/16\pi^2$).  
\section*{Acknowledgments}

We would like to thank K.~Agashe, Z.~Chacko, H.~Georgi, H.-S.~Goh, D.~Kim, D.~Krohn, M.~Toharia, J.~Mason, L.~McLerran, R.~Mohapatra, M.~Papucci, T.~Tait and especially S.~Chivukula, J.~Hewett, A.~E.~Nelson and E.~Simmons for valuable discussions and encouragement.  Special thanks to S.~Chivukula and E.~Simmons for detailed comments on the draft.  The author thanks the Aspen Center of Physics for their hospitality during which part of this work was completed.  This work was supported in part by a University of California Presidential Fellowship, a fellowship from the National Academies of Science and a grant from the LHC Theory Initiative.
\newline
\newline
During the completion of this work, \cite{Shelton:2010ta,Haba:2010bm,vonHarling:2012yn,Petraki:2011mv,Barr:2011cz,Frandsen:2011kt} appeared on the arXiv which shared the motivation to generate an asymmetry with a non-trivial dark gauge group and have the asymmetry communicated to the visible sector.  As we have emphasized, in our previous work \cite{Agashe:2010gt}, we first presented the idea of how new dark chiral gauge group could generate sphalerons at finite temperature, nucleate additional bubbles and may be relevant for electroweak baryogenesis.  The current work is to provide a specific model to the broad ideas described therein.  Other novel dark matter asymmetry scenarios that may be of interest to the reader can be found in~\cite{McDonald:2012vw,Kamada:2012ht,Cui:2011ab,Davoudiasl:2011aa,D'Eramo:2011ec,Chowdhury:2011ga,McDonald:2011sv,Graesser:2011vj,Kumar:2011np,Cui:2011qe,MarchRussell:2011fi,Cheung:2011if,Bell:2011tn,Heckman:2011sw,Falkowski:2011xh,Hall:2010jx,Behbahani:2010xa,Blennow:2010qp,Buckley:2010ui,McDonald:2011zz,Davoudiasl:2010am,Dutta:2010va}.



\section*{Appendix}

\section*{A. Messenger Fermion Effective Operators}
\label{app:effectiveops}

In this appendix, we list all the dimension six effective operators which involving the new particle content in Section~\ref{sec:model} and the SM.  Each operator implicitly has a prefactor of $c_i/\Lambda_i^2$ where $i$ labels the operator.  The operators with only one messenger fermion are
\begin{align}
\mathcal{O}_1 &= (q_L^c \, \sigma\, Q_L)\, (l_L^c\,\sigma\,\chi_L)           	&&  \mathcal{O}_2 = (e_R^c \, \sigma\,L_L)\, (l_L^c \, \sigma\, \chi_L) \nonumber  \\
\mathcal{O}_3 &= (u_R^c\,\sigma\,U) \,(e_R^c \,\sigma\,\chi_{R_1})         	&& \mathcal{O}_4 = (\chi_{R_1}^c\,\sigma\, E)\, (\l_L^c, \sigma\,n_R)  \nonumber  \\
\mathcal{O}_5 &= (d_R^c\, \sigma\,U)\, (\chi_{R_1}^c\, \sigma\, n_R)      	&& \mathcal{O}_6 = (\chi_{R_1}^c\,\sigma\,  N)\, (l_L^c\, \sigma\,  n_R)   \nonumber  \\
\mathcal{O}_7 &= (u_R^c \,  \sigma\,D)\, (e_R^c \,  \sigma\, \chi_{R_1})    	&&   \mathcal{O}_8 = (d_R^c\,  \sigma\, D)\, (\chi_{R_2}^c\, \sigma\, n_R)   \nonumber \\
 \mathcal{O}_9 &= Q_L\, q_L\, q_L\,\chi_L      						&&   \mathcal{O}_{10} = U\, u_L\, d_R\, \chi_{R_2} \nonumber \\
\mathcal{O}_{11} &= D\,u_R\,d_R\,\chi_{R_1}                          		     	&& \mathcal{O}_{12} =  Q_L\,u_R\,e_R\,\chi_L  \\
\mathcal{O}_{13} &= Q_L\,d_R\,n_R\,\chi_L      				     		&& \mathcal{O}_{14} = U\, q_L\, l_L\,\chi_{R_2} \nonumber \\  
\mathcal{O}_{15} &= D\, q_L\, l_L  \,\chi_{R_1}    					&&   \mathcal{O}_{16} = L_L\, q_L\,u_R\,\chi_L \nonumber   \\  
\mathcal{O}_{17} &= L_L\, l_L\, n_R\,\chi_L       						&&   \mathcal{O}_{18} = N\, l_L\,e_R\,\chi_{R_2} \nonumber  \\
 \mathcal{O}_{19} &= E_L\, l_L\, e_R\,\chi_{R_1}       					\nonumber
\end{align}
These operators were constructed only considering the symmetries of the particles involved.  For this appendix, all the lowercase, latin particles are the easily identifiable SM particles except for $n_R$ which is the right hand component of the SM neutrinos.  The important point here is the first eight operators are generated by new gauge bosons resulting from dark-electroweak unification.  These operators encode how the messenger fermions decay.  For completeness, we also list the operators involving two messenger fermions.  For ease, we restrict to operators resulting from scalar exchange.
\begin{align}
\mathcal{O}_{15} &= Q_L\, Q_L\, u_R\, n_R             &&   \mathcal{O}_{21} = D\, N\, d_R\, l_L \nonumber \\
\mathcal{O}_{16} &= Q_L\, U\, \chi_L\, \chi_{R_2}   &&   \mathcal{O}_{22} =  L_L\, L_L\, e_R\, n_R  \nonumber \\  
\mathcal{O}_{17} &= Q_L\, D\,  \chi_L\, \chi_{R_1}  &&   \mathcal{O}_{23} = L_L\, N\, \chi_L\, \chi_{R_2}  \\
\mathcal{O}_{18} &= Q_L\, L_L\, q_L\, n_R               &&   \mathcal{O}_{24} = L_L\, E\, \chi_L\, \chi_{R_1} \nonumber \\  
\mathcal{O}_{19} &= U\, D\, d_R\, d_R                       &&   \mathcal{O}_{25} =  N\, E\, l_L\, l_L \nonumber\\
\mathcal{O}_{20} &= U\, E\, d_R\, l_L                         &&   \hspace{0.1cm} \nonumber 
\end{align}
Finally, the operators with four messenger fermions are
\begin{align}
\mathcal{O}_{26} &= Q_L\, Q_L\, U\, D                   &&   \mathcal{O}_{28} = Q_L\, L_L\, N\, D\\
\mathcal{O}_{27} &= Q_L\, L_L\, U\, E                    &&   \mathcal{O}_{29} =  L_L\, L_L\, N\, E \nonumber 
\end{align}
There are no operators with three messenger and/or dark matter fermions.  We have omitted the trivial dimension six operators with four dark matter candidates. 

\section*{B.  Stabilizing the Dark and Electroweak Scales}
\label{sec:stable}

In this paper, we have discussed the effective description of this model at and below the dark and electroweak symmetry breaking scales.  This description features a very light dark higgs boson which facilitates efficient dark matter annihilation; as a result, this annihilation generates a small dark matter relic abundance.  Therefore, by construction, the observed dark matter abundance comes solely from the decay of decoupled messenger fermions.  There are some inconsistencies, however, that mar this picture.  It is well-known that radiative corrections to the electroweak (and dark) higgs boson masses are naively unsuppressed.  This would mean the mass of each of the higgses could be as large as the cutoff scale of new physics.  Since the only \textit{known} scale of new physics beyond the electroweak scale is the Planck scale at $10^{19}$ GeV, an incredible amount of fine-tuning of the higgs potential would be required to generate the dark and electroweak symmetry breaking scales.  
This is the well-known hierarchy problem.  With such a potentially large mass, the dark higgs would be integrated out of the low-energy effective theory and could not be used to help explain the ratio of the dark matter to baryon relic abundance.  Of course, the problems with a Planck scale electroweak higgs boson mass has been written about extensively.  As a final point to emphasize the magnitude of this problem, precision electroweak measurements generally set the scale of new physics to be around $5$ TeV~\cite{Nakamura:2010zzi,Barbieri:2000gf}.  Yet, the scattering of the longitudinal W bosons~\cite{Lee:1977eg} requires the SM higgs must have a mass at most around 1 TeV.  This deficit is called the little hierarchy problem.  For these reasons, an explanation outlining how the the dark and electroweak symmetry breaking scales can be stabilized 
must be germane to the goals of this publication.  In the following, we do just that by using approximate symmetries between the dark and electroweak sectors.  Interestingly, this symmetry can be broken in such a way so that the electroweak and dark higgs masses can be naturally of order the weak scale and around a GeV, respectively.  Finally, for completeness in the next section,  we briefly discuss a supersymmetric UV completion which can address the hierarchy problem as well.


\subsection*{B.1 A Motivation from Dark and Electroweak Unification} 
\label{subsec:unify}

It is well known the electroweak gauge group can be expanded to $SU(2)_L \otimes SU(2)_R$.  This gauge group is isomorphic to $SO(4)$.  At a high scale, it is assumed $SU(2)_R$ is broken to hypercharge thereby giving an origin for global custodial symmetry.    As previously discussed, the custodial symmetry is explicitly broken by hypercharge and the top\footnote{We neglect the other yukawa terms because they are extremely small relative to the top yukawa coupling.} yukawa coupling.     
Even with these explicit symmetry breaking terms, custodial symmetry is essential for the weak interactions and preserves the well-measured relation\footnote{Preservation of this relation means whatever new physics is associated with the weak scale is independent of electroweak symmetry breaking mechanism.}, 
\begin{equation}
M_W = M_Z \cos\theta_W.
\end{equation}
Here $\theta_W$ is the Weinberg angle.   
By analogy, we assume the dark gauge sector has a similar custodial structure.  This produces an additional dark $SO(4)$.  It is easy to unify the $SO(4) \otimes SO(4)$ dark and electroweak groups into the maximal subalgebra of $SO(8)$.  In the next section, we outline how a global chiral $SO(8)$ symmetry can stabilize the dark and weak scales.
\newline
\newline
Before moving on, in order to finish the unification story, we note the dark and weak gauge groups have four Cartan generators.
Given that $SU(3)$ color has an additional two, a simple dark and SM unification group would need at least six.  
This leaves the simple groups $E_7$, $E_8$, $SU(n + 1)$ and $SO(4\,n' + 2)$ where, respectively, $n \geq 7$ and $n' \geq 3$.  Here we made use of the fact that $SO(4\,n' + 2)$ are the only $SO(2N)$ groups with complex spinor representations; further, gauge anomaly freedom is automatic for $SO(2N)$ groups.  As an example, $SO(4\,n' + 2)$ can always be broken at the unification scale  to $SO(4\,n' + 2) \to SO(4\,n' - 6) \otimes SO(8)$ where $SO(8)$ contains the $SU(2)^4$ of interest.  This breaking is equivalent to the well-known Pati-Salam unification~\cite{Georgi:1974sy,Pati:1974yy}.

\subsection*{B.2 Stabilizing the Dark and Electroweak Scales with Overlapping Chiral \\  Symmetries}

In this section, we extend the model described in Section~\ref{sec:model} by employing additional chiral symmetries in order to stabilize the dark and electroweak symmetry breaking scales.  More explicitly, these symmetries protect both the dark and electroweak higgs masses.  Here we outline a scenario in which the electroweak and dark higgses generate masses at one- and three-loops, respectively.   This is needed as model relies on a light dark higgs bosons in order to obtain the correct the relic dark matter abundance.  The work in this section is an extension of Intermediate Higgs (IH) models~\cite{Katz:2005au} which, in turn, take inspiration from the Little Higgs (LH) mechanism~\cite{ArkaniHamed:2002qy,Katz:2003sn}.  
\newline
\newline
Loop corrections to the electroweak and dark higgses' masses produce quadratically divergent corrections.  Running in the loops are the fermions, gauge and scalar bosons described in Section~\ref{sec:model}.  For simplicity's sake, we apply the IH philosophy to this section:  To generate naturally light higgses, it is enough to cancel out the quadratic divergences from the top (fermion) sector using global symmetries to protect the higgs mass.  This rationale is due to the fact that the top yukawa coupling is so much larger than, say the electroweak gauge couplings, and therefore contributes the most to the radiative corrections of the electroweak higgs.  To make this section as simple as possible, we will assume the dark fermions have a much larger yukawa coupling than the gauge and scalar bosons.
%
%

\subsubsection*{B.2.1 Preliminaries}  

In this section, we detail some essentials to show a mechanism for how the electroweak and dark higgses generate masses at one- and three-loops respectively.  For simplicity, we follow the philosophy of the Intermediate Higgs (IH) models~\cite{Katz:2005au} and cancel the divergences only in the fermion sector.  
\newline
\newline
We focus on a model with the coset space 
\begin{equation}
G/H = SO(8)_1 \otimes SO(8)_2/SO(8)_V.
\end{equation}
This coset space generates 28 massless Nambu-Goldstone bosons whose mass is protected by the $SO(8)_1 \otimes SO(8)_2$ chiral symmetry.
For simplicity, we gauge both the $SU(2)_L \otimes SU(2)_D \otimes U(1)_Y \otimes U(1)_D$ subgroup in \textit{both} $SO(8)_{1}$ and $SO(8)_2$.  The (now) pseudo-Nambu-Goldstone bosons (PNGBs) transform as
\begin{align} 
h = (2,1)_{(1/2,0)} && h^\dagger = (2,1)_{(-1/2,0)} \\
\phi =  (1,2)_{(0,1/2)} && \phi^\dagger = (1,2)_{(0,-1/2)}. 
\end{align}
We identify these  PNGBs as the dark and electroweak higgses in Section~\ref{sec:model}.  As before, the entries in parenthesis label the $(SU(2)_L,SU(2)_D)$ representations.  The subscript parenthesis label the $(U(1)_Y,U(1)_D)$ representations.    The rest of the 28 PNGBs transform as
\begin{eqnarray}
S_h &=& (1,1)_{(0,0)} \oplus (1,1)_{(0,0)}    \label{eq:singlets1} \\
 S_\phi &=& (1,1)_{(0,0)} \oplus (1,1)_{(0,0)}   \label{eq:singlets2} \\
A &=& (2,2)_{(0,0)} \oplus (2,1)_{(0,1/2)}  \oplus (2,1)_{(0,-1/2)} \oplus (1,2)_{(1/2,0)}  \oplus (1,2)_{(-1/2,0)} \nonumber  \\
&&\oplus \, (1,1)_{(1/2,1/2)} + (1,1)_{(1/2,-1/2)} + (1,1)_{(-1/2,1/2)} + (1,1)_{(-1/2,-1/2)}.
\end{eqnarray}
We are most interested in how the PNGBs transform under the $SU(2)^2 \otimes U(1)^2$ gauged subgroup.  We therefore can 
describe the vacuum orientation for the $SO(8)_1 \otimes SO(8)_2$ symmetry breaking as a unitary matrix $\Sigma$ which transforms as
\begin{equation}
\Sigma \to V_1 \,\Sigma\, V_2^\dagger.
\end{equation}
Here $V_{1,2}$ are the representations of the full $SU(2)^2 \otimes U(1)^2$ subgroup.  The background field $\Sigma_0$ has the orientation
\begin{equation}
\Sigma_0 = \begin{pmatrix} \mathbb{1} & & &  \\  & \mathbb{1} & &  \\ & & \mathbb{1} & \\ & & & \mathbb{1}  \end{pmatrix}.
\end{equation}
The Nambu-Goldstone bosons are fluctuations around this background in the direction of the broken generators, $\Pi = \pi^a X^a$.  They can be parametrized as~\cite{Callan:1969sn}
\begin{equation}
\Sigma = e^{i\Pi/f}\, \Sigma_0\,e^{i\Pi^\dagger/f} = e^{2i\Pi/f}\,\Sigma_0. \label{eq:sigmafield}
\end{equation}
The generators of the gauged $SU(2)_L \otimes SU(2)_D \otimes U(1)_Y \otimes U(1)_D$ subgroup are
\begin{align}
Q_L^a =  {1 \over 2}\left(
\begin{array}{c}
\scalebox{.8}
{\begin{picture}(84,84)
\put(14,70){\makebox(0,0){\Large}}
\put(10.5,73.5){\makebox(0,0){$\sigma^a$}}
\put(35,49){\makebox(0,0){}}
\put(0,42){\line(1,0){84}}   
\put(42,0){\line(0,1){84}} 
\put(0,63){\line(1,0){42}} 
\put(21,42){\line(0,1){42}}
\put(63,0){\line(0,1){42}} 
\put(42,21){\line(1,0){42}}
\end{picture}}
\end{array}
\right) 
 && Y = \mathrm{diag}(0,0,1,-1,0,0,0,0)/2  \\
 Q_D^a = {1 \over 2}\left(
\begin{array}{c}
\scalebox{.8}
{\begin{picture}(84,84)
\put(14,70){\makebox(0,0){\Large}}
\put(52.5,31.5){\makebox(0,0){$\sigma^a$}}
\put(35,49){\makebox(0,0){}}
\put(0,42){\line(1,0){84}}   
\put(42,0){\line(0,1){84}} 
\put(0,63){\line(1,0){42}} 
\put(21,42){\line(0,1){42}}
\put(63,0){\line(0,1){42}} 
\put(42,21){\line(1,0){42}}
\end{picture}}
\end{array}
\right)   && Y_D = \mathrm{diag}(0,0,0,0,0,0,1,-1)/2
\end{align}
where the lower row of matrices are the generators for the dark gauge group.  The custodial $SU(2)$ matrices are  
 \begin{align}
 R^a =  {1 \over 2}\left(
\begin{array}{c}
\scalebox{.8}
{\begin{picture}(84,84)
\put(14,70){\makebox(0,0){\Large}}
\put(31.5,52.5){\makebox(0,0){$\sigma^a$}}
\put(35,49){\makebox(0,0){}}
\put(0,42){\line(1,0){84}}   
\put(42,0){\line(0,1){84}} 
\put(0,63){\line(1,0){42}} 
\put(21,42){\line(0,1){42}}
\put(63,0){\line(0,1){42}} 
\put(42,21){\line(1,0){42}}
\end{picture}}
\end{array}
\right)  && R_D^a ={1 \over 2}\left(
\begin{array}{c}
\scalebox{.8}
{\begin{picture}(84,84)
\put(14,70){\makebox(0,0){\Large}}
\put(73.5,10.5){\makebox(0,0){$\sigma^a$}}
\put(35,49){\makebox(0,0){}}
\put(0,42){\line(1,0){84}}   
\put(42,0){\line(0,1){84}} 
\put(0,63){\line(1,0){42}} 
\put(21,42){\line(0,1){42}}
\put(63,0){\line(0,1){42}} 
\put(42,21){\line(1,0){42}}
\end{picture}}
\end{array}
\right)  
 \end{align}
of which $Y$ and $Y_D$ are subgroups.  Notice, $\Sigma_0$ does not break the $SU(2)^2 \otimes U(1)^2$ subgroup of $SO(8)_1 \otimes SO(8)_2$.  
The PNGB matrix can be written as
 \begin{equation}
 \Pi = {1 \over 2}\left(
\begin{array}{c}
{\begin{picture}(84,84)
\put(14,70){\makebox(0,0){\Large}}
\put(31.5,73.5){\makebox(0,0){$h$}}
\put(35,49){\makebox(0,0){}}
\put(10.5,52.5){\makebox(0,0){$h^\dagger$}}
\put(0,42){\line(1,0){84}}   
\put(42,0){\line(0,1){84}} 
\put(0,63){\line(1,0){42}} 
\put(21,42){\line(0,1){42}}
\put(63,0){\line(0,1){42}} 
\put(42,21){\line(1,0){42}}
\put(63,63){\makebox(0,0){\large $A^\dagger$}}
\put(21,21){\makebox(0,0){\large $A$}}
\put(52.5,10.5){\makebox(0,0){$\phi^\dagger$}}
\put(73.5,31.5){\makebox(0,0){$\,\phi$}}
\end{picture}}
\end{array}
\right),
\label{eq:PNGB}
\end{equation}
where we have omitted the $SU(2)$ singlets in equations~\ref{eq:singlets1} and~\ref{eq:singlets2}.  They are irrelevant to the rest of the arguments in this section.  
\newline
\newline
We are now able to make the most important points of this section:  
\begin{enumerate}
\newcounter{enum_saved2}
\item In the limit where $A$ is heavy and integrated out, the dark and electroweak higgses are protected by \textit{different} $SO(4) \otimes SO(4)$ chiral symmetries.  Both higgses are protected by the full $SO(8) \otimes SO(8)$ chiral symmetry when $A$ is not integrated out.  
\item  The background field $\Sigma_0$ preserves both the ``overlapping" $SO(4) \otimes SO(4)$ and the full $SO(8) \otimes SO(8)$ chiral symmetries.
\setcounter{enum_saved2}{\value{enumi}}
\end{enumerate}
This last point a major difference between the models in this paper and LH theories.  As we will see, the preservation of additional ``overlapping" chiral symmetries allows the higgses in the theory to generate a mass at a larger loop order.  
To emphasize in this example, the $SO(8) \otimes SO(8)$ and $SO(4) \otimes SO(4)$ chiral symmetries are ``overlapping" because they both protect both the dark and electroweak higgses.  We now outline how this mechanism can be applied to the fermion (top) sector in the next section.  It is well-known the top sector generates the largest contribution to the electroweak higgs' mass.  For simplicity, we apply the same criteria~\cite{Katz:2005au} in this work.  
We briefly discuss  the scalar and gauge boson sectors in section following the next.  Further work to fully explore this mechanism is postponed for~\cite{postpone}.  Although we have developed this mechanism for this baryogenesis model, it can also be used to explain a light higgs and the absence of other new physics at the LHC~\cite{postpone}.  We also make some comments toward this research direction in the section following the next.

\subsubsection*{B.2.2 Dark Fermion and Top Quark Sector}   

As discussed in previous sections, we want a dark higgs that is naturally of order the GeV scale to facilitate dark matter annihilation.  To accomplish this in the fermion sector, we do the following:
\begin{enumerate}
\setcounter{enumi}{\value{enum_saved2}}
\item We explicitly break the chiral $SO(4) \otimes SO(4)$ symmetry under which the SM higgses transforms non-trivially as a goldstone boson.  This applies the LH mechanism to the electroweak higgs which gets a mass at one-loop.  This breaking also simultaneously breaks the $SO(8) \otimes SO(8)$ chiral symmetry.
\item The dark higgs only knows about the explicit $SO(8) \otimes SO(8)$ chiral symmetry breaking via the additional higgs particle, $A$.  Thus dark higgs gets a mass at three-loops due to diagrams which exchange $A$.  
\setcounter{enum_saved2}{\value{enumi}}
\end{enumerate}
In the dark fermion/top sector we have an approximate $SO(8)$ multiplet of fermions (see equation~\ref{eq:chiralquarks}) which include new fermions, $q_L$ and $q_R$.  The latter fermions transform like the left- and right-handed SM top quarks.  The fermion multiplet in the approximate 8 of SO(8) is
\begin{align}
\psi = \begin{pmatrix} Q_L \\ Q_R \\ q_L \\ q_R  \end{pmatrix}
\end{align}
where 
\begin{align}
Q_R = \begin{pmatrix} D^c \\ U^c \end{pmatrix}   && q_R  = \begin{pmatrix} b^c_R \\ t^c_R \end{pmatrix}.
\end{align}
To be explicit under the approximate global symmetries, the fermions transform as
\begin{align}
\overline{\psi} \to \overline{\psi} \,V_1^\dagger  && \psi \to V_2 \,\psi \\ \nonumber \\
\overline{Q} \to \overline{Q}\,U_4^\dagger  && Q \to V_4\,Q  \\
\overline{q} \to \overline{q}\,U_4^{'\, \dagger} &&  q \to V'_4\,q
\end{align}
where we have defined
\begin{align}
 Q = \begin{pmatrix} Q_L \\ Q_R \end{pmatrix} && q = \begin{pmatrix} q_L \\ q_R \end{pmatrix}.
\end{align}
$V_1$, $V_2$ were defined above.  $U_4$, $V_4$, $U'_4$ and $V'_4$ are the generators of the $SO(4)_1$, $SO(4)_2$, $SO(4)'_1$ and $SO(4)'_2$ global symmetries, respectively.  Following~\cite{Katz:2003sn,Katz:2003sn} and the requirements in Items 3 and 4 in the above list, we write the following 
\begin{equation}
\mathcal{L'} = \lambda_1 f\, \overline{\psi}\,\Sigma\,\psi - \lambda_2 f\,\overline{q}_L\,q'_L    - \lambda_3 f\,\overline{q}'_R\,q_R.
\label{eq:yuk}
\end{equation}
Here $\lambda_i$ are not related to the $\lambda_i$ couplings in Section~\ref{sec:model}.  $f$ is the scale of new physics which generates these couplings.  $q'_L$ and $q'_R$ are additional fermions that transform under the SM gauge group as top-quarks.  They  are not in the $SO(8)$ multiplet and explicitly break the global chiral symmetries.  It is clear the first term in equation~\ref{eq:yuk} preserves \textit{all} of the chiral symmetries.  As prescribed by the IH/LH mechanism, the $\lambda_2$ and $\lambda_3$ terms break only \textit{parts} of the $SO(4)'_1 \otimes SO(4)'_2$ chiral symmetry that acts on $q_L$ and $q_R$.  More explicitly, the $\lambda_2$ term  breaks the $SO(4)'_1$ global symmetry while the $\lambda_3$ term breaks  $SO(4)'_2$.  Thus, both terms are needed to completely break the $SO(4)'$ chiral symmetry.  Since $A$ is included in the higgs multiplet, $\lambda_2$ and $\lambda_3$ are both needed to break the overall $SO(8)_1 \otimes SO(8)_2$ chiral symmetry. 
\newline
\newline
\textbf{Effective Higgs Potential:}  We can eliminate the unphysical PNGBs from the dark and electroweak higgses by performing a dark and electroweak  gauge transformation.  As discussed in Section~\ref{sec:model}, the vacuum misalignment with respect to the dark and electroweak gauge interactions can be parametrized by the angles
\begin{align}
\alpha =  2\,\langle \eta \rangle/f  && \beta = 2\,\langle \xi \rangle/f &&  \theta =  2\,\langle \phi_0 \rangle/f
\end{align}
where we have required $\langle \eta \rangle = v_1$, $\langle \xi \rangle = v_2$ and $\langle \phi_0 \rangle = v_\mathrm{ew}$.  The fermion mass matrix can be written as
\begin{eqnarray}
M(\Sigma) =  \lambda_1 \left(\begin{array}{cccccccccccc}
1 & 0 & 0 & 0 & 0 & 0 & 0 & 0 & 0 & 0 & 0 & 0\\
0 & c' & i\,s' & i\,s''  & 0 & 0 & 0 & 0 & 0 & 0 & 0 & 0 \\
0 & -i\,s'^{\,*} &  c'' & -2\,s''' & 0 & 0 & 0 & 0 & 0 & 0 & 0 & 0\\
0 & -i\,s''^{\,*} & -2\,s'''^{\,*}  &  c''' & 0 & 0 & 0 & 0 & 0 & 0& 0 & 0 \\
0 & 0 & 0 & 0 & c & 0 & 0 & i  \,s & \lambda_{12} & 0 & 0 & 0\\
0 & 0 & 0 & 0 & 0 &  c & i\,s & 0 & 0 &  \lambda_{12} & 0 & 0\\
0 & 0 & 0 & 0 & 0 & i \,s &  c & 0 & 0 & 0 &  \lambda_{13} & 0\\
0 & 0 & 0 & 0 & i\,s & 0 & 0 & c & 0 & 0 & 0 &  \lambda_{13}\\
0 & 0 & 0 & 0 & \lambda_{12} & 0 & 0 & 0 & 0 & 0 & 0 & 0\\
0 & 0 & 0 & 0 & 0 & \lambda_{12} & 0 & 0 & 0 & 0 & 0 & 0 \\
0 & 0 & 0 & 0 & 0 & 0 & \lambda_{13} & 0 & 0 & 0 & 0 & 0\\
0 & 0 & 0 & 0 & 0 & 0 & 0 & \lambda_{13} & 0 & 0 & 0 & 0
\end{array}\right)&& \nonumber
\end{eqnarray}
We have evaluated the sigma field in equation~\ref{eq:sigmafield} and defined
\begin{align}
c &= \cos \theta && s' = \alpha\,\sin \theta'/\theta' && s'' = \beta \sin \theta'/\theta' && s''' = \alpha\,\beta^*\,s'^{\,2}/\theta'^{\,2}  \\
s &= \sin \theta && c' = \cos \theta' && c'' = \alpha^*\alpha + \beta^*\beta\,\,c'/\theta'^{\,2} && c''' = \beta^*\beta + \alpha^*\alpha\,\,c'/\theta'^{\,2}   
\end{align}
and $\theta' = \sqrt{\alpha^2 + \beta^2}$.  As well,  we also defined $\lambda_{12} = \lambda_2/\lambda_1$ and $\lambda_{13} = \lambda_3/\lambda_1$.  It is easy to verify
\begin{align}
&{\partial \over \partial \theta}\,\mathrm{tr}\,(M^\dagger M )^n= 0 && {\partial \over \partial \alpha}\,\mathrm{tr}\,(M^\dagger M)^{n'} = 0 && {\partial \over \partial \beta}\,\mathrm{tr}\,(M^\dagger M)^{n'} = 0 
\end{align}
where $n \leq 2 < n'$.  Because $n, n' \leq 2$ the quadratic and logarithmically divergent pieces of the one-loop Coleman-Weinberg potential
\begin{align}
-{3\Lambda^2 \over 8\,\pi^2}\,\mathrm{tr}M^\dagger M && -{3 \over 16\,\pi^2}\,\mathrm{tr}(M^\dagger M)^2\,\mathrm{ln}(M^\dagger M/\Lambda^2).
\end{align}
for the SM higgs automatically vanish.  The logarithmic two-loop divergent term is given in~\cite{} for various renormalization schemes.  However, we were unable to find the quadratically divergent piece for the two-loop Coleman-Weinberg potential in the literature.  Despite this, in the next paragraph, we make an estimate.
\newline
\begin{figure}[t]
\centering
\includegraphics[width=10.1truecm,height=11.1truecm,clip=true]{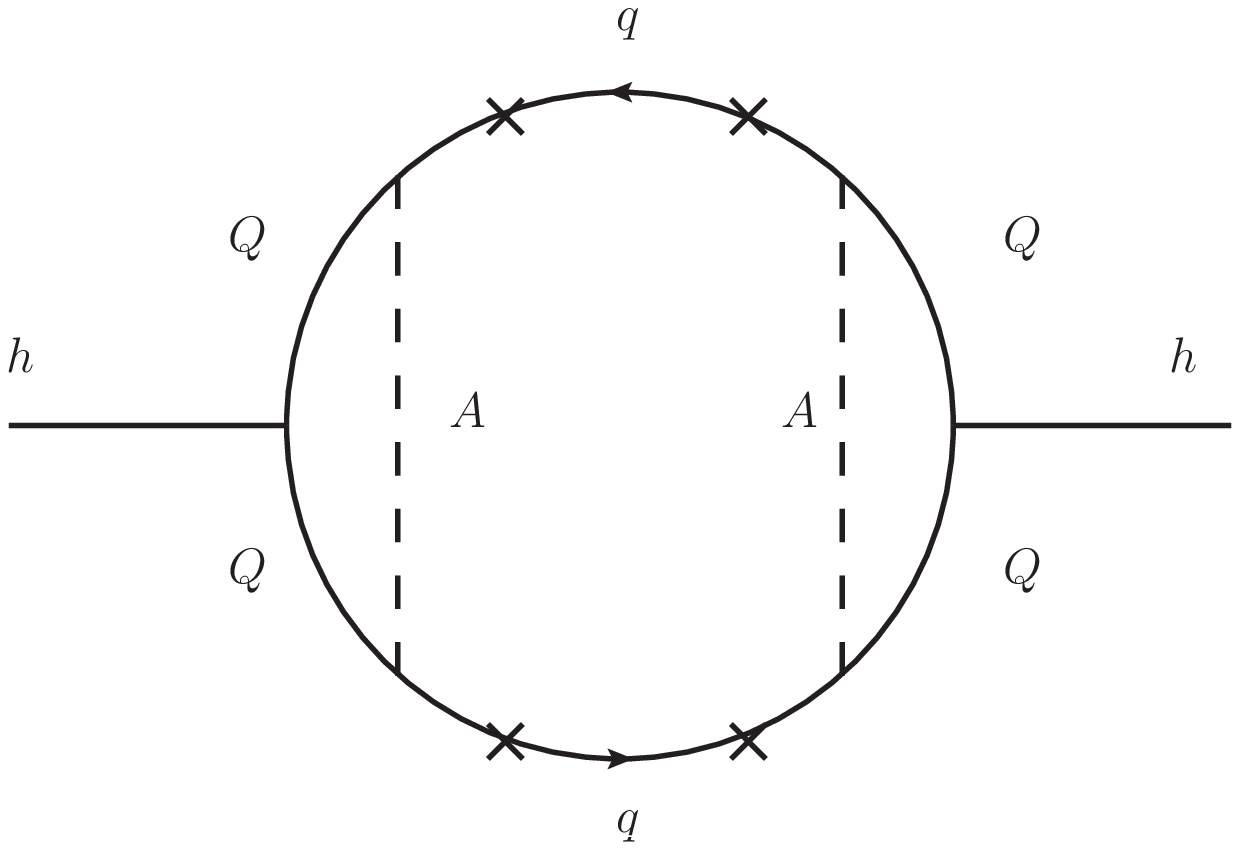} 
\hspace{-4cm}
\includegraphics[width=10.1truecm,height=11.1truecm,clip=true]{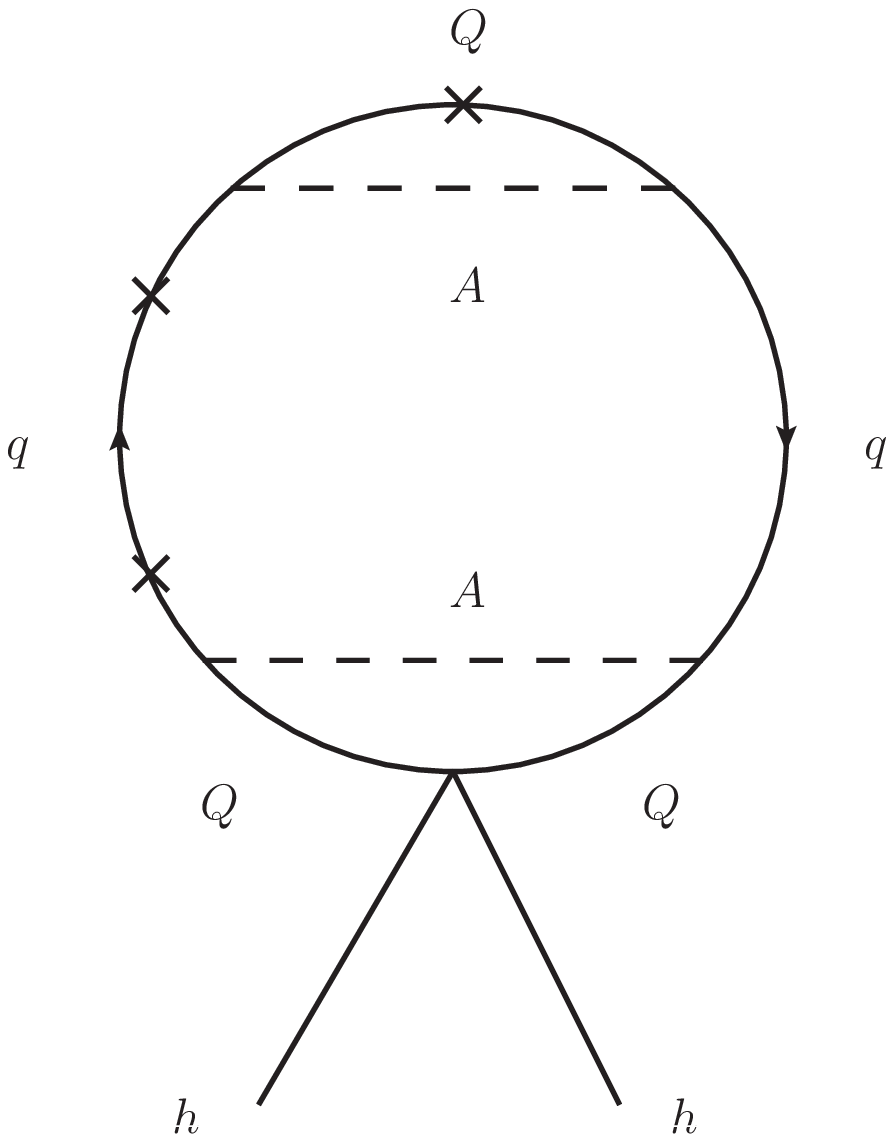} 
\vspace{-5.1cm}
\caption{Impartial cancellation between the three-loop diagrams for the dark higgs.}
\label{fig:higgsmass}
\end{figure}
\newline
\noindent
We can estimate the contribution of the fermion loops to the two-loop Coleman-Weinberg potential by checking the two-loop cancellations diagram-by-diagram.  Of course, we limit our attention to diagrams that involve the dark fermions and scalars which originate from equation~\ref{eq:yuk}.  We expand $\Sigma$ in equation~\ref{eq:yuk} to determine the couplings which could radiatively generate a higgs mass.  The important terms are
\begin{eqnarray}  
\mathcal{L'} &\supset& \lambda_1f\,\overline{Q}\,Q +  \lambda_1f\,\overline{q}\,q - \lambda_2 f\,\overline{q}_L\,q'_L    - \lambda_3 f\,\overline{q}'_R\,q_R \\
&+& 2i\,\lambda_1\,\biggl(\overline{Q}\, H \,Q + \overline{Q}\,A^\dagger q + \overline{q}\,A\,Q + \overline{q}\,\Phi\,q \biggr)  \nonumber\\
&-& {2\lambda_1 \over f}\,\biggl(\overline{Q}\,(H\,H^\dagger + A^\dagger A)\,Q + \overline{Q}\,(H^\dagger H + A^\dagger \Phi)\,q \nonumber \\
&+& \overline{q}\,(A H + \Phi A)\,Q + \overline{q}(A^\dagger A + \Phi^\dagger \Phi) q\biggr) + \ldots \nonumber
\end{eqnarray}
where we have defined 
\begin{align}
H = \begin{pmatrix} & h \\ h^\dagger & \end{pmatrix} && \Phi = \begin{pmatrix} & \phi \\ \phi^\dagger & \end{pmatrix}. 
\end{align}
Using this equation, it is straightforward to show all of the coefficients to the 1-loop SM higgs mass exactly cancel expect for those diagrams that involve \textit{both} $\lambda_2$ and $\lambda_3$.  This, again, was the point of the LH mechanism and verified by the previous paragraph.  However, all of the 1- and 2-loop corrections to the dark higgs mass cancel.  The 3-loop diagram (shown in the Figure 6 above) that involves both $\lambda_2$ and $\lambda_3$ is incompletely canceled.  Thus the fermion sector in this model generates natural masses of order
\begin{align}
m_\phi \sim \lambda^4 f/(4\pi) && m_h \sim \lambda^8 f/(4\pi)^2. \label{eq:osmasses}
\end{align}
%
In the next section we discuss the correction to the higgs potential by the gauge and scalar bosons.

\subsubsection*{B.2.3 Gauge and Scalar Boson Sectors}

\textbf{Gauge Boson Sector:}  As previously mentioned, the $SU(2)_L \otimes SU(2)_R \otimes SU(2)_D \otimes U(1)_Y \otimes U(1)_D$ is gauged.  The unique two derivative term for the $\Sigma$ field is
\begin{equation}
{f^2 \over 4} \mathrm{tr}|D\Sigma|^2
\end{equation}
where 
\begin{equation}
D\Sigma = \partial\Sigma - i \sum_{i =1,2}g_i \,W_i^a(Q_i^a \,\Sigma + \Sigma\,Q_i^a) - i \sum_{i=1,2} g'_i B_i(Y_i \,\Sigma + \Sigma\,Y_i).
\end{equation}
To make the notation compact, we define $g_i = (g_\mathrm{ew},g_\mathrm{dark})$ and the generators $Q^a_i = (Q_L^a,Q_D^a)$, and $Y_i = (Y,Y_D)$.  We also define $W_i^a = (W^a,V^a)$, $B_i = (B,U)$ where $V$ and $U$ are dark gauge bosons.  
For simplicity, we have not implemented any symmetries to protect the dark or electroweak higgs potential from large radiative corrections from the gauge sector.  Thus, the higgses get masses of order 
\begin{align}
m_\phi \sim   g_1^2\,f && m_h \sim g_2^2\,f
\end{align}
at one loop where we have assumed a cutoff of $\Lambda = 4\pi f$.  Again,  $f$ is the scale of new physics and $g_i = (g_\mathrm{ew}, g_\mathrm{dark})$.  We assume the higgses' bare mass is tiny.  As discussed in~\cite{}, canceling only the top divergences results in a 6 TeV cutoff for new physics in the electroweak sector.  So the gauge boson corrections to the dark higgs sector are not larger than the cancellations in the dark fermion sector, we assume the gauge coupling is very weak
\begin{equation}
g_2 < 1/(4 \pi).
\end{equation}
This is with the assumption that $\lambda$ in equation~\ref{eq:osmasses} is approximately one.  Again this restriction is done to make this section as simple as possible. 
\newline
\newline
\textbf{Scalar Sector:}  Naively, the dark higgses gets a mass of order 
\begin{equation}
m_h \sim \lambda^2\, f \label{eq:rhiggsmass}
\end{equation}
from the dark and electroweak higgs' scalar potential.  Again, the coupling is unrelated to the $\lambda$s in other sections of the paper; and, we have assumed a cutoff of $\Lambda = 4\pi f$ and $f$ is the scale of new physics.  We note that the interaction terms in the scalar potential that can generate this mass also are the same interactions which allow dark matter annihilations through the higgs portal.  In particular the operator
\begin{equation}
\lambda' \phi^\dagger \phi\, h^\dagger h,
\end{equation}
is essential for both the radiative corrections and the dark matter annihilations.  Just like in IH/LH theories, the scalar potential is generated by the one- and two-loop Coleman-Weinberg potential.  The potential is produced by fermion and gauge boson corrections.  Above, we have required the gauge boson corrections to be parametrically suppressed.  However, as described above, the Coleman-Weinberg potential generated by the top interactions cancels (at minimum) one-loop radiative corrections.  These cancellations are preserved in the couplings of the dark higgs and electroweak higgs scalar potential thereby negating equation~\ref{eq:rhiggsmass}.

\subsubsection*{B.2.4 Generalized Mechanism/Additional Points}  
\label{sec:additionalpts}

In this section, we very briefly outline how to extend the mechanism of the overlapping chiral symmetries to the gauge sector.  Before doing so, we note the mechanism to give the dark higgs a mass at three-loops can be used for LHC model building; this is especially true if the LHC continues not to find any new physics beyond the SM.  In this case, the SM higgs can be the light higgs boson without finetuning.  Moreover, adding an additional discrete symmetry, \textit{e.g.} T-parity~\cite{Cheng:2003ju}, to the theory can decouple the SM higgs by an additional loop factor.  This further decouples new physics from the SM.  We hope these techniques will not be needed and the LHC finds new physics soon.
\newline
\newline
Intentionally, we did not apply the overlapping chiral symmetries to the gauge sector for simplicity.  It is, however, to imagine how this can be accomplished.  In LH scenarios, the higgs bosons have to be protected as goldstone bosons in the event on of the gauge couplings is set to zero.  In the LH case the global symmetry protecting the higgs is explicitly broken by to gauge couplings.  This eliminates one-loop divergences.  For the considered scenario, it seems reasonable one has to construct a model where the global symmetry protecting the goldstone bosons is broken by four gauge couplings.  Further work on this is postponed for~\cite{postpone}.

\section*{C.  Dark Matter Relic Abundance from Dark Photon}
\label{app:relicdecoup}

In this appendix, we simply show how efficient it is for the dark matter to annihilates via the dark photon.  This annihilation parametrically depends on the kinetic mixing parameter $\epsilon$.  We will show the available parameter space needed to ensure the correct relic abundance is small thereby demonstrating the need for a light dark higgs.  
\newline
\newline
As a quick reminder of Section 4, the low-velocity thermally averaged annihilation cross section is
\begin{eqnarray}
\langle \sigma |v| \rangle &=& a + {6\, T \over m_\chi}\, b + {60 \,T^2 \over m_\chi^2} \,c + \ldots
\end{eqnarray}
Explicit computation of $a$ and $b$ for annihilations through the kinetic mixing portal results in
\begin{eqnarray}
a_\mathrm{kinetic} &=& \sum_l\,\frac{8\, \pi\, \epsilon^2 \, \alpha \,\alpha_\mathrm{dark} \,m_\chi \sqrt{m_\chi^2-m_l^2} \left(m_l^2+2 \,m_\chi^2\right) }{\left(m_{\gamma'}^2\, m_\chi - 4\, m_\chi^3\right)^2}  \\
&+& \sum_q\,\frac{32\,\pi\,\epsilon^2 \alpha\,\alpha_\mathrm{dark} \,m_\chi \sqrt{m_\chi^2 -m_q^2} \left(m_q^2+2 m_\chi^2\right) }{3 \left(m_{\gamma'}^2 \,m_\chi - 4 \,m_\chi^3\right)^2},\nonumber \\
b_\mathrm{kinetic} &=& \sum_l \frac{\pi \epsilon^2 \,\alpha\, \alpha_\mathrm{alpha}\left(64\,m_\chi^6 +8 m_\chi^4(m_{\gamma'}^2-4 m_l^2) -4 m_l^2 m_\chi^2(m_{\gamma'}^2+17 m_l^2) + 5 m_{\gamma'}^2 m_l^4\right)}{3 m_\chi \sqrt{m_\chi^2-m_l^2} \left(m_{\gamma'}^2-4 m_\chi^2\right)^3}\\
&+& \sum_q \frac{4\pi\epsilon^2\, \alpha\,\alpha_\mathrm{dark} \left(64 \,m_\chi^6+8 m_\chi^4(m_{\gamma'}^2-4 m_q^2)-4 m_q^2 m_\chi^2\,(m_{\gamma'}^2+17 m_q^2 ) + 5 m_{\gamma'}^2 m_q^4\right) }{9 m_\chi \sqrt{m_\chi^2-m_q^2} \left(m_{\gamma'}^2-4 m_\chi^2\right)^3}.  \nonumber
\end{eqnarray}
where again $\epsilon$ is the kinetic mixing parameter.  $m_{\gamma'}$ is the mass of the heavy photon.  Implicit is the summation is over the final state quarks and leptons.  The dark matter relic abundance is 
 \begin{eqnarray}
\Omega_\chi\,h^2 \sim {1.04 \times 10^9 \over M_\mathrm{pl}}\,{x_F \over \sqrt{g_*}}\,{1 \over a + 3b/x_F}.
\end{eqnarray}
In Figure 8, we fix $\epsilon \sim 10^{-3}$ and vary the dark mass so that  $h^2\, \Omega_{\mathrm{relic\,\,DM}}  < 0.0034$. 
\newline
\begin{figure}[t]
\centering
\includegraphics[width=8.0truecm,height=6.1truecm,clip=true]{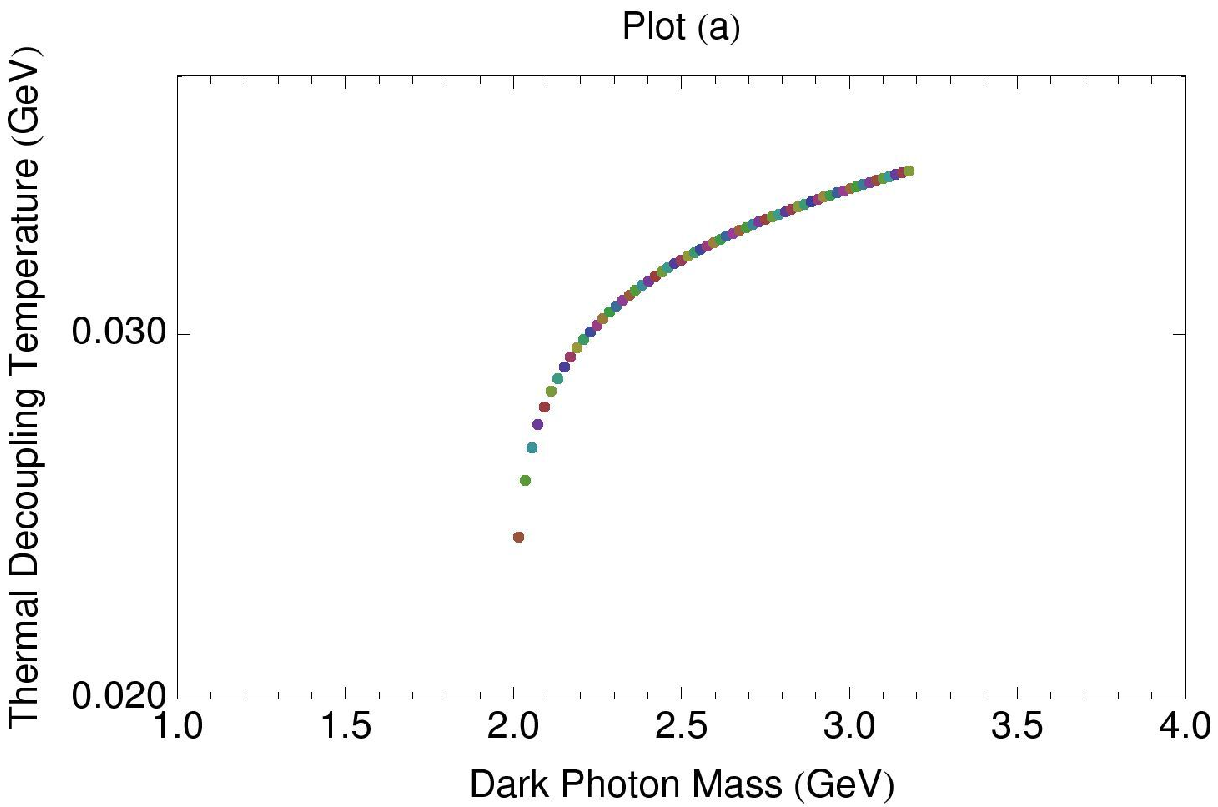}  \hspace{0.2cm}
\includegraphics[width=8.0truecm,height=6.1truecm,clip=true]{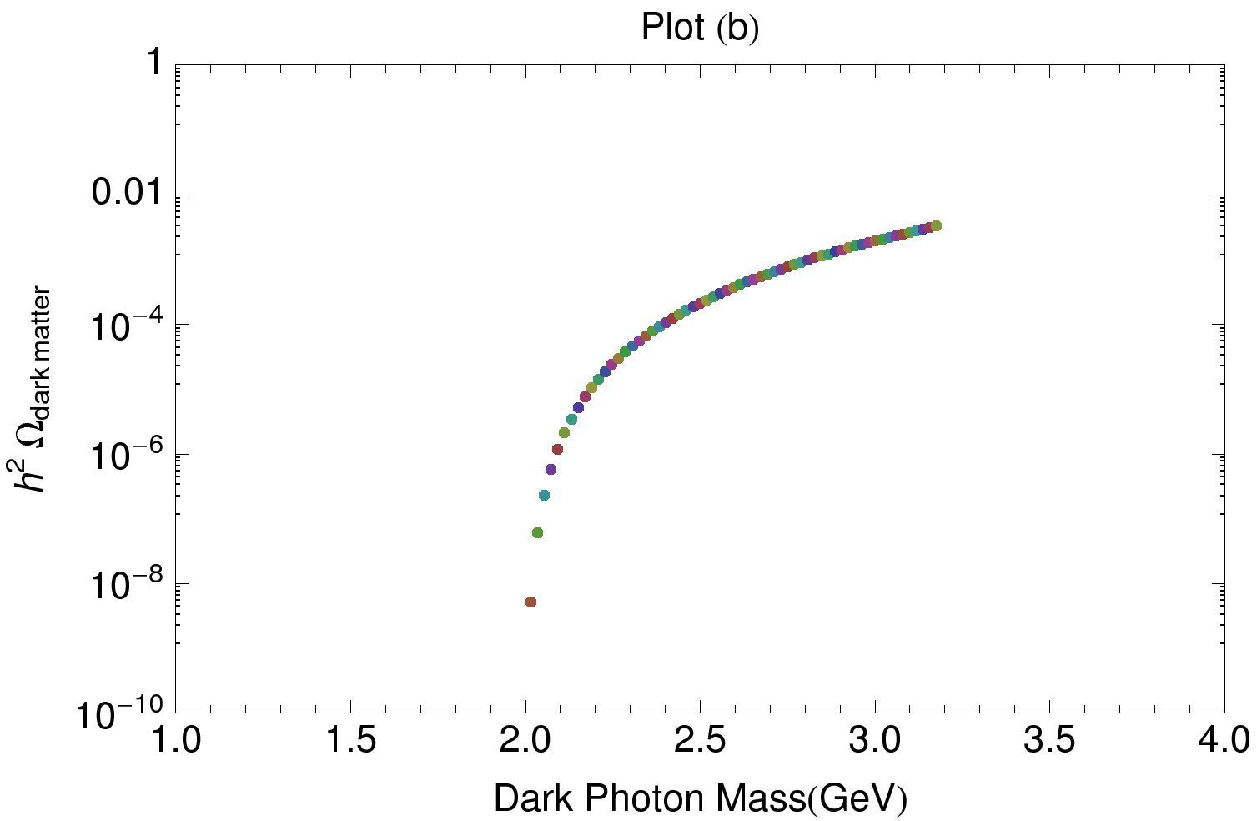}  \vspace{0.3cm} \\
\includegraphics[width=8.0truecm,height=6.0truecm,clip=true]{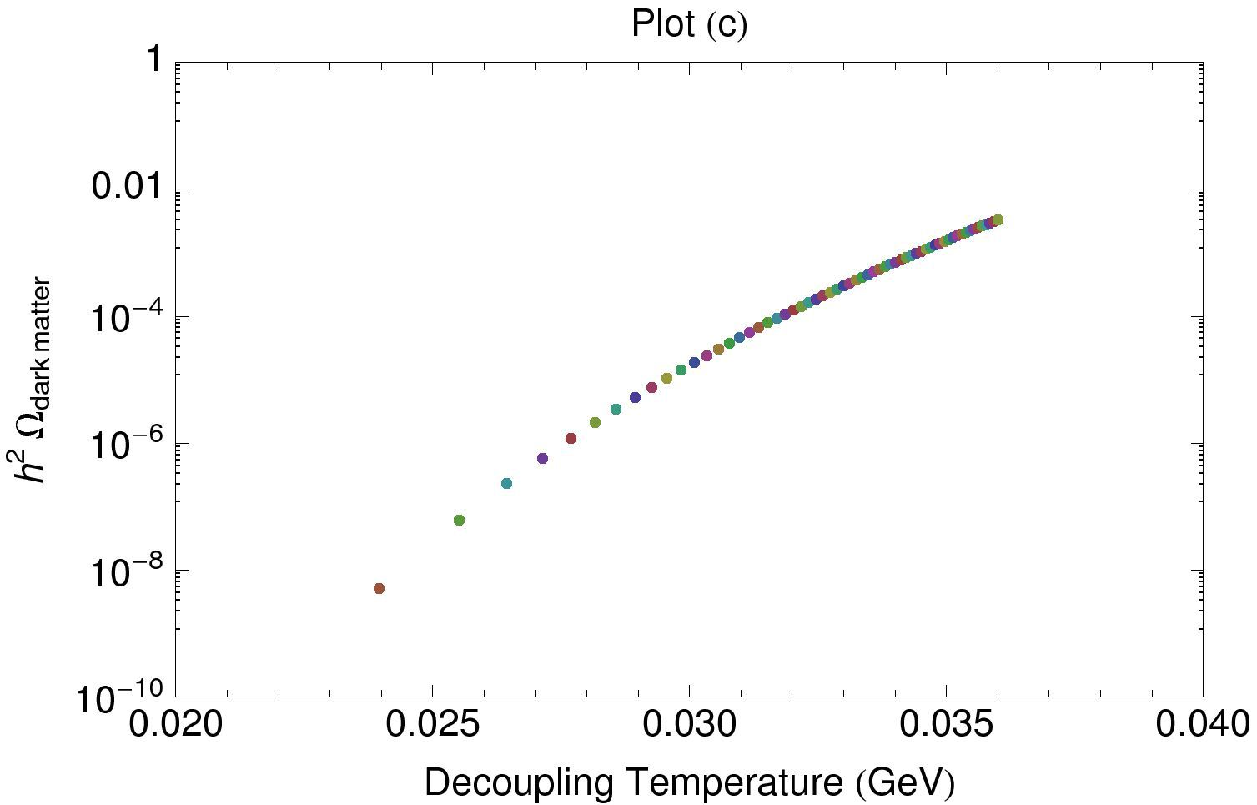} 
\caption{Allowed parameter points for the relic dark matter abundance that does not originate from the messenger fermion decays. Each point represents a point in parameter space that yields an abundance of $h^2 \Omega < 0.0034$.  The \textbf{ \textit{difference}} between this plot and the plots in Section 4 is the dark matter annihilates via the dark photon only.  It is clear this model requires dark matter annihilation via a dark higgs to be dominant.}
\end{figure}
\newline

\begin{thebibliography}{99}


\bibitem{Agashe:2010gt}
  K.~Agashe, D.~Kim, M.~Toharia and D.~G.~E.~Walker,
  Phys.\ Rev.\  D {\bf 82}, 015007 (2010).

\bibitem{Walker:2009en}
  D.~G.~E.~Walker,
  [arXiv:0907.3146 [hep-ph]]. 
  
\bibitem{Walker:2009ei}
  D.~G.~E.~Walker,
  [arXiv:0907.3142 [hep-ph]].
  
  \bibitem{Agashe:2010tu}
  K.~Agashe, D.~Kim, D.~G.~E.~Walker, L.~Zhu,
  [arXiv:1012.4460 [hep-ph]].

\bibitem{Bertone:2004pz}
  See, for example, G.~Bertone, D.~Hooper and J.~Silk,
  Phys.\ Rept.\  {\bf 405}, 279 (2005).

\bibitem{Hinshaw:2008kr}
  G.~Hinshaw {\it et al.}  [WMAP Collaboration],
  arXiv:0803.0732 [astro-ph].

\bibitem{Kolb:1990vq}
  E.~W.~Kolb and M.~S.~Turner,
  Front.\ Phys.\  {\bf 69}, 1 (1990).

  \bibitem{Barr:1990ca}
  S.~M.~Barr, R.~S.~Chivukula and E.~Farhi,
  Phys.\ Lett.\  B {\bf 241}, 387 (1990).
  
\bibitem{Kaplan:1991ah}
  D.~B.~Kaplan,
  Phys.\ Rev.\ Lett.\  {\bf 68}, 741 (1992).
 
\bibitem{Dodelson:1991iv}
  S.~Dodelson, B.~R.~Greene, L.~M.~Widrow,
  Nucl.\ Phys.\  {\bf B372}, 467-493 (1992).
  
  \bibitem{Cline:2006ts}
  J.~M.~Cline,
  arXiv:hep-ph/0609145.
  
  \bibitem{Sakharov:1967dj}
  A.~D.~Sakharov,
  Pisma Zh.\ Eksp.\ Teor.\ Fiz.\  {\bf 5}, 32 (1967)
  [JETP Lett.\  {\bf 5}, 24 (1967)]
  [Sov.\ Phys.\ Usp.\  {\bf 34}, 392 (1991)]
  [Usp.\ Fiz.\ Nauk {\bf 161}, 61 (1991)].

\bibitem{Kuzmin:1985mm}
  V.~A.~Kuzmin, V.~A.~Rubakov and M.~E.~Shaposhnikov,
  Phys.\ Lett.\  B {\bf 155}, 36 (1985).
  
  \bibitem{Cohen:1990it}
  A.~G.~Cohen, D.~B.~Kaplan and A.~E.~Nelson,
  Nucl.\ Phys.\  B {\bf 349}, 727 (1991).
  
  \bibitem{Cohen:1990py}
  A.~G.~Cohen, D.~B.~Kaplan and A.~E.~Nelson,
  Phys.\ Lett.\  B {\bf 245}, 561 (1990).
 
 \bibitem{Cohen:1993nk}
  A.~G.~Cohen, D.~B.~Kaplan and A.~E.~Nelson,
  Ann.\ Rev.\ Nucl.\ Part.\ Sci.\  {\bf 43}, 27 (1993).
 
 \bibitem{Pospelov:2005pr}
  M.~Pospelov and A.~Ritz,
  Annals Phys.\  {\bf 318}, 119 (2005).
 
\bibitem{Barate:2003sz}
  R.~Barate {\it et al.}  [LEP Working Group for Higgs boson searches and
                  ALEPH Collaboration and  and],
  Phys.\ Lett.\  B {\bf 565}, 61 (2003).

  \bibitem{atlashiggsresults}
   The ATLAS Collaboration, ATLAS-CONF-2011-163.

  \bibitem{CMShiggsresults}
    The CMS Collaboration], CMS-PAS-HIG-11-032.

\bibitem{Grojean:2004xa}
  C.~Grojean, G.~Servant and J.~D.~Wells,
  Phys.\ Rev.\  D {\bf 71}, 036001 (2005).

\bibitem{Fukugita:1986hr}
  M.~Fukugita and T.~Yanagida,
  Phys.\ Lett.\  B {\bf 174}, 45 (1986).
 
\bibitem{Carena:2008vj}
  M.~Carena, G.~Nardini, M.~Quiros and C.~E.~M.~Wagner,
  Nucl.\ Phys.\  B {\bf 812}, 243 (2009).
  
 \bibitem{CDFstopref}
 The CDF Collaboration, 
 CDF/PUB/EXOTIC/PUBLIC/9775.

\bibitem{Abazov:2008kz} 
   V.~M.~Abazov {\it et al.}  [D0 Collaboration],
  Phys.\ Lett.\  B {\bf 675}, 289 (2009).
 
 \bibitem{Affleck:1984fy}
  I.~Affleck and M.~Dine,
  Nucl.\ Phys.\  B {\bf 249}, 361 (1985).

\bibitem{Dine:2003ax}
  M.~Dine and A.~Kusenko,
  Rev.\ Mod.\ Phys.\  {\bf 76}, 1 (2004).
  
 \bibitem{Buchmuller:2004nz}
  W.~Buchmuller, P.~Di Bari and M.~Plumacher,
  Annals Phys.\  {\bf 315}, 305 (2005).

\bibitem{Enqvist:2003gh}
  K.~Enqvist and A.~Mazumdar,
  Phys.\ Rept.\  {\bf 380}, 99 (2003).

\bibitem{Dolgov:1991fr}
  A.~D.~Dolgov,
  Phys.\ Rept.\  {\bf 222}, 309 (1992).

\bibitem{Nussinov:1985xr}
  S.~Nussinov,
  Phys.\ Lett.\  B {\bf 165}, 55 (1985).

\bibitem{Chivukula:1989qb}
  R.~S.~Chivukula and T.~P.~Walker,
  Nucl.\ Phys.\  B {\bf 329}, 445 (1990).

\bibitem{Barr:1991qn}
  S.~M.~Barr,
  Phys.\ Rev.\  D {\bf 44}, 3062 (1991).
  
   \bibitem{Kaplan:2009ag}
  D.~E.~Kaplan, M.~A.~Luty and K.~M.~Zurek,
  Phys.\ Rev.\  D {\bf 79}, 115016 (2009).

 \bibitem{Shu:2006mm}
  J.~Shu, T.~M.~P.~Tait and C.~E.~M.~Wagner,
  Phys.\ Rev.\  D {\bf 75}, 063510 (2007)
  [arXiv:hep-ph/0610375].

\bibitem{Dutta:2010va} 
  B.~Dutta and J.~Kumar,
  Phys.\ Lett.\ B {\bf 699}, 364 (2011).
  

  \bibitem{ArkaniHamed:2002qy}
  N.~Arkani-Hamed, A.~G.~Cohen, E.~Katz, A.~E.~Nelson,
  JHEP {\bf 0207}, 034 (2002).
  
\bibitem{Chacko:2005pe}
  Z.~Chacko, H.~-S.~Goh, R.~Harnik,
  Phys.\ Rev.\ Lett.\  {\bf 96}, 231802 (2006).

\bibitem{Dimopoulos:1980hn}
  S.~Dimopoulos, S.~Raby and L.~Susskind,
  Nucl.\ Phys.\  B {\bf 173}, 208 (1980).
  
 \bibitem{Fradkin:1978dv}
  E.~H.~Fradkin, S.~H.~Shenker,
  Phys.\ Rev.\  {\bf D19}, 3682-3697 (1979).

\bibitem{FileviezPerez:2010gw}
  P.~Fileviez Perez and M.~B.~Wise,
  Phys.\ Rev.\  D {\bf 82}, 011901 (2010).
  
\bibitem{Dulaney:2010dj}
  T.~R.~Dulaney, P.~Fileviez Perez and M.~B.~Wise,
  arXiv:1005.0617 [hep-ph].

\bibitem{Batell:2010bp}
  B.~Batell,
  arXiv:1007.0045 [hep-ph].


  
\bibitem{Georgi:1978xz}
  H.~Georgi,
  Hadronic J.\  {\bf 1}, 155 (1978).

\bibitem{Georgi:1982jb}
  H.~Georgi,
  Front.\ Phys.\  {\bf 54}, 1 (1982).

\bibitem{Perl:2001xi}
  M.~L.~Perl, P.~C.~Kim, V.~Halyo, E.~R.~Lee, I.~T.~Lee, D.~Loomba and K.~S.~Lackner,
  Int.\ J.\ Mod.\ Phys.\  A {\bf 16}, 2137 (2001).
  
\bibitem{Nakamura:2010zzi}
  K.~Nakamura  [Particle Data Group],
  J.\ Phys.\ G {\bf 37}, 075021 (2010).

\bibitem{Sarkar:1995dd}
  S.~Sarkar,
  Rept.\ Prog.\ Phys.\  {\bf 59}, 1493 (1996).

\bibitem{Iocco:2008va}
  F.~Iocco, G.~Mangano, G.~Miele, O.~Pisanti and P.~D.~Serpico,
  Phys.\ Rept.\  {\bf 472}, 1 (2009).

\bibitem{Wells:1994qy}
See, for example,
  J.~D.~Wells,
 [hep-ph/9404219].


\bibitem{Aprile:2010um} 
  E.~Aprile {\it et al.}  [XENON100 Collaboration],
  Phys.\ Rev.\ Lett.\  {\bf 105}, 131302 (2010).

\bibitem{Aalseth:2011wp} 
  C.~E.~Aalseth, P.~S.~Barbeau, J.~Colaresi, J.~I.~Collar, J.~Diaz Leon, J.~E.~Fast, N.~Fields and T.~W.~Hossbach {\it et al.},
  Phys.\ Rev.\ Lett.\  {\bf 107}, 141301 (2011).
  
  \bibitem{Ackermann:2011wa}
  M.~Ackermann {\it et al.}  [Fermi-LAT collaboration],
  Phys.\ Rev.\ Lett.\  {\bf 107}, 241302 (2011).
  
  \bibitem{MiraldaEscude:2000qt} 
  J.~Miralda-Escude,
  astro-ph/0002050.

\bibitem{Markevitch:2003at} 
  M.~Markevitch, A.~H.~Gonzalez, D.~Clowe, A.~Vikhlinin, L.~David, W.~Forman, C.~Jones and S.~Murray {\it et al.},
  Astrophys.\ J.\  {\bf 606}, 819 (2004).

\bibitem{Randall:2007ph} 
  S.~W.~Randall, M.~Markevitch, D.~Clowe, A.~H.~Gonzalez and M.~Bradac,
  Astrophys.\ J.\  {\bf 679}, 1173 (2008).

\bibitem{Rajaraman:2011wf} 
  A.~Rajaraman, W.~Shepherd, T.~M.~P.~Tait and A.~M.~Wijangco,
  arXiv:1108.1196 [hep-ph].

\bibitem{Peccei:1977hh} 
  R.~D.~Peccei and H.~R.~Quinn,
  Phys.\ Rev.\ Lett.\  {\bf 38}, 1440 (1977).


  \bibitem{Weinberg:1989dx} 
  S.~Weinberg,
  Phys.\ Rev.\ Lett.\  {\bf 63}, 2333 (1989).

\bibitem{Barr:1990vd}
  S.~M.~Barr and A.~Zee,
  Phys.\ Rev.\ Lett.\  {\bf 65}, 21 (1990)
  [Erratum-ibid.\  {\bf 65}, 2920 (1990)].
  
  \bibitem{Baker:2006ts}
    C.~A.~Baker {\it et al.},
  Phys.\ Rev.\ Lett.\  {\bf 97}, 131801 (2006)
  [arXiv:hep-ex/0602020].
  
  \bibitem{Regan:2002ta}
  B.~C.~Regan, E.~D.~Commins, C.~J.~Schmidt and D.~DeMille,
  Phys.\ Rev.\ Lett.\  {\bf 88}, 071805 (2002).
  
 \bibitem{Griffith:2009zz}
  W.~C.~Griffith, M.~D.~Swallows, T.~H.~Loftus, M.~V.~Romalis, B.~R.~Heckel and E.~N.~Fortson,
  Phys.\ Rev.\ Lett.\  {\bf 102}, 101601 (2009).
  
\bibitem{Aad:2011hz} 
  G.~Aad {\it et al.}  [ATLAS Collaboration],
  Phys.\ Lett.\ B {\bf 703}, 428 (2011).

\bibitem{Khachatryan:2011ts} 
  V.~Khachatryan {\it et al.}  [CMS Collaboration],
  JHEP {\bf 1103}, 024 (2011).
  
  \bibitem{Khachatryan:2010uf} 
  V.~Khachatryan {\it et al.}  [CMS Collaboration],
  Phys.\ Rev.\ Lett.\  {\bf 106}, 011801 (2011).
  
 
\bibitem{Aad:2011mb} 
  G.~Aad {\it et al.}  [ATLAS Collaboration],
  Phys.\ Lett.\ B {\bf 698}, 353 (2011).
  
  \bibitem{Fairbairn:2006gg} 
  M.~Fairbairn, A.~C.~Kraan, D.~A.~Milstead, T.~Sjostrand, P.~Z.~Skands and T.~Sloan,
  Phys.\ Rept.\  {\bf 438}, 1 (2007).
  
  


\bibitem{Aad:2009wy}
  G.~Aad {\it et al.}  [The ATLAS Collaboration],
  arXiv:0901.0512 [hep-ex].
  
  \bibitem{CMScuts}
  A. Rizzi et al. [CMS Collaboration], CMS-AN-2007/049.
  
   \bibitem{privateZ}
  Private discussion with P. Zalewski.
  
  \bibitem{Barbieri:2000gf} 
  R.~Barbieri and A.~Strumia,
  hep-ph/0007265.
  
  \bibitem{Lee:1977eg} 
  B.~W.~Lee, C.~Quigg and H.~B.~Thacker,
  Phys.\ Rev.\ D {\bf 16}, 1519 (1977).
  
  
\bibitem{Georgi:1974sy}
  H.~Georgi, S.~L.~Glashow,
  Phys.\ Rev.\ Lett.\  {\bf 32}, 438-441 (1974).
  
  \bibitem{Pati:1974yy}
  J.~C.~Pati, A.~Salam,
  Phys.\ Rev.\  {\bf D10}, 275-289 (1974).
  
  \bibitem{Katz:2005au} 
  E.~Katz, A.~E.~Nelson and D.~G.~E.~Walker,
  JHEP {\bf 0508}, 074 (2005).
  
  \bibitem{Katz:2003sn} 
  E.~Katz, J.~-y.~Lee, A.~E.~Nelson and D.~G.~E.~Walker,
  JHEP {\bf 0510}, 088 (2005).

  
  \bibitem{Callan:1969sn} 
  C.~G.~Callan, Jr., S.~R.~Coleman, J.~Wess and B.~Zumino,
  Phys.\ Rev.\  {\bf 177}, 2247 (1969).
  
  \bibitem{postpone} 
  D.~G.~E.~Walker, et.~.al., \textit{to appear}. 
  
  \bibitem{Cheng:2003ju} 
  H.~-C.~Cheng and I.~Low,
  JHEP {\bf 0309}, 051 (2003)
  [hep-ph/0308199].
  

\bibitem{Kirzhnits:1972ut}
  D.~A.~Kirzhnits and A.~D.~Linde,
  Phys.\ Lett.\  B {\bf 42}, 471 (1972).

\bibitem{Weinberg:1974hy}
  S.~Weinberg,
  Phys.\ Rev.\  D {\bf 9}, 3357 (1974).





  \bibitem{Kado:2002er}
  M.~M.~Kado and C.~G.~Tully,
  Ann.\ Rev.\ Nucl.\ Part.\ Sci.\  {\bf 52}, 65 (2002).

  
  \bibitem{Riotto:1999yt}
  A.~Riotto and M.~Trodden,
  Ann.\ Rev.\ Nucl.\ Part.\ Sci.\  {\bf 49}, 35 (1999).
  
  \bibitem{Trodden:1998ym}
  M.~Trodden,
  Rev.\ Mod.\ Phys.\  {\bf 71}, 1463 (1999).
 
 \bibitem{Georgi:1989xz}
  H.~Georgi, E.~E.~Jenkins and E.~H.~Simmons,
  Nucl.\ Phys.\  B {\bf 331}, 541 (1990).
  
  \bibitem{Georgi:1989ic}
  H.~Georgi, E.~E.~Jenkins and E.~H.~Simmons,
  Phys.\ Rev.\ Lett.\  {\bf 62}, 2789 (1989)
  [Erratum-ibid.\  {\bf 63}, 1540 (1989)].





  
  \bibitem{Dolan:1973qd}
  L.~Dolan and R.~Jackiw,
  Phys.\ Rev.\  D {\bf 9}, 3320 (1974).
  

\bibitem{Shelton:2010ta}
  J.~Shelton and K.~M.~Zurek,
  arXiv:1008.1997 [hep-ph].


\bibitem{Haba:2010bm} 
  N.~Haba and S.~Matsumoto,
ÊÊProg.\ Theor.\ Phys.\  {\bf 125}, 1311 (2011).
ÊÊ

\bibitem{vonHarling:2012yn} 
  B.~von Harling, K.~Petraki and R.~R.~Volkas,
ÊÊarXiv:1201.2200 [hep-ph].
ÊÊ

\bibitem{Petraki:2011mv} 
  K.~Petraki, M.~Trodden and R.~R.~Volkas,
ÊÊarXiv:1111.4786 [hep-ph].
ÊÊ

\bibitem{Barr:2011cz} 
  S.~M.~Barr,
ÊÊPhys.\ Rev.\ D {\bf 85}, 013001 (2012).
Ê
ÊÊ

\bibitem{Frandsen:2011kt} 
  M.~T.~Frandsen, S.~Sarkar and K.~Schmidt-Hoberg,
ÊÊPhys.\ Rev.\ D {\bf 84}, 051703 (2011).
Ê
ÊÊ

\bibitem{McDonald:2012vw} 
  J.~McDonald,
ÊÊarXiv:1201.3124 [hep-ph].
ÊÊ

\bibitem{Kamada:2012ht} 
  K.~Kamada and M.~Yamaguchi,
ÊÊarXiv:1201.2636 [hep-ph].
ÊÊ

\bibitem{Cui:2011ab} 
  Y.~Cui, L.~Randall and B.~Shuve,
ÊÊarXiv:1112.2704 [hep-ph].
ÊÊ

\bibitem{Davoudiasl:2011aa} 
  H.~Davoudiasl and I.~Lewis,
ÊÊarXiv:1112.1939 [hep-ph].
ÊÊ

\bibitem{D'Eramo:2011ec} 
  F.~D'Eramo, L.~Fei and J.~Thaler,
ÊÊarXiv:1111.5615 [hep-ph].
ÊÊ

\bibitem{Chowdhury:2011ga} 
  T.~A.~Chowdhury, M.~Nemevsek, G.~Senjanovic and Y.~Zhang,
ÊÊarXiv:1110.5334 [hep-ph].
ÊÊ

\bibitem{McDonald:2011sv} 
  J.~McDonald,
ÊÊarXiv:1108.4653 [hep-ph].
ÊÊ

\bibitem{Graesser:2011vj} 
  M.~L.~Graesser, I.~M.~Shoemaker and L.~Vecchi,
ÊÊarXiv:1107.2666 [hep-ph].
ÊÊ

\bibitem{Kumar:2011np} 
  P.~Kumar and E.~Ponton,
ÊÊJHEP {\bf 1111}, 037 (2011).
ÊÊ

\bibitem{Cui:2011qe} 
  Y.~Cui, L.~Randall and B.~Shuve,
ÊÊJHEP {\bf 1108}, 073 (2011).
ÊÊ
ÊÊ

\bibitem{MarchRussell:2011fi} 
  J.~March-Russell and M.~McCullough,
ÊÊarXiv:1106.4319 [hep-ph].
ÊÊ

\bibitem{Cheung:2011if} 
  C.~Cheung and K.~M.~Zurek,
ÊÊPhys.\ Rev.\ D {\bf 84}, 035007 (2011).
ÊÊ

\bibitem{Bell:2011tn} 
  N.~F.~Bell, K.~Petraki, I.~M.~Shoemaker and R.~R.~Volkas,
ÊÊPhys.\ Rev.\ D {\bf 84}, 123505 (2011).
Ê
ÊÊ


\bibitem{Heckman:2011sw} 
  J.~J.~Heckman and S.~-J.~Rey,
ÊÊJHEP {\bf 1106}, 120 (2011).
Ê
ÊÊ


\bibitem{Falkowski:2011xh} 
  A.~Falkowski, J.~T.~Ruderman and T.~Volansky,
ÊÊJHEP {\bf 1105}, 106 (2011).
ÊÊ


\bibitem{Hall:2010jx} 
  L.~J.~Hall, J.~March-Russell and S.~M.~West,
ÊÊarXiv:1010.0245 [hep-ph].
ÊÊ

\bibitem{Behbahani:2010xa} 
  S.~R.~Behbahani, M.~Jankowiak, T.~Rube and J.~G.~Wacker,
ÊÊAdv.\ High Energy Phys.\  {\bf 2011}, 709492 (2011).
ÊÊ

\bibitem{Blennow:2010qp} 
  M.~Blennow, B.~Dasgupta, E.~Fernandez-Martinez and N.~Rius,
ÊÊJHEP {\bf 1103}, 014 (2011).
ÊÊ

\bibitem{Buckley:2010ui} 
  M.~R.~Buckley and L.~Randall,
ÊÊJHEP {\bf 1109}, 009 (2011).
ÊÊ

\bibitem{McDonald:2011zz} 
  J.~McDonald,
ÊÊPhys.\ Rev.\ D {\bf 83}, 083509 (2011).
ÊÊ

\bibitem{Davoudiasl:2010am} 
  H.~Davoudiasl, D.~E.~Morrissey, K.~Sigurdson and S.~Tulin,
ÊÊPhys.\ Rev.\ Lett.\  {\bf 105}, 211304 (2010).
ÊÊ

\end{thebibliography}
\end{document}